\documentclass[fleqn,twocolumns,traditabstract,longauth]{aa}
\usepackage[english]{babel}
\usepackage{graphicx,amsmath}
\usepackage{epsf,color}
\usepackage[mathscr]{eucal}
\usepackage{amsmath}
\usepackage{amssymb,amsfonts}
\usepackage{natbib}
\usepackage{multicol}
\usepackage{txfonts}
\usepackage[labelfont=bf]{caption}
\usepackage{dsfont} 
\usepackage{verbatim}
\definecolor{Mygreen}{rgb}{0.00, 0.72, 0.0}
\definecolor{Mypink}{rgb}{1.0, 0.0, 0.5}
\usepackage[breaklinks, citecolor=blue, linkcolor=Mygreen, urlcolor=Mypink, colorlinks=true, debug, baseurl=' ']{hyperref}
\usepackage{float}
\usepackage{ulem}
\usepackage{color}
\usepackage{widetext}
\usepackage{appendix}
\usepackage{colortbl}
\usepackage{stfloats}
\usepackage{placeins}

\usepackage{subcaption}

\definecolor{pastelorange}{rgb}{1.0, 0.7, 0.28}

\bibpunct{(}{)}{;}{a}{}{,}
\begin{document}
\title{NIKA2 Cosmological Legacy Survey}
\subtitle{Blind detection of galaxy clusters in the COSMOS field via the Sunyaev-Zel’dovich effect}
\titlerunning{Blind detection of galaxy clusters with NIKA2}
\authorrunning{D.~Chérouvrier, J.~F.~Mac\'ias-P\'erez et al.}
\author{D.~Chérouvrier\inst{\ref{LPSC}}\thanks{Corresponding author: \url{cherouvrier@lpsc.in2p3.fr}} 
     \and   J.~F.~Mac\'ias-P\'erez\inst{\ref{LPSC}}
     \and   F.~X.~D\'esert\inst{\ref{IPAG}}
     \and   R.~Adam \inst{\ref{OCA}}
     \and  P.~Ade \inst{\ref{Cardiff}}
     \and  H.~Ajeddig \inst{\ref{CEA}}
     \and  S.~Amarantidis \inst{\ref{IRAME}}
     \and  P.~Andr\'e \inst{\ref{CEA}}
     \and  H.~Aussel \inst{\ref{CEA}}
     \and  R.~Barrena \inst{\ref{tenerife1}, \ref{tenerife2}}
     \and  A.~Beelen \inst{\ref{LAM}}
     \and  A.~Beno\^it \inst{\ref{Neel}}
     \and  S.~Berta \inst{\ref{IRAMF}}
     \and  M. B\'ethermin \inst{\ref{Strasbourg}, \ref{LAM}}
     \and  A.~Bongiovanni \inst{\ref{IRAME}}
     \and  J.~Bounmy \inst{\ref{LPSC}}
     \and  O.~Bourrion \inst{\ref{LPSC}}
     \and  L.~-J.~Bing \inst{\ref{Sussex}}
     \and  M.~Calvo \inst{\ref{Neel}}
     \and  A.~Catalano \inst{\ref{LPSC}}
     \and  M.~De~Petris \inst{\ref{Roma}}
     \and  S.~Doyle \inst{\ref{Cardiff}}
     \and  E.~F.~C.~Driessen \inst{\ref{IRAMF}}
     \and  G.~Ejlali \inst{\ref{Teheran}}
     \and  A.~Ferragamo \inst{\ref{Roma}}
     \and  M.~Fern\'andez-Torreiro\inst{\ref{LPSC}}
     \and  A.~Gomez \inst{\ref{CAB}} 
     \and  J.~Goupy \inst{\ref{Neel}}
     \and  C.~Hanser \inst{\ref{LPSC}}
     \and  S.~Katsioli \inst{\ref{AthenObs}, \ref{AthenUniv}}
     \and  F.~K\'eruzor\'e \inst{\ref{Argonne}}
     \and  C.~Kramer \inst{\ref{IRAMF}}
     \and  B.~Ladjelate \inst{\ref{IRAME}} 
     \and  G.~Lagache \inst{\ref{LAM}}
     \and  S.~Leclercq \inst{\ref{IRAMF}}
     \and  J.-F.~Lestrade \inst{\ref{LERMA}}
     \and  S.~C.~Madden \inst{\ref{CEA}}
     \and  A.~Maury \inst{\ref{Barcelona1}, \ref{Barcelona2}, \ref{CEA}}
     \and  F.~Mayet \inst{\ref{LPSC}}
     \and  J.~-B.~Melin \inst{\ref{IRFU}}
     \and  A.~Monfardini \inst{\ref{Neel}}
     \and  A.~Moyer-Anin \inst{\ref{LPSC}}
     \and  M.~Mu\~noz-Echeverr\'ia \inst{\ref{IRAP}}
     \and  I.~Myserlis \inst{\ref{IRAME}}
     \and  R.~Neri \inst{\ref{IRAMF}}
     \and  A.~Paliwal \inst{\ref{Roma2}}
     \and  L.~Perotto \inst{\ref{LPSC}}
     \and  G.~Pisano \inst{\ref{Roma}}
     \and  E.~Pointecouteau \inst{\ref{IRAP}}
     \and  N.~Ponthieu \inst{\ref{IPAG}, \ref{LAM}}
     \and  G.~W.~Pratt \inst{\ref{CEA}}
     \and  V.~Rev\'eret \inst{\ref{CEA}}
     \and  A.~J.~Rigby \inst{\ref{Leeds}}
     \and  A.~Ritacco \inst{\ref{LPSC}}
     \and  H.~Roussel \inst{\ref{IAP}}
     \and  F.~Ruppin \inst{\ref{IP2I}}
     \and  M.~S\'anchez-Portal \inst{\ref{IRAME}}
     \and  S.~Savorgnano \inst{\ref{LPSC}}
     \and  K.~Schuster \inst{\ref{IRAMF}}
     \and  A.~Sievers \inst{\ref{IRAME}}
     \and  C.~Tucker \inst{\ref{Cardiff}}
     \and  R.~Zylka \inst{\ref{IRAMF}}
     }
   \institute{
     Univ. Grenoble Alpes, CNRS, Grenoble INP, LPSC-IN2P3, 53, avenue des Martyrs, 38000 Grenoble, France
     \label{LPSC}
     \and
     Univ. Grenoble Alpes, CNRS, IPAG, 38000 Grenoble, France 
     \label{IPAG}
     \and
     Universit\'e C\^ote d'Azur, Observatoire de la C\^ote d'Azur, CNRS, Laboratoire Lagrange, France 
     \label{OCA}
     \and
     School of Physics and Astronomy, Cardiff University, Queen’s Buildings, The Parade, Cardiff, CF24 3AA, UK 
     \label{Cardiff}
     \and
     Universit\'e Paris-Saclay, Universit\'e Paris Cit\'e, CEA, CNRS, AIM, F-91191, Gif-sur-Yvette, France
     \label{CEA}
     \and
     Institut de Radioastronomie Millim\'etrique (IRAM), Avenida Divina Pastora 7, Local 20, E-18012, Granada, Spain
     \label{IRAME} 
     \and 
     Instituto de Astrofísica de Canarias, C/ Vía Láctea s/n, E-38205 La Laguna, Tenerife, Spain
     \label{tenerife1}
     \and
     Universidad de La Laguna, Departamento de Astrofísica, E-38206 La Laguna, Tenerife, Spain
     \label{tenerife2}
     \and	
     Aix Marseille Univ, CNRS, CNES, LAM (Laboratoire d'Astrophysique de Marseille), Marseille, France
     \label{LAM}
     \and
     Institut N\'eel, CNRS, Universit\'e Grenoble Alpes, France
     \label{Neel}
     \and
     Institut de Radioastronomie Millim\'etrique (IRAM), 300 rue de la Piscine, 38400 Saint-Martin-d’Hères, France
     \label{IRAMF}
     \and 
     Université de Strasbourg, CNRS, Observatoire astronomique de Strasbourg, UMR 7550, 67000 Strasbourg, France
     \label{Strasbourg}
     \and
     Astronomy Centre, Department of Physics and Astronomy, University of Sussex, Brighton BN1 9QH
     \label{Sussex}
     \and 
     Dipartimento di Fisica, Sapienza Universit\`a di Roma, Piazzale Aldo Moro 5, I-00185 Roma, Italy
     \label{Roma}
     \and
     Institute for Research in Fundamental Sciences (IPM), School of Astronomy, Tehran, Iran
     \label{Teheran}
     \and
     Centro de Astrobiolog\'ia (CSIC-INTA), Torrej\'on de Ardoz, 28850 Madrid, Spain
     \label{CAB}
     \and
     National Observatory of Athens, Institute for Astronomy, Astrophysics, Space Applications and Remote Sensing, Ioannou Metaxa
     and Vasileos Pavlou GR-15236, Athens, Greece
     \label{AthenObs}
     \and
     Department of Astrophysics, Astronomy \& Mechanics, Faculty of Physics, University of Athens, Panepistimiopolis, GR-15784
     Zografos, Athens, Greece
     \label{AthenUniv}
     \and
     High Energy Physics Division, Argonne National Laboratory, 9700 South Cass Avenue, Lemont, IL 60439, USA
     \label{Argonne}
     \and  
     LERMA, Observatoire de Paris, PSL Research University, CNRS, Sorbonne Universit\'e, UPMC, 75014 Paris, France  
     \label{LERMA}
     \and
    Institute of Space Sciences (ICE), CSIC, Campus UAB, Carrer de Can Magrans s/n, E-08193, Barcelona, Spain
    \label{Barcelona1}
    \and
    ICREA, Pg. Lluís Companys 23, Barcelona, Spain
    \label{Barcelona2}
    \and
     Universit{\'e} Paris-Saclay, CEA, D{\'e}partement de Physique des Particules, 91191, Gif-sur-Yvette, France
     \label{IRFU}
     \and
     IRAP, CNRS, Université de Toulouse, CNES, UT3-UPS, (Toulouse), France 
     \label{IRAP}
     \and
     Dipartimento di Fisica, Universit\`a di Roma ‘Tor Vergata’, Via della Ricerca Scientifica 1, I-00133 Roma, Italy	
     \label{Roma2}
     \and
     School of Physics and Astronomy, University of Leeds, Leeds LS2 9JT, UK
     \label{Leeds}
     \and
     Laboratoire de Physique de l’\'Ecole Normale Sup\'erieure, ENS, PSL Research University, CNRS, Sorbonne Universit\'e, Universit\'e de Paris, 75005 Paris, France 
     \label{ENS}
     \and
     INAF-Osservatorio Astronomico di Cagliari, Via della Scienza 5, 09047 Selargius, IT
     \label{INAF}
     \and    
     Institut d'Astrophysique de Paris, CNRS (UMR7095), 98 bis boulevard Arago, 75014 Paris, France
     \label{IAP}
     \and
     University of Lyon, UCB Lyon 1, CNRS/IN2P3, IP2I, 69622 Villeurbanne, France
     \label{IP2I}
     \and
     University Federico II, Naples, Italy
     \label{Naples}}

\date{Received March 10, 2025 / Accepted June 2, 2025}

\abstract{
Clusters of galaxies, formed in the latest stages of structure formation, are unique cosmological probes. With the advent of large CMB surveys like those from the \textit{Planck} satellite, the ACT and SPT telescopes, we now have access to a large number of galaxy clusters detected at millimeter wavelengths via the thermal Sunyaev-Zel'dovich (tSZ) effect. Nevertheless, it is interesting to complement them with high-angular-resolution (tens of arcseconds) observations to target the lowest-mass and highest-redshift clusters. This is the case of observations with the NIKA2 camera, which is installed on the IRAM 30--m telescope in Pico Veleta, Spain.
We used the existing 150~GHz (2\,mm) data from the NIKA2 Cosmological Legacy Survey (N2CLS) Large Program to blindly search for galaxy clusters in the well-known COSMOS field, across a 877 arcmin$^2$ region centered on (R.A., Dec.)$_{J2000}$ = (10h00m28.81s, +02d17m30.44s).
We first developed a dedicated data reduction pipeline to construct NIKA2 maps at 2\,mm. We then used a matched-filter algorithm to extract cluster candidates assuming a universal pressure profile to model the expected cluster tSZ signal. We computed the purity and completeness of the sample by applying the previous algorithm to simulated maps of the sky signal in the COSMOS field, including tSZ contribution, point sources and instrumental noise.
We find a total of 16 cluster candidates at S/N $>4$, from which eight have either an optical or X-ray cluster (or group of galaxies) counterpart. This is the first blind detection of clusters of galaxies at mm wavelengths at 18$\arcsec$ angular resolution. For candidates with available redshift estimates, we derived their mass by modeling the cluster tSZ signal with a universal pressure profile via a MCMC analysis. From this analysis, we confirm that NIKA2 and the IRAM 30--m telescope should be sensitive to low-mass clusters at intermediate and high redshift, complementing current and planned large tSZ-based cluster surveys. 
}

\maketitle
\section{Introduction}
\label{sec:introduction}

\indent Clusters of galaxies are formed in the latest stages of structure formation and are unique cosmological probes to study structure formation and its evolution \citep{voit05,allen11}. In particular, cluster number counts as a function of mass and redshift are sensitive to the universe matter content and root-mean-square (rms) fluctuations via the $\Omega_{m}$ and $\sigma_8$ cosmological parameters \citep[see][]{Planck_XXVII}.
Clusters are intricate systems comprising dark matter, galaxies, and hot-ionized gas known as the intracluster medium (ICM), making them observable across multiple wavelengths of the electromagnetic spectrum. Catalogs of clusters of galaxies have thus been built by surveys at different wavelengths, primarily in the optical and infrared \citep{Abell, Wen_2012, Wen_2018, Bleem_2015, Oguri_2018, Gonzalez_2019}, through the detection of galaxy overdensities, but also in X-ray \citep{EMSS_survey, XMM_survey, Adami_2018, Klein_2019} via the bremsstrahlung emission of hot electrons in the ICM. Furthermore, the last decade has seen a major advance in the study of clusters at millimeter wavelengths \citep{Planck_XXVII, ACT, SPT_new} via the Sunyaev-Zel'dovich effect \citep[SZ,][]{SZ_paper}.\\
\indent The thermal Sunyaev-Zel'dovich (tSZ) effect, resulting from the inverse Compton scattering of cosmic microwave background (CMB) photons by hot-thermal electrons in the ICM, is a powerful tool for detecting high-redshift clusters. Unlike other observables, it is not impacted by cosmological dimming, with the cluster's size being the only limiting factor. Furthermore, the cluster size is related to the cluster mass as clusters are expected to be self-similar objects \citep{A10, allen11}. Therefore, we expect low-mass, high-redshift objects to be small in size but detectable via the tSZ effect.\\
\indent The first successful targeted measurements of the tSZ effect were obtained in the 1970s \citep{Pariiskii_SZ, Gull_SZ}, with the first blind detection of tSZ clusters following three decades later \citep{first_blind}. Current blind cluster catalogs of order 10$^3$-10$^4$ clusters have been obtained using instruments dedicated to CMB observations -- the \textit{Planck} satellite \citep{Planck_2011_VIII, Planck_2013_XXIX, Planck_2013_XXXII, Planck_XXVII}, the ground-based South Pole Telescopes \citep[SPT,][]{SPT_2011, SPT_2015, SPT-ECS, SPT_new, SPT_2025}, and the Atacama Cosmology Telescope \citep[ACT,][]{ACT_2011, ACT_2013, ACT_2018, ACT}. However, they have relatively poor angular resolutions: 5 arcmin for \textit{Planck}, and about 1 arcmin for SPT and ACT.
As a consequence, there is a need for high-angular-resolution instruments (tens of arcseconds) in order to map the tSZ signal for high-redshift and/or low-mass clusters of galaxies, which are expected to be of great cosmological interest \citep[see for example][]{Tinker_hmf}. In this respect, observations of the tSZ effect with the NIKA2 camera are particularly interesting.\\
\indent The New IRAM KID Array \citep[NIKA2, ][]{Bourrion_NIKA2, Calvo_NIKA2, Adam_NIKA2} is a dual-band millimeter camera consisting of three arrays of kinetic inductance detectors (KIDs) installed at the IRAM 30--m telescope in Granada, Spain. NIKA2 operates at 150 (2\,mm) and 260~GHz (1.2\,mm), where the tSZ signal is strongly negative and slightly positive, respectively.
NIKA2 has a field of view of 6.5$\arcmin$ and offers a high angular resolution of 17.6$\arcsec$ at 2\,mm and 11.1$\arcsec$ at 1.2\,mm \citep[for a full review of the performance of the instrument see][]{Perotto_Pipeline}. As part of the NIKA2 guaranteed time, the NIKA2 Cosmological Legacy Survey (N2CLS) acquired 300 hours of observations distributed between the GOODS-N and COSMOS fields \citep{N2CLS_count, ponthieu2025}. The main goal of the survey is to study dust-obscured galaxies up to very high redshift, but thanks to the depth and very high angular resolution of the observations, the detection of galaxy clusters via the tSZ effect is also possible. The N2CLS catalog of dusty star-forming galaxies will be published in Béthermin et al. (in prep.).\\
\indent Here we focus on the COSMOS field \citep{COSMOS_survey, COSMOS2015, COSMOS2020}, where the NIKA2 observations cover a larger area and a large number of ancillary data sets are available. 
This part of the sky is particularly well adapted for this project as it is widely visible, and also uncontaminated by bright X-ray, UV, and radio sources as well as massive tSZ clusters. The field has been thoroughly observed by both space and ground-based experiments such as the XMM-\textit{Newton}, HST, \textit{Chandra}, Spitzer, Herschel, Keck, Subaru, ESO-VLT, and CFHTLS telescopes \citep[see][]{COSMOS_survey}. This wealth of data enabled the assembly of cluster catalogs in COSMOS, such as the ALHAMBRA catalog \citep{ALH_survey}, the CES catalog \citep{CES_catalog}, the HIROCS survey \citep{HIROCS}, and other studies \citep{Bellagamba_2011}. Clusters have also been detected in this region by larger surveys like the SDSS \citep{Wen_2012}, DESI \citep{DESI_survey}, XMM \citep{XMM_survey}, CFHTLS \citep{Adami_2011}, KiDS \citep{KIDS_survey}, and Subaru Weak-Lensing \citep{Subaru_survey} surveys. However, at the time of writing this article, there are no detections of SZ clusters in the NIKA2 COSMOS field. The new N2CLS data are thus quite promising, as we are able to match any cluster candidate with the available catalogs and identify optical or X-ray counterparts.\\
\indent In this paper, we present the blind detection of clusters via the SZ effect with the NIKA2 camera in the COSMOS field. The paper is structured as follows. In Section \ref{Observations}, we present the data used and the data reduction pipeline. Section \ref{Detection_method} describes the blind cluster detection algorithm and the cluster sample identification process.
In Section \ref{Simulations}, we discuss the simulation framework used to characterize our sample. Finally, Sections \ref{Cluster_candidates} and \ref{sec:candidates_properties} present the candidate clusters compared to other multiwavelength cluster catalogs and their properties.
We summarize and draw conclusions in Section \ref{Conclusion}.\\
\indent Throughout this article, we assume a flat $\Lambda$CDM cosmology model \citep[e.g.,][]{Planck18_cosmo} with $\Omega_m$ = 0.31, $\Omega_{\Lambda}$ = 0.69, and Hubble constant H$_0$ = 67.7 km\,s$^{-1}$\,Mpc$^{-1}$. All masses are expressed using the quantity $M_{500}$, which represents the mass contained within a radius $R_{500}$, where the average density equals 500 times the critical density at the cluster redshift.  

\section{The NIKA2 maps of COSMOS}
\label{Observations}
\subsection{Observations}

\indent The N2CLS data set of the COSMOS field corresponds to a total of 195 hours of observations carried out from October 2017 to January 2023. Two groups of 27.0\arcmin×34.7\arcmin\,and 35.0\arcmin×28.0\arcmin\,raster scans, centered on (R.A., Dec.)$_{J2000}$ = (10h00m28.81s, +02d17m30.44s), were executed for a total area of $\sim$ 1400 arcmin$^2$. The two groups of scans were performed with a speed of 60 \arcsec /s at position angles of 0 and 90 degrees in the RA–Dec coordinate system of the telescope \citep{N2CLS_count}. In this paper, we concentrate on the 2\,mm data, where the signal-to-noise ratio (S/N) for the tSZ signal is larger. The 1.2\,mm data were only used for robustness tests and are not discussed in this paper. \\

\subsection{Data reduction pipeline}
\label{Data_reduction_pipeline}

\begin{figure}[H]
	\centering
	\includegraphics[width=0.5\textwidth]{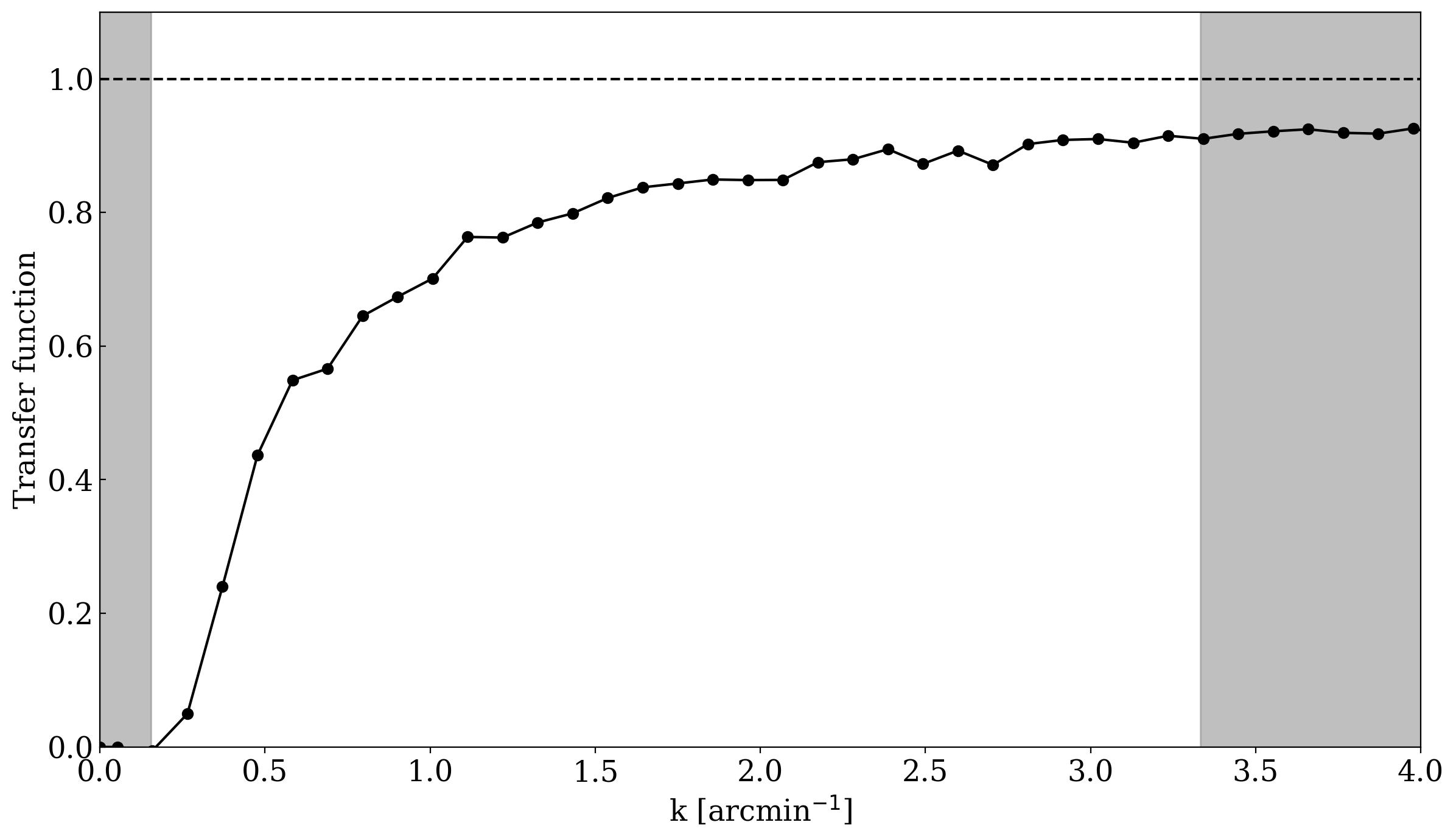}
	\caption{Transfer function as a function of angular scale for the pipeline used to construct the NIKA2 2\,mm COSMOS maps discussed in this paper. The shaded gray regions represent the NIKA2 FoV (left) and instrumental beam (right).}
\label{fig:TF_plot}
\end{figure}

\indent The raw NIKA2 data per detector array were reduced using the data reduction pipeline described in \citet{ponthieu2025}. It is an evolution of that of \citet{Perotto_Pipeline} and is optimized for the detection of faint and small angular size objects. We here summarize its main steps.
The raw time-ordered information (TOI) data were first calibrated following the baseline procedure outlined in \citet{Perotto_Pipeline}. We first transformed the NIKA2 raw data into KID resonance frequency shift \citep{Calvo_NIKA2}, and then Uranus was used as a primary calibrator accounting for line-of-sight opacity absorption. Furthermore, the TOIs were corrected from detector correlated atmospheric and electronic noise using a decorrelation procedure. For each 45s subscan, we fitted the TOI of each detector, $k$, to the following model:
\begin{multline}
D_k(t) = S^{sky}_k(t) + a_k \cdot S^{atm}(t) + b_k \cdot S^{atm'}(t) + d_k \cdot (S^{atm}(t))^2 \\
      + \sum_{i} e_{k,i} \cdot T^{elec}_{i}(t) + \sum^{N_{harms}}_{h=0} \left(A_{h,k} \cos(w_h t) + B_{h,k} \sin(w_h t) \right) + c_k + c'_{k} \cdot t\,,
\end{multline}

\noindent where $S^{sky}_{k}$ is the astronomical signal at the position observed by detector \textit{k}. $S^{atm}$ is a template of the atmospheric signal built from all detectors of a given array, and $S^{atm'}$ its time derivative to account for gradients across the array. The square of the atmospheric model was also added to account for possible nonlinearities of a given KID. To account for electronic noise, we used off-resonance measurements \citep[see][]{Catalano_NIKA_off_res}. Off-resonance measurements are sampled as the detector signal and come from reading tone frequencies set apart from any KID resonance. As the electronic noise is strongly correlated between detectors and the off-resonance measurements, the electronic noise templates, $T^{elec}_i$, were constructed by coadding all off-resonances that belong to the same electronic box $i$. The harmonic term consists of a series of cosine and sine terms with $N_{harms}=8$, which for a 27 second subscan is equivalent to a high-pass filtering above 1/3 Hz. The final terms $c_k + c'_{k} \cdot t$ allow a linear baseline, where $t$ is the time. For each KID $k$, a single linear regression was performed per subscan, to determine the coefficients $a_k$, $b_k$, $c_k$, $c'_k$, $d_k$, $e_{k,i}$, $A_{h,k}$, and $B_{h,k}$, once the templates $S^{atm}$, $S^{atm'}$, and $T^{elec}_i$ were computed.\\
\indent We observed some residuals in maps that are projected in Nasmyth coordinates, with striping corresponding to the electronic box orientation at 1.2\,mm. We decided to correct for this pattern with one template per array.\\
\indent The cleaned TOIs were then projected on a grid and coadded with weights depending on their noise level, using an inverse variance weighting scheme, to produce the final maps per array. The decorrelation method needs to be performed iteratively because the sky signal can impact the procedure. By subtracting from $S^{sky}_k$ the high S/N point source signal from the previous iteration, we can improve the quality of the final maps. We have found that only few iterations are needed for this process to converge.\\
\indent The processing parameters (basically the high-pass filtering) were tuned to optimize the detection of point and compact sources, but the filtering is mild enough to allow slightly extended sources to be detected, in particular the SZ effect of compact clusters or groups. We found that the decorrelation procedure induces a filtering at large scales on the final maps (due to the final harmonic term) that we need to take into account in our analysis (see Sect.~\ref{Match_filtered}). We ran a white noise simulation through the pipeline and computed the transfer function as the ratio between the power spectra of the input white noise map and of the pipeline output. The transfer function is shown in Fig.~\ref{fig:TF_plot} as a function of the inverse of the angular scale, $k$. We observe significant filtering for $k<0.5$ arcmin$^{-1}$.

\subsection{Final sky maps}
\label{Final_sky_maps}
\indent The final 2\,mm map was constructed combining the signal from all available observation scans for the final iteration step of the pipeline. In addition, null maps, or jackknife \citep[JK, ][]{JK} maps, were constructed by alternately multiplying scan pairs by +1, -1 and coadding them. This effectively subtracts the astrophysical signal to obtain realistic noise maps. These noise maps are used in our cluster detection algorithm presented in Sect.~\ref{Detection_method}.\\
\indent We decided to limit our analysis to the region inside the black line shown in Fig.~\ref{Map_with_candidates}, discarding the very noisy edges of the map. We chose a region of 877 arcmin$^2$ where the noise is below 1.2 times the median noise value over the full map. 

In \cite{N2CLS_count}, the same COSMOS data were reduced using the PIIC software \citep{PIIC_soft}, and optimized for positive point sources detection. We thus have access to two final sky maps, one produced by PIIC, and one produced using the pipeline described in Sect.~\ref{Data_reduction_pipeline}. 
In the following, we use the map produced with the IDL pipeline described above, which is better adapted for the detection of extended negative signal. Furthermore, we extensively used simulations (see Sect.~\ref{Simulations}) that required full access to the code, which was more difficult to achieve with the PIIC software.\\
\indent The 2\,mm signal map will be published in a forthcoming publication (Carvajal et al. in prep.). Here we present only zoom-in maps of patches around cluster candidates (see Sect.~\ref{sec:candidates_properties} and Append.~\ref{app:clustercandidatesample}). \\

\section{Cluster detection}
\label{Detection_method}
\subsection{tSZ signal modeling}
\label{subsec:spatialclustermodel}

\indent The tSZ cluster contribution to the NIKA2 2\,mm maps can be modeled as 
\begin{equation}
S_{2 \ \mathrm{mm}}^{tSZ} (x)= C_{ 2 \ \mathrm{mm}}^{tSZ}  \  y(x) \,,
\label{eq:tSZdef}
\end{equation}
where $x$ is the spatial dependency and $C_{ 2 \ \mathrm{mm}}^{tSZ}$ accounts for the tSZ spectrum in Jy/beam units for the 2\,mm NIKA2 band \citep[see ][]{2018A&A...615A.112R,2020A&A...644A..93K,2023A&A...671A..28M}.
The Compton parameter, $y$ is given by the integration along the line of sight, $l$, of the cluster electron pressure profile $P_{e}$:
\begin{equation}
y = \frac{\sigma_{T}}{m_{e} \ c^2} \int P_{e}\, \mathrm{dl} \,,
\label{eq:comptonparameter}
\end{equation}
where $\sigma_{T}$, $m_{e}$, and $c$ are the Thomson cross-section, the electron rest mass, and the speed of light. \\

In the following, we assume spherically symmetric clusters and the universal pressure profile (UPP) model defined in \cite{A10} as
\begin{flalign}
\label{eq:A10}
  \begin{aligned}
P_e(r) = & 1.65 \: \times \: 10^{-3} \; h(z)^{8/3} \; \left[\frac{M_{500}}{3\times10^{14} \; h_{70}^{-1}M_{\odot}}\right]^{2/3+0.12} \\
& \times \; p(x) \; h_{70}^2 \; \mathrm{keV \, cm^{-3}} \,,
  \end{aligned}&&&
\end{flalign}
where $h_{70} = \frac{H_0}{70\;\mathrm{km\;s^{-1}\;Mpc^{-1}}}$ , $h(z)$  the dimensionless Hubble parameter, and $p(x)$ is the generalized Navarro-Frenk-White (gNFW) model proposed by \cite{Nagai_gnfw}:

\begin{equation}
p(x) = \frac{P_0}{(c_{500}x)^\gamma \; [1 + (c_{500}x)^\alpha]^{(\beta-\gamma)/\alpha}}\,,
\end{equation}
with $P_0$ a normalization constant, $\gamma$ and $\beta$ the inner and outer slopes of the profile, respectively, $c_{500} = \frac{R_{500}}{r_s}$ the concentration where $r_s$ is the transition radius between the inner and outer regimes of the profile, $\alpha$ the slope of the transition, and $x = \frac{r}{R_{500}}$ a normalized radius. 
We considered as free parameters of the model only the cluster redshift, $z$, and the mass $M_{500}$. The rest of the parameters were fixed to the values derived in \citet{A10}:
   $$ [P_0, c_{500}, \gamma, \alpha, \beta] = [8.403 \; h_{70}^{-3/2}, 1.177, 0.3081, 1.0510, 5.4905] $$
We also assume the following scaling relation between the integrated cluster mass and the integrated Compton parameter at $R_{500}$:
\begin{equation}
\label{eq:y500}
    Y_{500} = 2.925\times10^{-5} I(1) \; h(z)^{2/3} \left[\frac{M_{500}}{3\times10^{14}h_{70}^{-1} M_{\odot}}\right]^{5/3+0.12} \; h_{70}^{-1} \; \mathrm{Mpc^2} \,,
\end{equation}
where $I(1) = 0.6145$, and define the angular size of the cluster as
\begin{equation}
\label{eq:theta500}
    \theta_{500} = \arctan \left(\frac{R_{500}}{\mathscr{D}_A}\right) \,,
\end{equation}
where ${\mathscr{D}_A}$ is the angular diameter distance. \\
\indent The pressure model was integrated along the line of sight and projected on a two-dimensional (2D) grid to obtain a Compton parameter map, $y(\theta)$. We converted it to surface brightness (see Eq. \ref{eq:tSZdef}) using the $C_{ 2 \ \mathrm{mm}}^{tSZ}$ value discussed in \citet{2018A&A...615A.112R}. We convolved the model map with the NIKA2 2\,mm beam and corrected by the transfer function computed in Sect.~\ref{Data_reduction_pipeline} to account for the filtering at large angular scales. 

\subsection{Matched-filtering}
\label{Match_filtered}

\indent We searched for galaxy cluster candidates in the N2CLS data using a matched-filter technique \citep{MMF, Match_filter_technique}, to optimally extract the tSZ signal from the map. This method has been used by previous large SZ surveys (see Sect.~\ref{sec:introduction}) to blindly detect galaxy clusters, without prior knowledge on their positions.
As general spatial features of clusters and the spectral characteristics of the tSZ signal are well known, it is possible to build a source filter that maximizes the cluster signal after filtering. This filter can be applied to a set of multifrequency maps or to a single map. In this analysis, we only used the 2\,mm data, as the tSZ signal is expected to be well below the noise in the 1.2\,mm map, and the contamination due to point sources, mainly dusty galaxies, is too high. 
The signal contained in a single frequency map can be described as $I(x) = f \cdot S(x) + N(x)$ where $f$  accounts for the tSZ spectral distortion at 2\,mm, $S(x)$ the spatial template of galaxy clusters and $N(x)$ the noise map. 
To extract the signal at maximum significance, \cite{optimal_SZ} have shown that the optimal filter is expressed as $\Psi = \left[ \tau^T \; C^{-1} \; \tau \right]^{-1} \; \tau \; C^{-1}$
with $\tau$ being the Fourier transform of the spatial template $S$ convolved with the pipeline's transfer function (see Sect.~\ref{Data_reduction_pipeline}), and $C$ the noise power spectrum $C=$\,diag$(|N(k)|^2)$.\\
\indent We used the PyMF Python package described in \cite{Pymf} as our matched-filter algorithm. For the spatial template $S$ we used the tSZ model presented above and given by Eq.~\ref{eq:tSZdef}.
To cover a wide range of possible cluster angular sizes, we produced a set of 15 cluster templates, with typical angular sizes $\theta_{500}$ $\in$ [18.5 $\arcsec$, 185 $\arcsec$]. For simplicity, we kept the redshift constant and varied the cluster's mass to obtain different cluster sizes.
The SZ cluster templates were obtained using the minot Python package \citep{minot}.\\

\begin{figure}[h]
	\centering
	\includegraphics[width=0.5\textwidth]{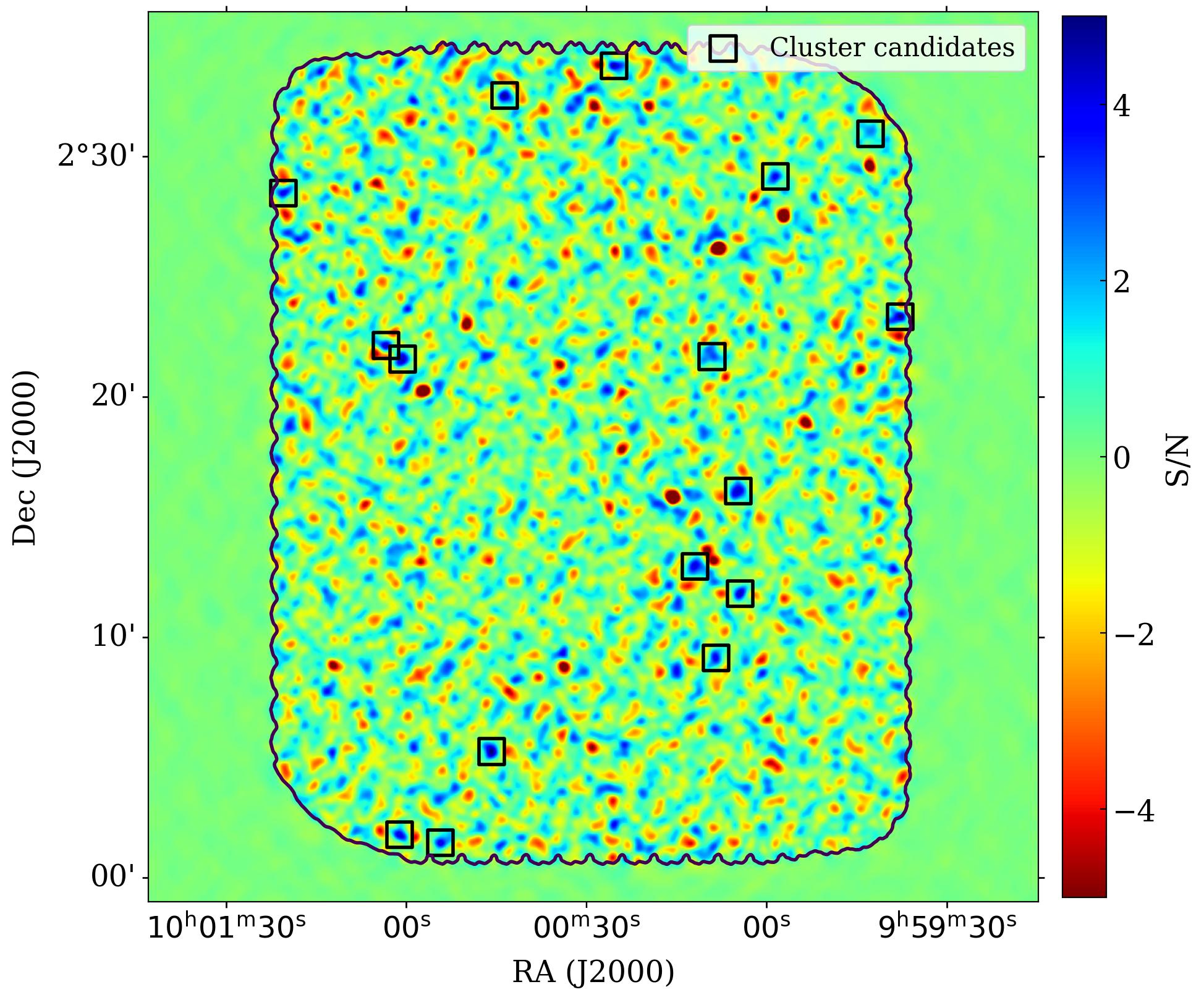}
	\caption{S/N map of the matched-filtered 2\,mm COSMOS map. Positive S/N values corresponding to cluster candidates are shown in blue, matching the colors in Figs.~\ref{fig:MCMC_fit_z} and \ref{cutout}, as expected for tSZ emission. The angular size of the applied matched filter is equal to the size of the NIKA2 beam at 2\,mm (18.5$\arcsec$). The black contour shows the high-quality region discussed in Sect.~\ref{Final_sky_maps}. Cluster candidates are highlighted in black squares. Sixteen candidates are detected above a S/N threshold of 4 (see Sect.~\ref{Sample identification}).}
	\label{Map_with_candidates}
\end{figure}

\subsection{Identification of cluster candidates}
\label{Sample identification}

\begin{figure}[h!]
	\centering
	\includegraphics[width=0.5\textwidth]{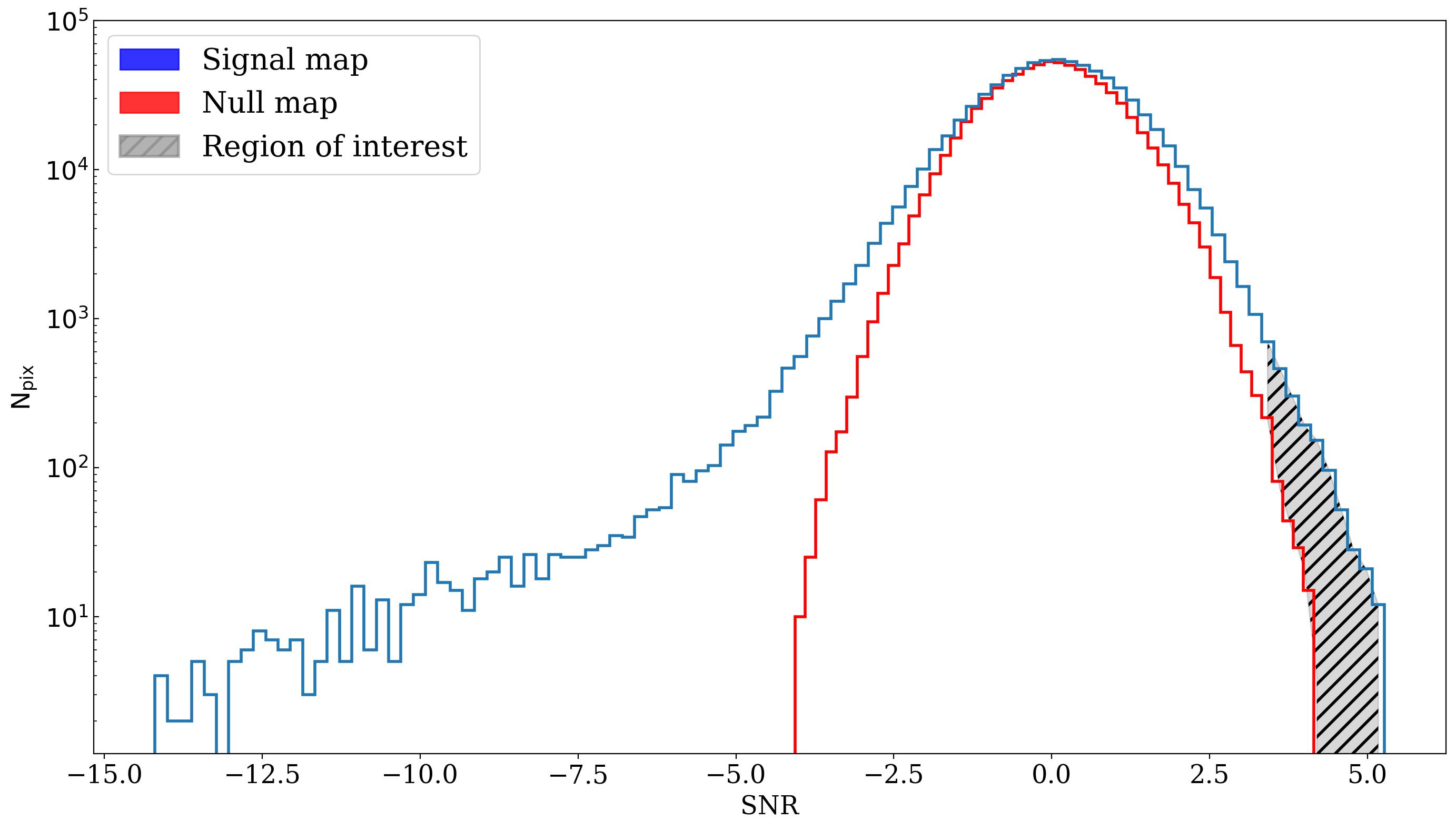}
	\caption{Pixel negative S/N distribution in the match-filtered N2CLS 2\,mm signal map (blue) and null map (red). The high negative S/N tail is due to point sources. The hatched gray area shows a small positive S/N tail up to S/N $\sim$ 5, indicating strong negative signal in the 2\,mm signal map, as expected for the SZ sources.}
	\label{fig:COSMOS_pixel distrib}
\end{figure}

\indent Using the method detailed in Sect.~\ref{Match_filtered}, we obtained a set of match-filtered maps where the tSZ signal is enhanced. Using the find\_peak routine in the photutils package \citep{photutils}, we identified cluster candidates as peaks above a S/N threshold of 4 in the match-filtered maps. The extraction box size was set to 9 pixels, to only extract one candidate per beam. When a candidate was detected in multiple filtered maps, its center position and S/N were taken from the map where the detection had the highest S/N value.\\
\indent During the match-filtering process, we also carefully masked bright positive point sources in the map. Indeed, the filtering of these sources can produce ringing artifacts with negative signal, and thus impact cluster detection. Fig.~\ref{Map_with_candidates} shows, as an example, one of the S/N maps, filtered to optimally extract objects with $\theta_{500}=18.5\arcsec$. The cluster's S/N is positive, as the negative sign of the tSZ signal at 2\,mm has already been accounted for in the filter. Positive peaks in this map, highlighted in black, can be identified as cluster candidates. To validate the significance of the S/N estimates, we also applied the matched-filtering technique to the null maps (see Sect.~\ref{null_maps_detection}).
Fig.~\ref{fig:COSMOS_pixel distrib} shows the pixel S/N distribution for both the match-filtered data and null maps. For the null map (red), we obtain a Gaussian-like distribution consistent with the noise distribution in the data. For the filtered data (blue), we clearly observe a large negative S/N tail corresponding to point sources in the original map. We also observe a positive tail, which corresponds to the cluster candidates.\\
\indent As presented in Table~\ref{tbl:candidates_table}, we detect 16 cluster candidates in the low noise area defined in Sect.~\ref{Final_sky_maps}. The mean S/N of our candidate sample is S/N $\sim$ 4.51, with detections spanning from a S/N of 4 to 5.31. Fig.~\ref{cutout} shows a cutout of the 2\,mm signal map around each of the cluster candidates. In this case, they appear as compact negative objects in the map, due to the filtering of the signal at large angular scales.\\

\begin{table*}[ht]
\caption{Galaxy cluster candidates, sorted by decreasing S/N, with S/N > 4.}
\centering
\resizebox{2\columnwidth}{!}{
\begin{tabular}{ |c|c|c|c|c|c|c|c|c|c| }
\hline  
\rule{0pt}{2ex} Candidate Name & RA & DEC & S/N & z  & $\theta_{500}$ & $Y_{500}$ & M$_{500}$  & Matching cluster name & Reference(s) \\[3pt]
 & ° & ° &   &  & arcmin & 10$^{-5}$ arcmin$^2$ & 10$^{14}$ M$_{\odot}$  & (distance [\arcsec]) &  \\ \hline
\rule{0pt}{12pt} 
NK2-CL J100045.8+020514.3 & 150.1907 & 2.0873 & 5.31 & -- & $0.74_{-0.20}^{+0.40}$ & $1.60_{-0.56}^{+1.45}$ & -- &--&--\\[6pt]
NK2-CL J095937.7+022320.4 & 149.9071 & 2.3890 & 5.00 & 0.74$\pm 0.03$ (p) & $1.38_{-0.13}^{+0.12}$ & $3.35_{-1.34}^{+1.77}$  & $1.51_{-0.38}^{+0.40}$ & ALH J0959.38+0223.03 (17.8\arcsec) & (1)\\[6pt]
NK2-CL J100004.7+021604.4 & 150.0194 & 2.2679 & 4.97 & -- & $0.92_{-0.29}^{+0.62}$ & $1.95_{-0.88}^{+2.65}$ & -- & --&--\\[6pt]
NK2-CL J100043.6+023232.4 & 150.1818 & 2.5423 & 4.87 & -- & $0.65_{-0.13}^{+0.31}$ & $1.11_{-0.30}^{+0.75}$ & -- & --&--\\[6pt]

NK2-CL J100025.3+023346.4 & 150.1056 & 2.5629 & 4.67 & 0.72$\pm 0.02$ (p) & $1.37_{-0.14}^{+0.12}$ & $2.90_{-1.24}^{+1.57}$ & $1.38_{-0.37}^{+0.38}$ & [BMH2011] 124 (11.5\arcsec) & (2), (3), (4) \\[6pt]
NK2-CL J100100.6+022134.4 & 150.2524 & 2.3596 & 4.67 & 0.77$\pm 0.01$ (p) & $1.36_{-0.11}^{+0.09}$ & $3.47_{-1.25}^{+1.42}$ & $1.56_{-0.34}^{+0.33}$ & [SCC2012] 0788 (9.1\arcsec) & (5)\\[6pt]

NK2-CL J100004.4+021148.4 & 150.0183 & 2.1968 & 4.60 & 0.94$\pm 0.05$ (s) & $1.10_{-0.10}^{+0.09}$ & $2.10_{-0.78}^{+0.91}$ & $1.21_{-0.27}^{+0.27}$ & [KLI2009] 146\tablefootmark{*} (8.7\arcsec) & (3)\\[6pt]
NK2-CL J100103.4+022208.4 & 150.2641 & 2.3690 & 4.54 & --  & $0.59_{-0.09}^{+0.23}$ & $0.85_{-0.20}^{+0.35}$ & -- & -- & -- \\[6pt]
NK2-CL J100011.9+021256.5 & 150.0494 & 2.2157 & 4.48 & 0.24$\pm$0.08\tablefootmark{**} (p) & $3.79_{-1.01}^{+2.11}$ & $28.20_{-16.79}^{+69.45}$ & $2.83_{-0.88}^{+1.55}$ & XMMXCS J100012.3+021246.7 (11.7\arcsec) & (6)\\[6pt]
NK2-CL J100101.1+020146.6 & 150.2546 & 2.0296 & 4.30 & -- & $0.71_{-0.16}^{+0.44}$ & $0.96_{-0.29}^{+0.83}$ & -- & -- & --\\[6pt]
NK2-CL J100054.3+020126.4 & 150.2262 & 2.0240 & 4.28 & 1.42$\pm$0.01 (p) & $0.81_{-0.06}^{+0.05}$ & $1.56_{-0.53}^{+0.62}$ & $1.00_{-0.21}^{+0.21}$ & [SCC2012] 1517 (16.0\arcsec) & (5) \\[6pt]
NK2-CL J100009.1+022140.3 & 150.0378 & 2.3612 & 4.27 & -- & $2.56_{-1.16}^{+2.96}$ & $14.30_{-10.55}^{+81.99}$ & -- & -- & --\\[6pt]
NK2-CL J095942.6+023056.5 & 149.9277 & 2.5157 & 4.08 & 0.73$\pm 0.02$ (s) & $1.45_{-0.13}^{+0.11}$ & $4.10_{-1.55}^{+1.83}$ & $1.69_{-0.39}^{+0.39}$ & DESI 2353000051 (12.0\arcsec) & (7) \\[6pt]
NK2-CL J100120.5+022828.2 & 150.3353 & 2.4745 & 4.07 & --  & $0.71_{-0.15}^{+0.44}$ & $0.93_{-0.28}^{+0.75}$ & -- & -- &--\\[6pt]
NK2-CL J100008.4+020908.3 & 150.0350 & 2.1523 & 4.04 & --  & $0.77_{-0.21}^{+0.65}$ & $1.20_{-0.44}^{+1.72}$ & -- &-- &--\\[6pt]
NK2-CL J095958.5+022910.4 & 149.9938 & 2.4862 & 4.01 & 0.40$\pm$0.01 (p) & $2.52_{-0.23}^{+0.22}$ & $12.24_{-4.77}^{+6.22}$ & $2.38_{-0.57}^{+0.61}$ & [SCC2012] 0270 (14,8\arcsec) & (5)\\[6pt]
\hline
\end{tabular}}
\tablefoot{The first column gives the name of the candidate. The right ascension (J2000) and declination (J2000) are presented in the second and third columns, respectively. The S/N of each candidate is given in the fourth column. The fifth column lists the candidates' estimated redshift, $z$, associated uncertainties and redshift type (photometric (p) or spectroscopic (s)). In the sixth and seventh columns we present the cluster angular size $\theta_{500}$  and tSZ flux, $Y_{500}$ (see Sect.~\ref{sec:candidates_properties}). For candidates with known redshift, we list their estimated mass $M_{500}$ and uncertainties in the eighth column (see Sect.~\ref{mass_redshift}). The last column gives the name of the matched counterpart in external catalogs with its distance to the NIKA2 cluster candidate center in arcsec and the reference. A $^{*}$ indicates that the object is a group of galaxies. A $^{**}$ indicates a possible superposition of a low and a high-redshift cluster (see Sect.~\ref{sec:spec_photo_redshift}).}
\tablebib{(1)~\cite{ALH_survey}; (2)~\cite{Bellagamba_2011}; (3)~\cite{zCOSMOS_Group}; (4)~\cite{COSMOS_Wall}; (5)~\cite{CES_catalog}; (6)~\cite{XMM_survey}; (7)~\cite{DESI_survey}}
\label{tbl:candidates_table}
\end{table*}

\subsection{Random detection rate}
\label{null_maps_detection}

\indent To estimate the expected number of random detections in the survey area, we generated 1000 noise-only maps without any tSZ signal, and ran the detection algorithm. We find an average of 2.36 random detections above a S/N of 4 per null map. We thus expect an overall purity of 85\% for our sample, excluding point source effect. The overall purity increases to 93\% above a S/N of 4.2, where we only expect $\sim 1$ random detection. A more detailed assessment of the purity using tSZ simulations is given in Sect.~\ref{Simulations}.

\section{Cluster detection characterization via simulations}
\label{Simulations}

\subsection{Cluster detection on simulations}
\label{Detection_simulations}

\begin{figure*}[ht!]
\centering
\begin{minipage}{.48\textwidth}
\begin{subfigure}{\textwidth}
\includegraphics[width=\textwidth]{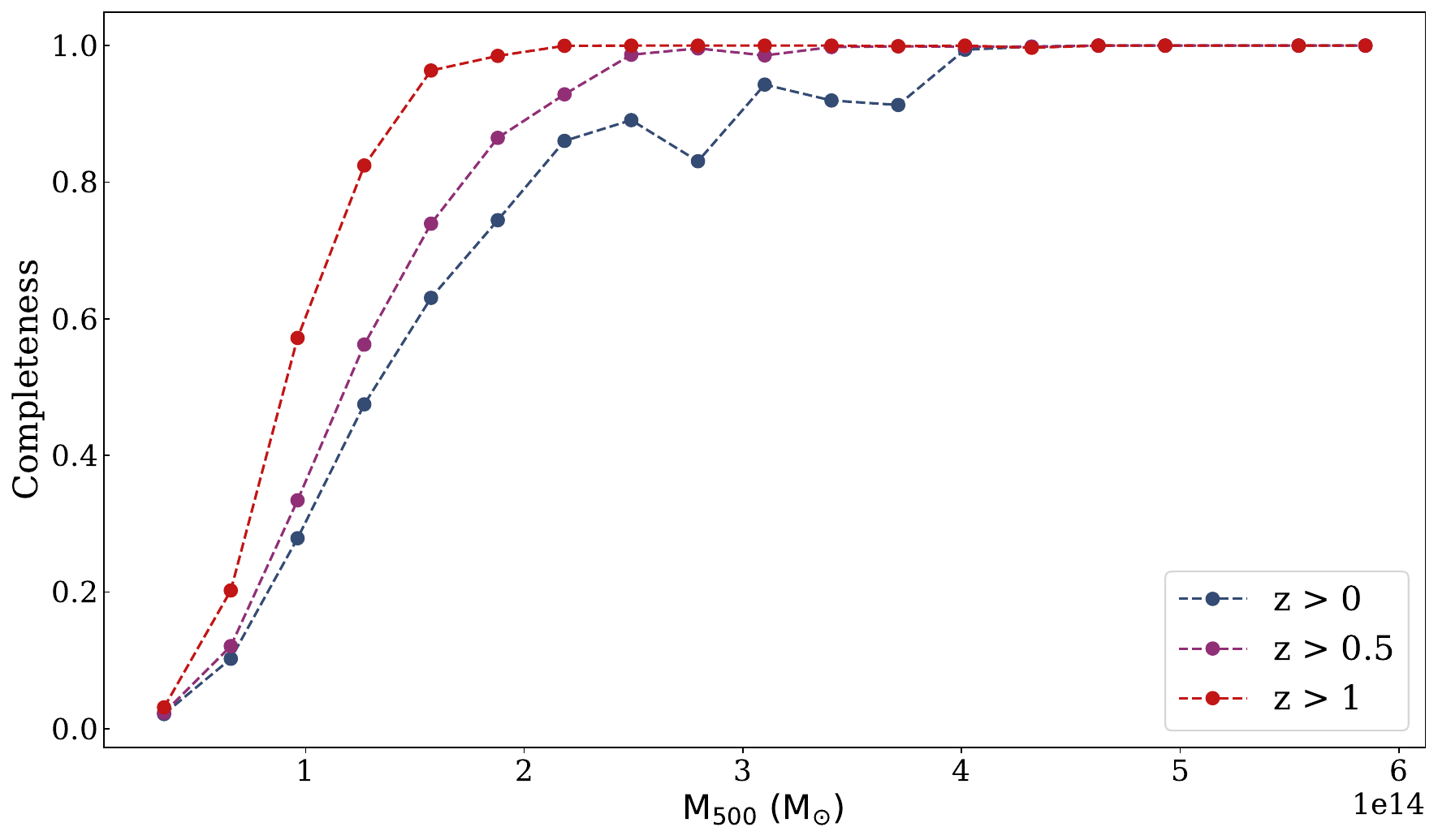}
\end{subfigure}
\end{minipage}
\begin{minipage}{.48\textwidth}
\begin{subfigure}{\textwidth}
\includegraphics[width=\textwidth]{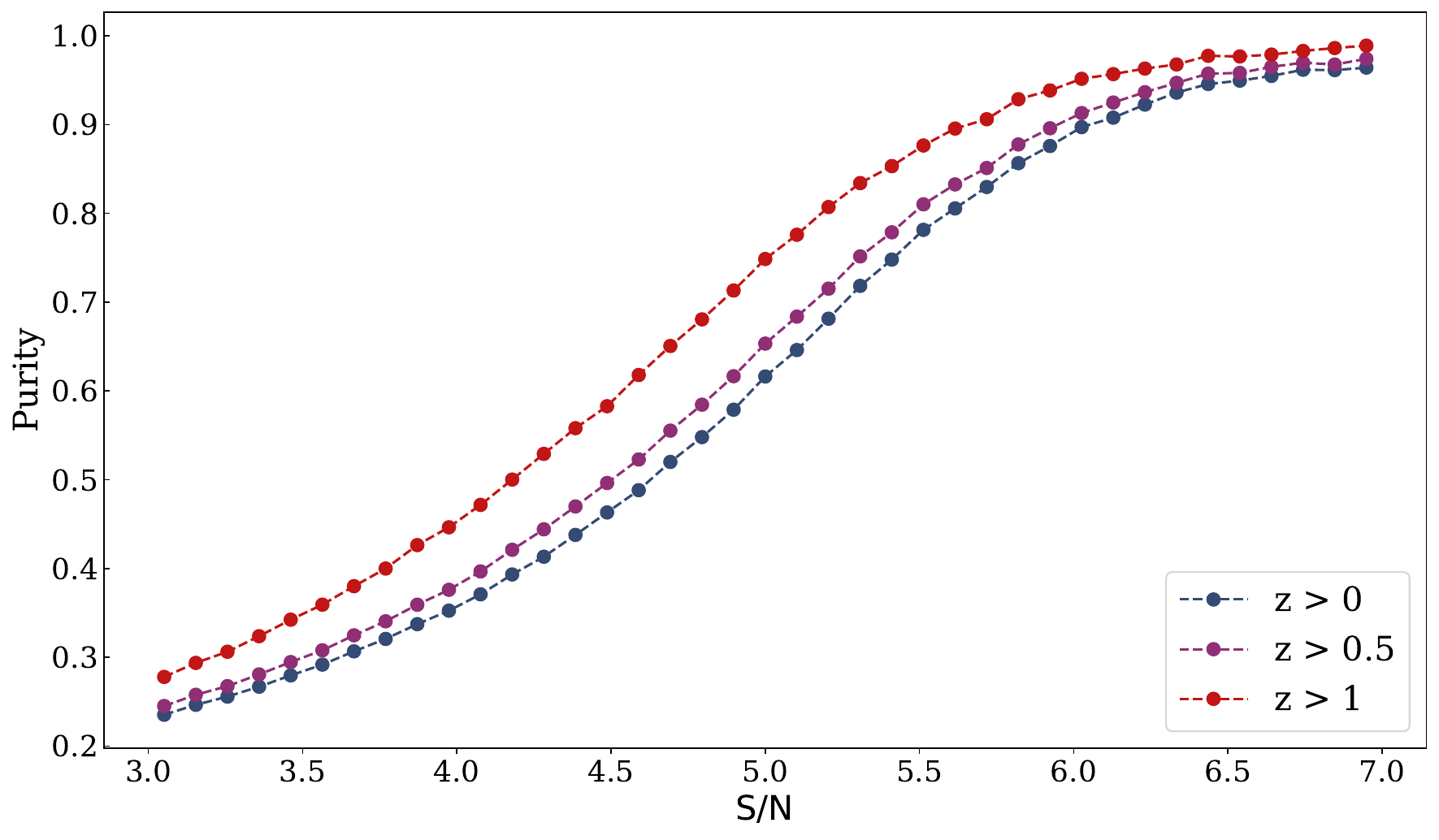}
\end{subfigure}
\end{minipage}
\caption{(Left) Completeness as a function of the mass, $M_{500}$, for different redshift ranges. We reach 80\% completeness at all redshift for $M_{500}> \ 2\times10^{14}M_{\odot}$.
(Right) Purity as a function of S/N for different redshift ranges. The purity rises quickly with S/N and reaches 60\% at S/N $\sim$ 5.}
\label{Completeness_purity_line}
\end{figure*}

\indent To characterize our cluster detection algorithm, we used a set of realistic simulations of the COSMOS field. 
We first produced 1000 large mock cluster catalogs. We estimated the expected number of clusters in the area considered in our analysis (877 arcmin$^2$), by integrating the \cite{Tinker_hmf} mass function in bins of redshift ($0 < z < 3$) and mass ($2\times10^{13}M_{\odot}$ < $M_{500}$ < $3\times10^{15}M_{\odot}$).
For each catalog, we drew the number of clusters from a Poisson distribution, assigned to each cluster a mass and redshift drawn from the mass function, and a random position in the survey area. To model the cluster signal, we used the model presented in Sect.~\ref{subsec:spatialclustermodel}.
To this thermal SZ map, we added residual atmospheric and instrumental noise by using the null maps described in Sect.~\ref{Observations}.  \\
\indent Finally, we added a sky model, the simulated infrared dusty extragalactic sky \citep[SIDES, ][]{SIDES_2017, SIDES_2022}. SIDES includes galaxy clustering, and produces a catalog of galaxies with a large set of physical parameters. We used the SIDES light cone from the Uchuu simulation \citep{Uchuu, SIDES_Gkogkou}.
Contributions from radio sources and the spatial correlation between simulated components were not taken into account. For each simulated cluster catalog, we produced 104 simulated maps using independent tiles from the SIDES sky model. This prevented any overlapping of bright point sources and massive clusters from impacting the completeness and purity of the sample (see Sect.~\ref{Completeness} and \ref{Purity}). \\
\indent We estimated the CMB anisotropies contribution to the NIKA2 2\,mm map. We note that measurements of the CMB power spectrum, $C^{\mathrm{CMB}}_{\ell}$, at the very high $\ell$ multipoles sampled by NIKA2 ($\ell > 10^4$) are not available. Furthermore, theoretical models become inaccurate at such small angular scales. Thus, we fitted a power law to the current \textit{Planck} best-fit CMB power spectrum and extrapolated it to very large $\ell$ multipoles. Finally, we computed the expected CMB rms signal as

\begin{equation}
\sigma^2_{\mathrm{CMB}} = \langle T_1 T_2 \rangle = \frac{1}{4\pi} \sum_{\ell} (2\ell+1) C^{\mathrm{CMB}}_{\ell} P_{\ell}(\cos{(0)}) \,,
\end{equation}

\noindent where $P_{\ell}$ are the Legendre polynomials. We find that, accounting for the pipeline transfer function, the CMB induced noise is negligible at the scales probed in this analysis. Thus, it is not accounted for in the following. \\
\indent We first ran the detection algorithm presented in Sect.~\ref{Match_filtered} on the full set of simulated maps, using a 2D Gaussian as template, to detect very bright positive sources in the maps (above S/N = 10). The filtering of these bright sources could produce ringing artifacts with negative signal and impact cluster detection. The detected bright sources were fitted with a 2D Gaussian and subtracted from the simulations. We then convolved these simulated maps with the NIKA2 beam and corrected by the pipeline's transfer function shown in Fig.~\ref{fig:TF_plot}. The bright sources were added back after filtering. This method was also used in the data reduction pipeline described in Sect.~\ref{Data_reduction_pipeline}. Using the transfer function directly is a good compromise, as processing all simulated maps through the pipeline is very long and computationally expensive. There are no big differences between the two processes, as bright sources have been subtracted before filtering.\\ 
\indent We then ran the detection algorithm on the simulated maps using the cluster template presented in Sect.~\ref{subsec:spatialclustermodel}.
We identified S/N peaks with S/N > 4 in the resulting match-filtered maps. The position and S/N of each cluster candidate were stored in an output catalog. We defined the matching radius to 20$\arcsec$ -- the size of the NIKA2 beam slightly extended for uncertainties as in previous studies \citep[see e.g.,][]{Planck_XXVII}. We first matched each input cluster with the closest detection in the output catalog using the astropy function match\_to\_catalog\_sky. Then, we matched the output catalog with the input catalog. We note that we require two-way matches for the following studies. If there were multiple associated input clusters with a single output cluster in a radius the size of the NIKA2 beam around the detection, we chose the best matching cluster based on two criteria: the $Y_{500}$ value and the cluster-detection separation. This prevented detections from being matched with very faint clusters, which can introduce biases on the sample's completeness (see Sect.~\ref{Completeness}). From the results, we computed the average number of detections. Taking the median across simulations, we find an average of $26 \pm 5$ detections for S/N > 4. This is within two Poisson errors from what we have found on the NIKA2 COSMOS data as shown in Sect.~\ref{Sample identification}. Therefore, it is compatible with the distribution found from the simulations.

\subsection{Completeness}
\label{Completeness}

\indent The completeness of a survey is a crucial information to characterize the sample and the performance of the detection algorithm. We define it as the fraction of true cluster detected with respect to the clusters in the input mock catalog, $$C = \frac{N^{true}_{det}}{N_{input}} \,,$$

\begin{figure}[h!]
\centering
\includegraphics[width=0.5\textwidth]{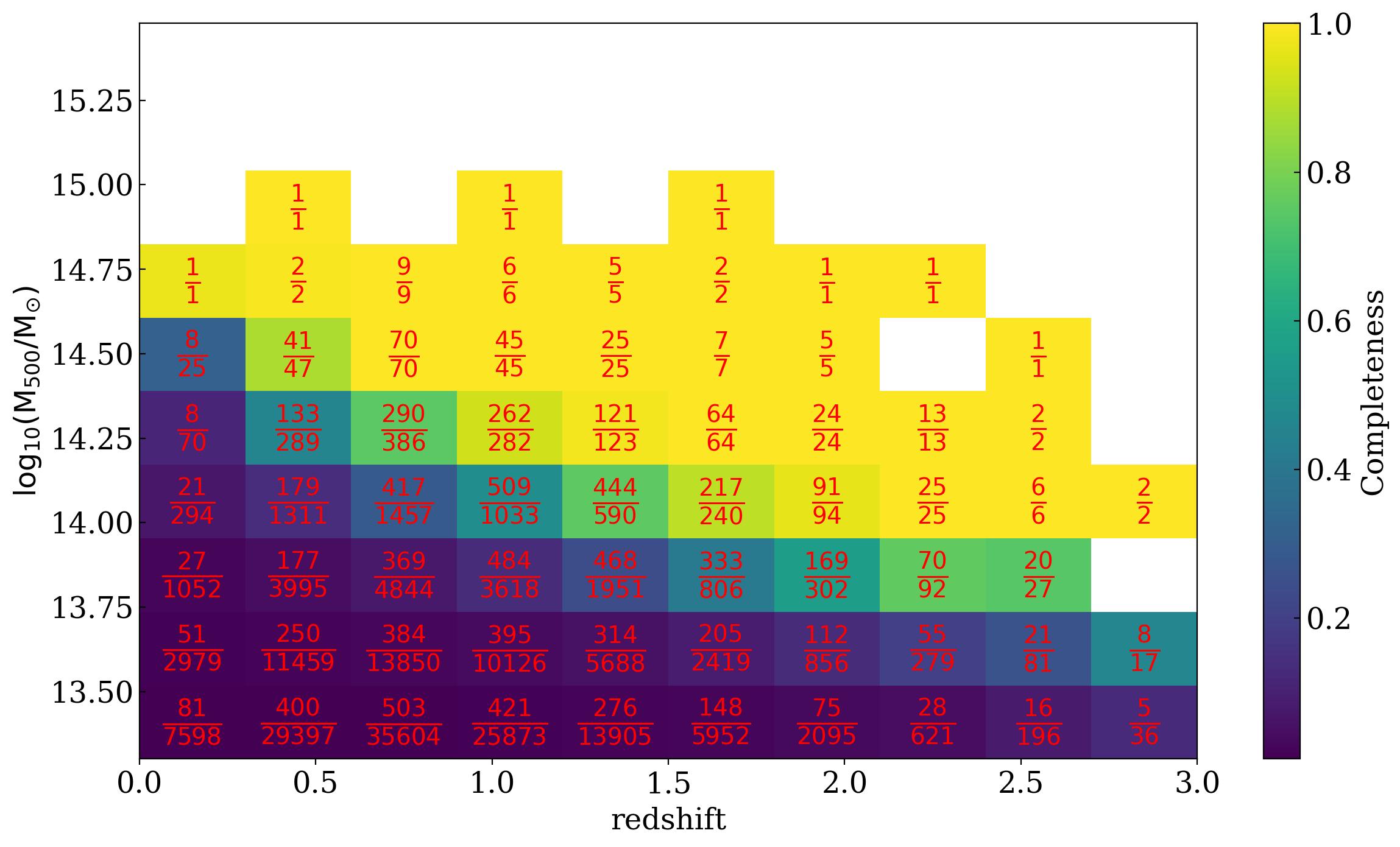}
\caption{Color-coded completeness defined as the probability for a cluster of mass $M_{500}$ and redshift $z$ to be detected. The number of true clusters detected with respect to the total number of simulated clusters is given for each bin in mass and redshift.}
\label{Completeness_purity_bins}
\end{figure}

In Fig.~\ref{Completeness_purity_bins} we present, color-coded, the completeness in bins of mass and redshift for the NIKA2 sample as derived from the simulations described above.
We observe, as expected, that the detection algorithm performance is limited at low mass and low redshift, for which clusters are faint and extended and thus filtered out by the pipeline transfer function. (see Sect.~\ref{Data_reduction_pipeline}). However, at higher redshift, $z>1$, we reach a completeness above 80\% for $M_{500} > \ 2\times10^{14}M_{\odot}$.  As the redshift increases, we can effectively detect clusters with lower and lower masses, down to $M_{500}$ $\sim$ $1\times10^{14}M_{\odot}$ at $z\sim1.5$.
This is also what we observe in the left panel of Fig.~\ref{Completeness_purity_line}, which shows the completeness as a function of $M_{500}$ for different redshift ranges. The whole sample is complete, $C > 80$\%, for $M_{500} > 2\times10^{14}M_{\odot}$ in all redshift ranges. We expect to detect most clusters above this mass threshold in the survey area.\\
\indent These results highlight the interest of high-resolution instruments to target the detection via the tSZ effect of high-redshift and low-mass clusters, which are not accessible with more traditional CMB experiments. We stress also that the limit in cluster mass is mainly imposed by the depth of the survey, which in the case of the COSMOS field was not optimized for cluster detection.

\subsection{Purity}
\label{Purity}

\indent Another key information that we can draw from the simulations is the purity, which we define as the percentage of true detections in the sample
$$P = \frac{N^{true}_{det}}{N_{det}} \,,$$
We show in Fig.~\ref{Completeness_purity_line} the purity of the sample as a function of the S/N for different redshift ranges. We find that the purity increases significantly with redshift. For $z>1$ the purity is above 70\%  (90 \%) for S/N > 5 (> 6). 
From the simulations and the results presented in Sect.~\ref{null_maps_detection}, we have observed that the purity of the sample is much better if point sources are not included in the simulations. Sect.~\ref{null_maps_detection} describes a best case scenario where the detection is noise limited. Adding point sources shifts the map's zero-level and introduces ringing artifacts after filtering.

\section{Cluster candidates' validation}
\label{Cluster_candidates}

We performed an extensive validation of the cluster candidates presented in Sect.~\ref{Sample identification} using existing datasets and archives. 

\subsection{Cross-matching in NED, SIMBAD and VizieR databases}
\label{sec:Candidates_matching}

\begin{figure}[ht]
\centering
\includegraphics[width=0.35\textwidth]{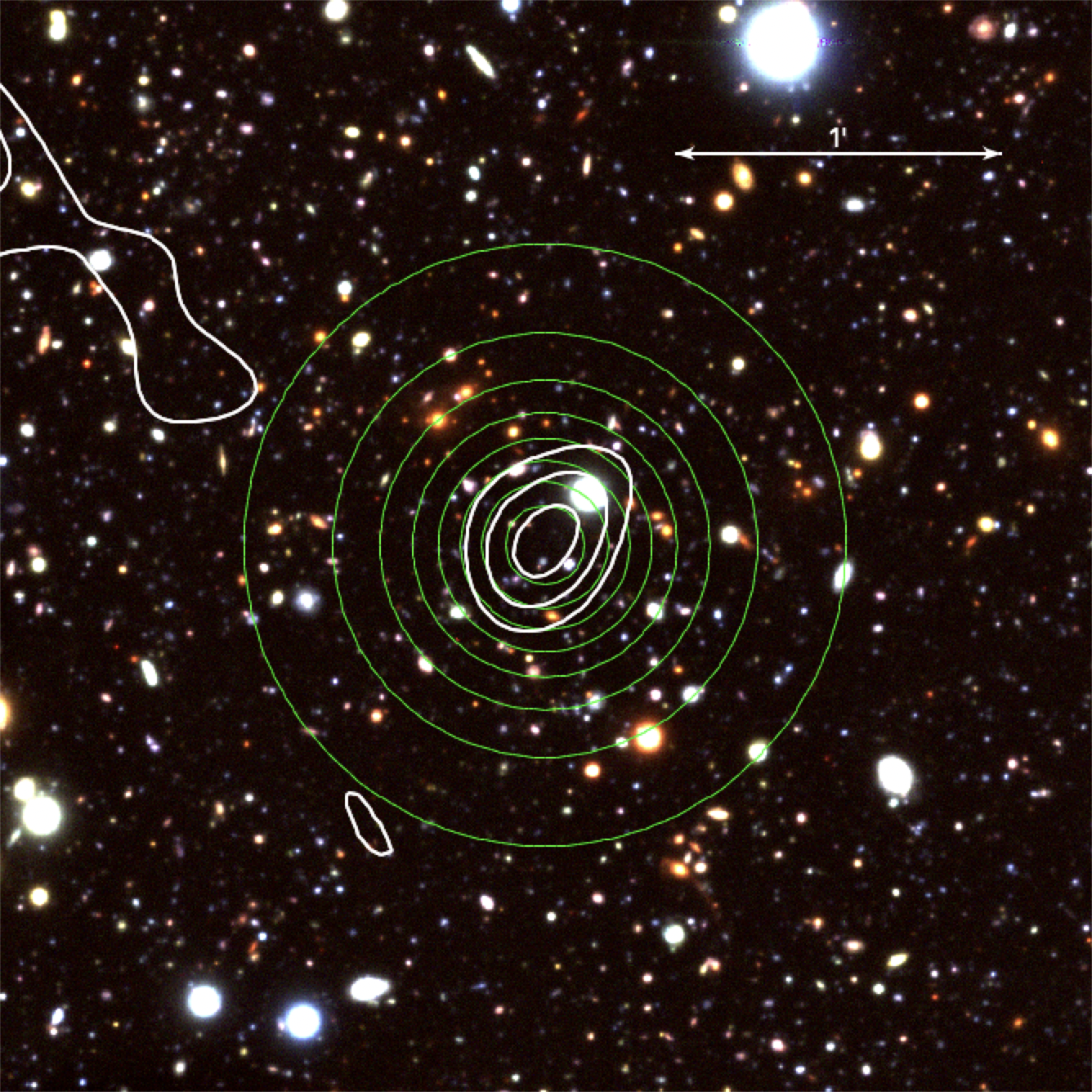}
\caption{Hyper Suprime-Cam (HSC) \textit{gri} image of NK2-CL J100004.4+021148.4. The image is $200\arcsec$ on a side, with north at the top and east at the left. The contours in white show S/N levels in the match-filtered map from Fig.~\ref{Map_with_candidates}, starting at 2$\sigma$ and spaced by 1$\sigma$. The green contours show the unfiltered tSZ cluster model, as seen in the third panel of Fig.~\ref{fig:MCMC_fit_z}.}
\label{fig:hsc}
\end{figure}

\indent We searched for known clusters around our candidates in the NED\footnote{\url{https://ned.ipac.caltech.edu/}} and SIMBAD\footnote{\url{https://simbad.u-strasbg.fr/simbad/}} databases, where most of the detected clusters are cataloged. We adopted a search radius of 20 arcsecond around each candidate for both databases. In SIMBAD, we only queried objects listed as galaxy clusters or group of galaxies. For NED, after querying all objects in the vicinity of the candidates, we only considered objects listed as a cluster or group of galaxies for the matching. 
We also performed a systematic query in the VizieR\footnote{\url{https://vizier.cds.unistra.fr/}} database to avoid missing any association with recent cluster catalogs \citep[e.g., DESI,][]{DESI_survey}. Our candidates were matched with VizieR catalogs classified as 'Clusters of galaxies'.\\
\indent We find that, from our sample of 16 cluster candidates, eight are matched with a known cluster. From those, seven were identified in the SIMBAD and NED databases, and one in VizieR. 
For each cluster candidate, the name of the closest counterpart, the distance to it, and its redshift are given in Table~\ref{tbl:candidates_table}. The median redshift of the sample is $z\sim0.74$, with two candidates at high redshift, $z>0.9$ .
We have no match with already detected tSZ clusters from traditional large CMB surveys when querying the SZMC database \footnote{\url{http://szcluster-db.ias.u-psud.fr/}}.
We have seven matches in optical and NIR cluster catalogs like the ALHAMBRA \citep{ALH_survey},  BMH \citep{Bellagamba_2011}, DESI \citep{DESI_survey}, COSMOS 10k \citep{COSMOS_survey}, COSMOS Wall \citep{COSMOS_Wall}, and SCC \citep{CES_catalog} catalogs. 
One candidate matches with an X-ray detected cluster in the XMMXCS catalog \citep{XMM_COSMOS_2007}. We notice that the highest S/N cluster candidate,  NK2-CL J100045.8+020514.3, does not match any known cluster, making it an interesting prospect for follow-up observations. Fig.~\ref{fig:hsc} shows an example optical image of the candidate NK2-CL J100004.4+021148.4, which has an optical counterpart at $z = 0.94$.\\

\vspace{-4pt}

\indent To quantify the impact of random associations between our cluster candidates and clusters in the literature, we used our cross-matching procedure described in Sect.~\ref{sec:Candidates_matching} on our detected simulated clusters from Sect.~\ref{Simulations}. We find that 24.6\% of them are matched with a cluster in the literature, compared to 50\% in our observed sample.
We find that $\sim$ 0.28\% are randomly matched with an X-ray cluster, compared to 6.5\% for our sample. The SCC catalog \citep{CES_catalog} has a lot of entries, and thus a non negligible 6.9\% chance to match at random. In our sample, 19\% are matched with a cluster in the SCC catalog. We tested other matching radii, but found that this one gives the best ratio between true and random associations.\\

\subsection{Spectroscopic and photometric redshift catalogs}
\label{sec:spec_photo_redshift}

\indent To complete the database search, we also had access to a semi-public spectroscopic redshift catalog \citep{Khostovan} and a public photometric catalog \citep{COSMOS2020}. For both catalogs, we searched for all objects labeled as galaxies, with an available redshift measurement, within a radius of 1 arcmin around each cluster candidate. This search radius is consistent with the average $\theta_{500}$ value we have found for the NIKA2 candidates (see Sect.~\ref{integrated_quantities} and Table~\ref{tbl:candidates_table}). For each of the NIKA2 cluster candidates, we show histograms of spectroscopic (blue) and photometric (red) galaxy redshifts in Fig.~\ref{redshift_hist}. When available, redshift estimates of the cluster candidates are shown as a dashed black line. We observe that in some cases -- see for example panels (b), (g), (m), and (p) --  the candidates' redshift estimates are found at the peak of the galaxy redshift distribution. Furthermore, we observe that for some of the unmatched NIKA2 cluster candidates, well-defined redshift peaks are found at relatively high redshift. In some cases, we find multiple well-defined peaks, suggesting a superposition of a low and a high-redshift cluster. A more detailed analysis of these redshift distributions is out of the scope of this paper. 

\begin{figure*}[ht!]
\centering
\begin{minipage}{.23\textwidth}
\begin{subfigure}{\textwidth}
\includegraphics[width=\textwidth]{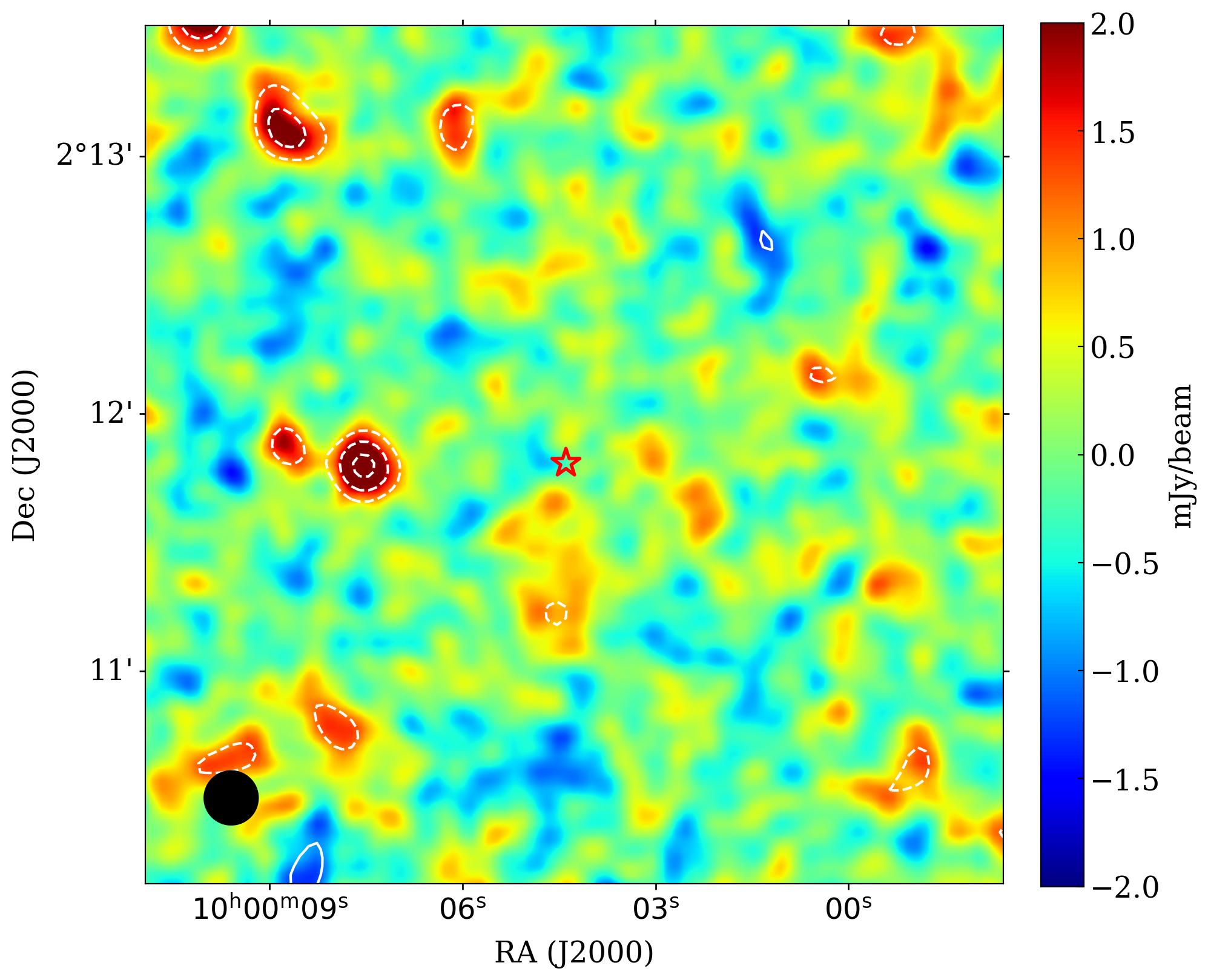}
\end{subfigure}
\end{minipage}
\begin{minipage}{.23\textwidth}
\begin{subfigure}{\textwidth}
\includegraphics[width=\textwidth]{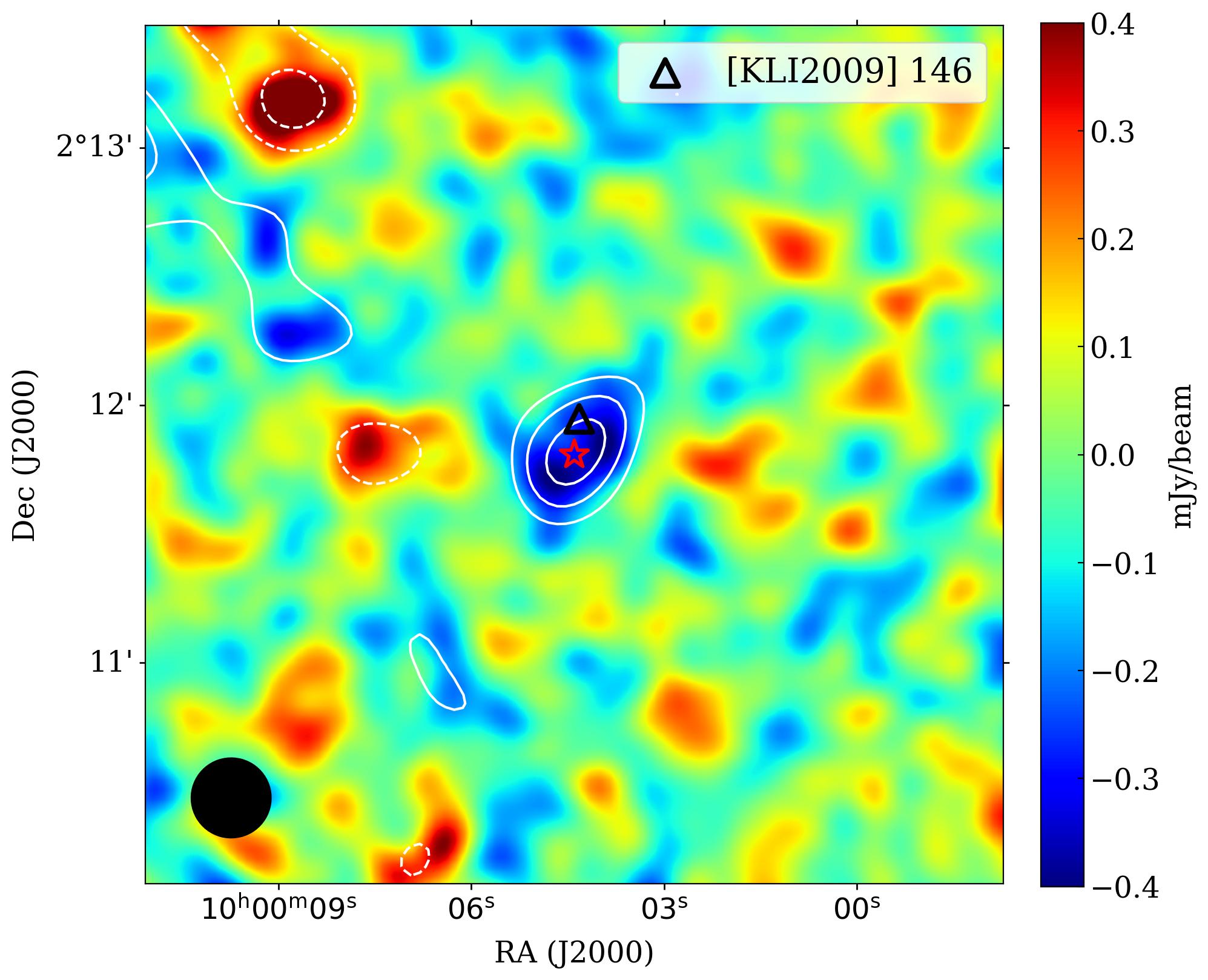}
\end{subfigure}
\end{minipage}
\begin{minipage}{.23\textwidth}
\begin{subfigure}{\textwidth}
\includegraphics[width=\textwidth]{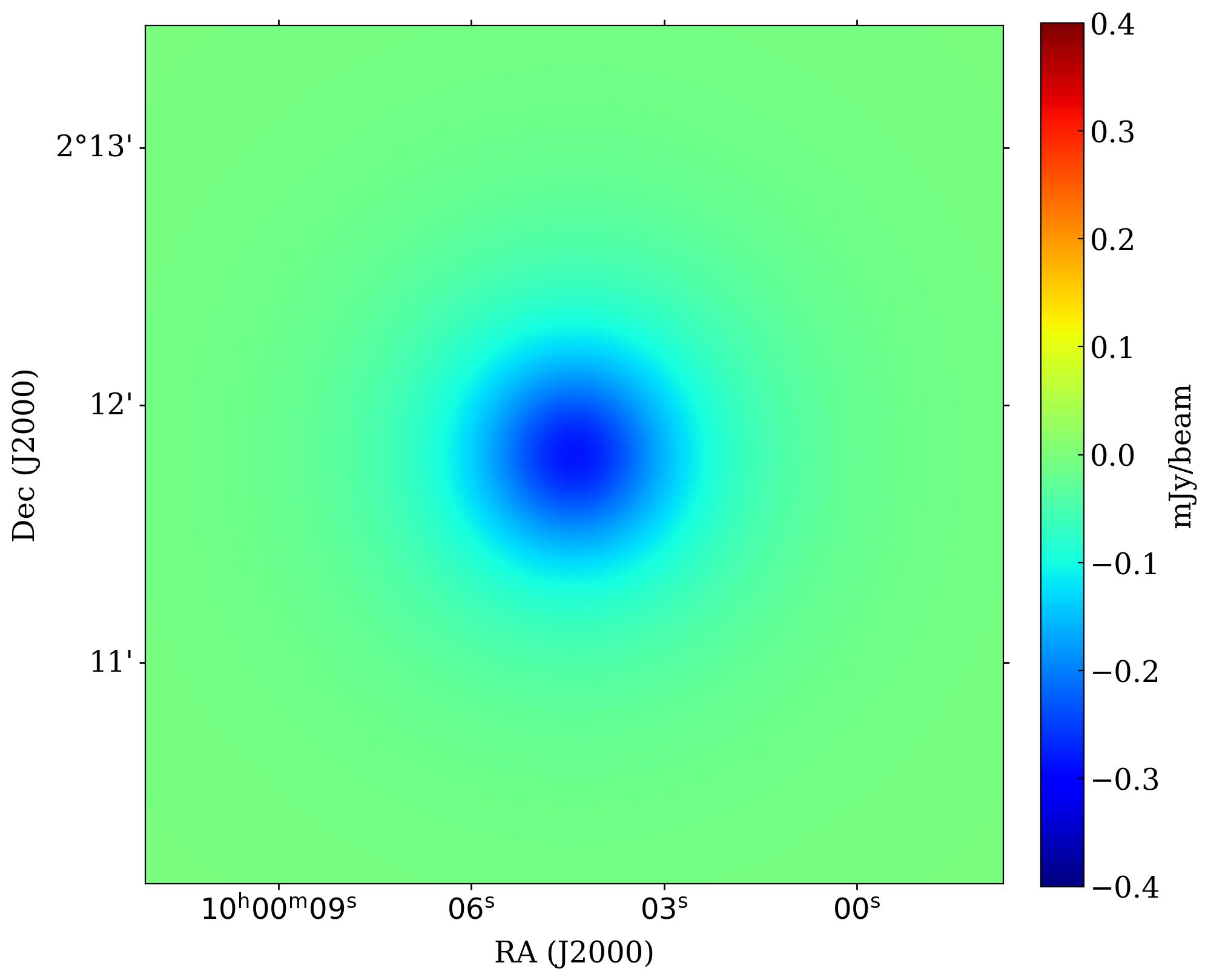}
\end{subfigure}
\end{minipage}
\begin{minipage}{.23\textwidth}
\begin{subfigure}{\textwidth}
\includegraphics[width=\textwidth]{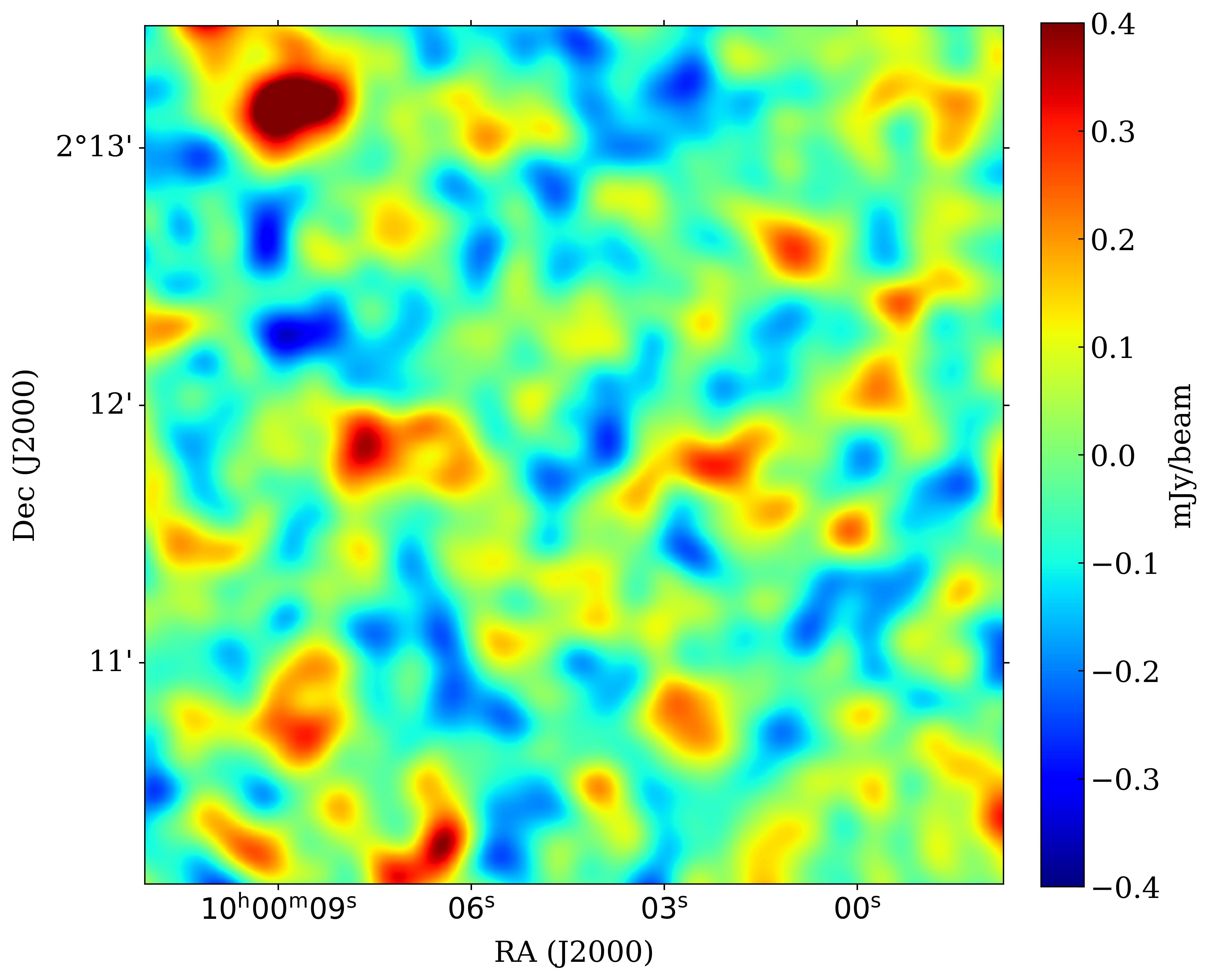}
\end{subfigure}
\end{minipage}
\begin{minipage}{.23\textwidth}
\begin{subfigure}{\textwidth}
\includegraphics[width=\textwidth]{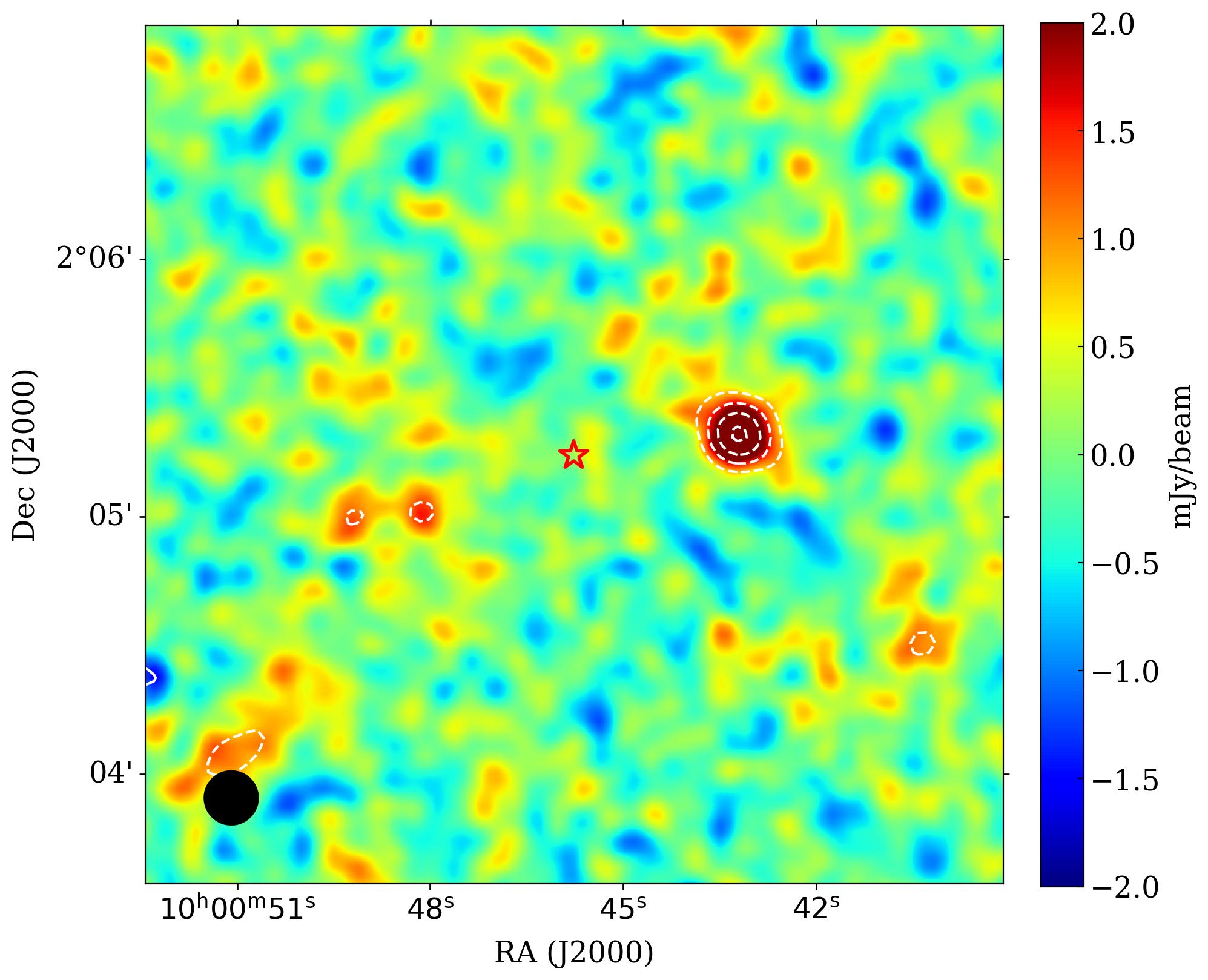}
\end{subfigure}
\end{minipage}
\begin{minipage}{.23\textwidth}
\begin{subfigure}{\textwidth}
\includegraphics[width=\textwidth]{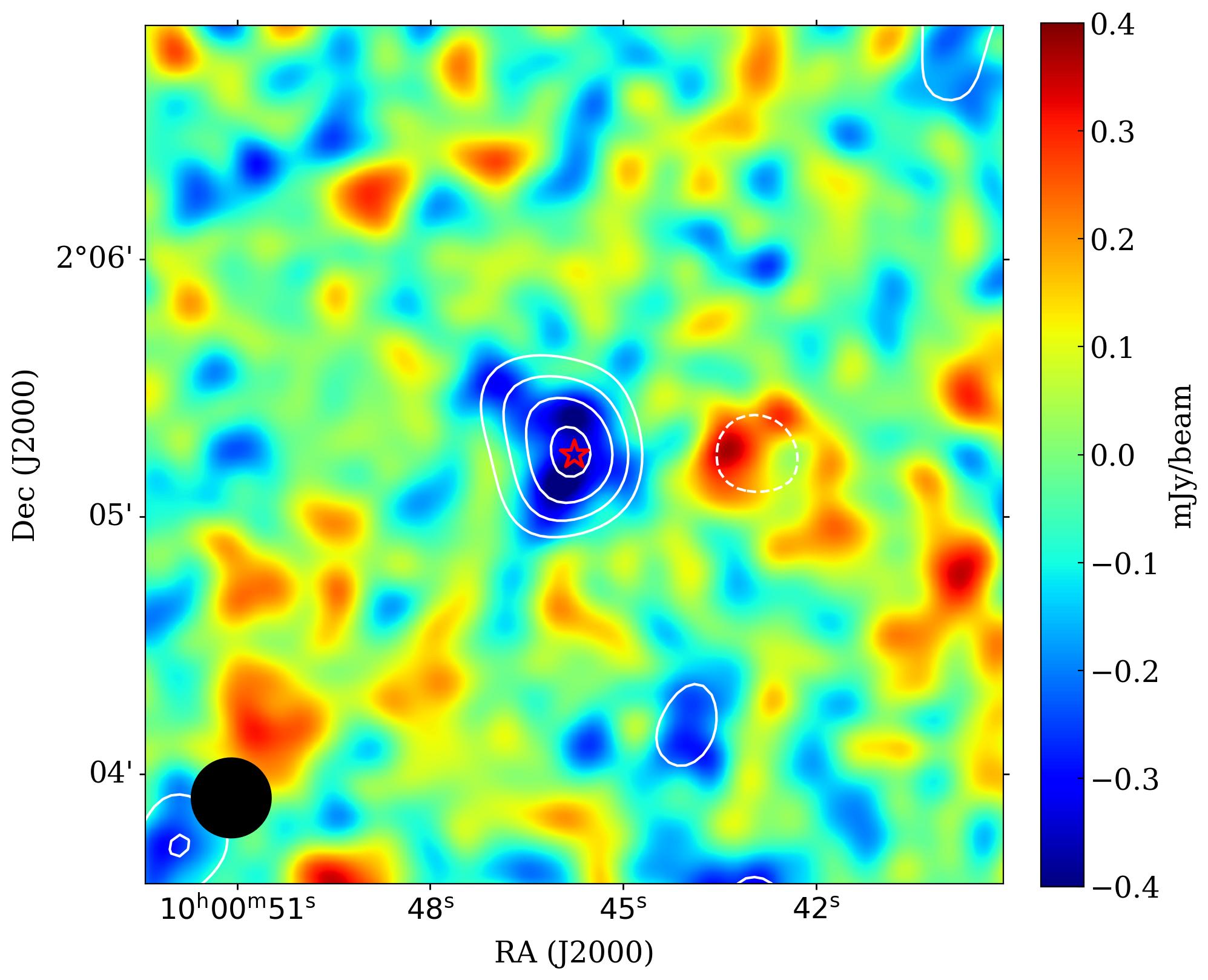}
\end{subfigure}
\end{minipage}
\begin{minipage}{.23\textwidth}
\begin{subfigure}{\textwidth}
\includegraphics[width=\textwidth]{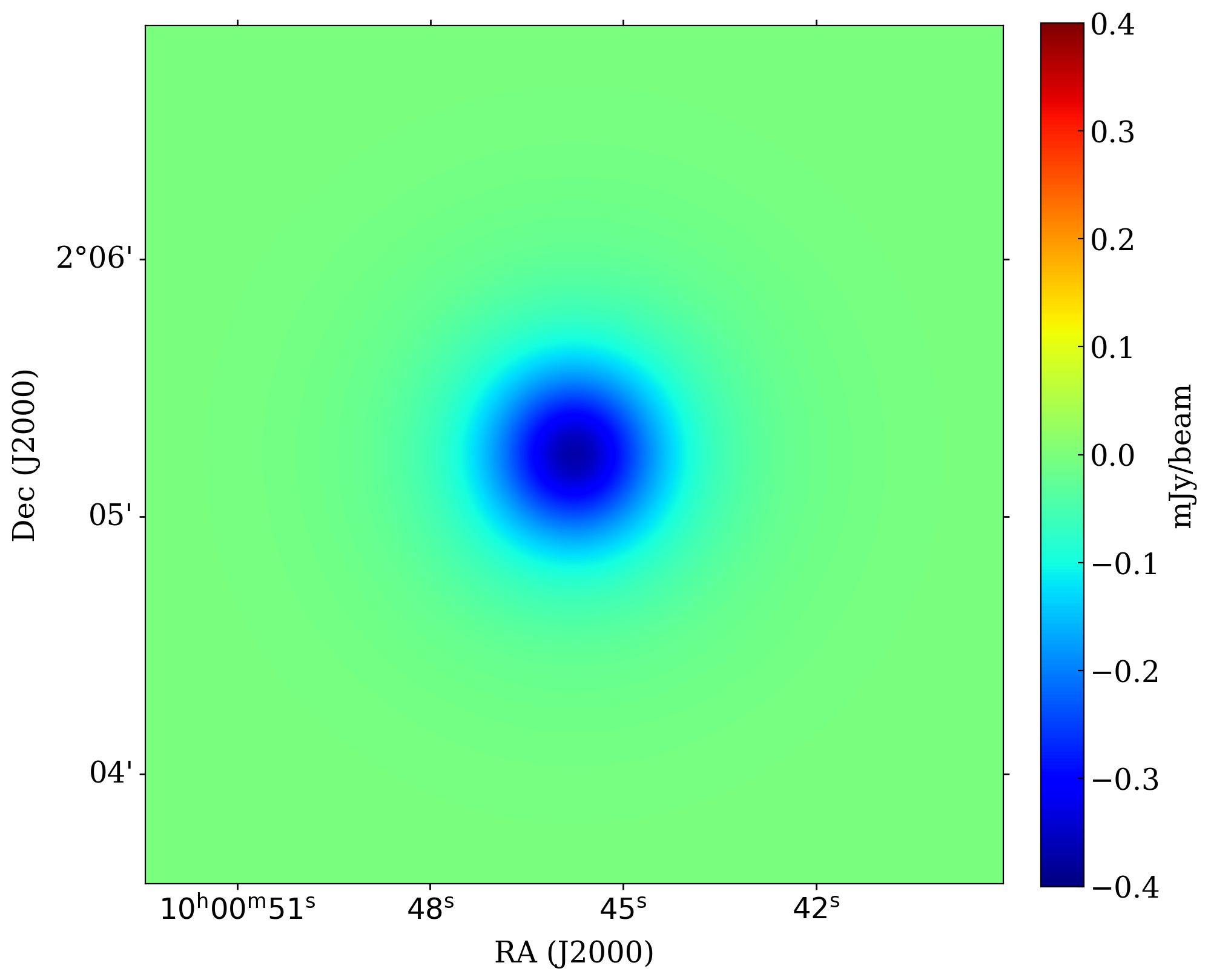}
\end{subfigure}
\end{minipage}
\begin{minipage}{.23\textwidth}
\begin{subfigure}{\textwidth}
\includegraphics[width=\textwidth]{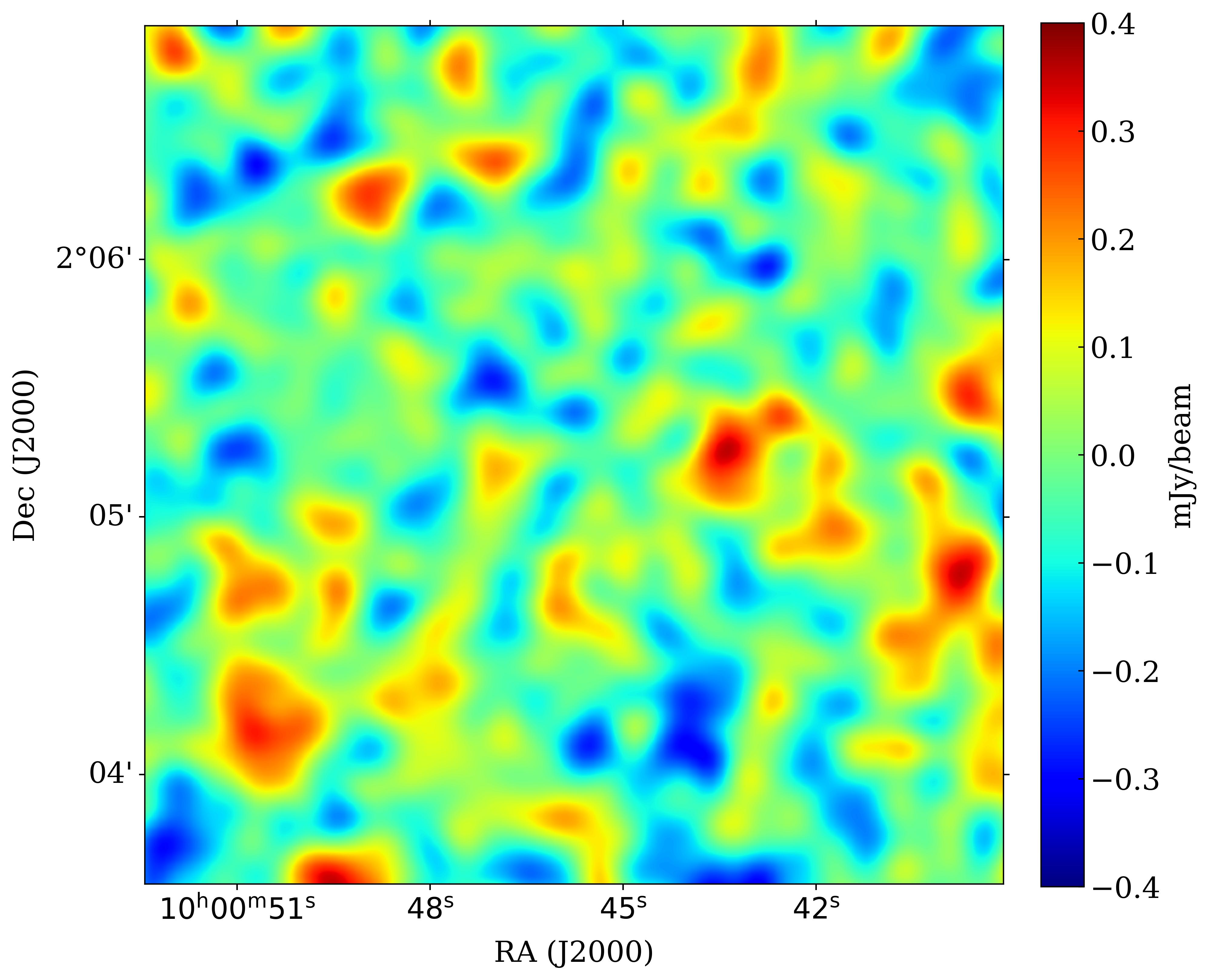}

\end{subfigure}
\end{minipage}
\caption{Results of the fit of $200\arcsec\times200\arcsec$ cutouts of the NIKA2 COSMOS 2\,mm map centered on two cluster candidates. NK2-CL J100004.4+021148.4 (top) has a known redshift $z=0.94$.  In contrast, for NK2-CL J100045.8+020514.3 (bottom) we found no counterpart. The left map corresponds to the 1.2\,mm map centered on the cluster candidate. Then, from left to right, we show the 2\,mm cutout map for the cluster candidate, the best-fit tSZ model, and residuals. For visualization purposes, the maps have been smoothed with a Gaussian 2D kernel of $6\arcsec$ and $10\arcsec$ for the 1.2\,mm and 2\,mm maps, respectively. The effective FWHM for the 1.2\,mm and 2\,mm maps is represented as a black disk in the bottom left corner. The white contours have the same scaling as in Fig.~\ref{fig:hsc}. We observe that the tSZ effect at the cluster position is negative in the 2\,mm band, and not observed at the 1.2\,mm band, as expected.}
\label{fig:MCMC_fit_z}
\end{figure*}

\section{Properties of cluster candidates}
\label{sec:candidates_properties}

\subsection{Integrated tSZ emission}
\label{integrated_quantities}

\indent In order to characterize the tSZ contribution in each cluster candidate, we performed an MCMC analysis by fitting the millimeter emission in the cluster candidate cutout maps (see Append.~\ref{app:clustercandidatesample}) to the spherical model of the tSZ emission presented in Sect.~\ref{Detection_simulations}. We considered a Gaussian likelihood function as in \cite{2023A&A...671A..28M}, with the cluster redshift, $z$, and mass $M_{500}$ as free parameters. We used flat priors similar to the redshift and mass limits presented in Sect.~\ref{subsec:spatialclustermodel}. For the cluster candidates with known counterparts, we considered a Gaussian prior on the redshift, based on the values and uncertainties from the literature. We used the emcee package \citep{emcee} to run the  MCMC analysis. The convergence of the chains was checked using the $\hat{R}$ test of \cite{Gelman_rubin_test}. The mass and redshift posterior distributions were then converted to relevant cluster quantities, the integrated tSZ flux $Y_{500}$ and the cluster's angular size $\theta_{500}$, using Eqs.~\ref{eq:y500} and \ref{eq:theta500}.\\

\begin{figure}[ht]
	\centering
	\includegraphics[width=0.5\textwidth]{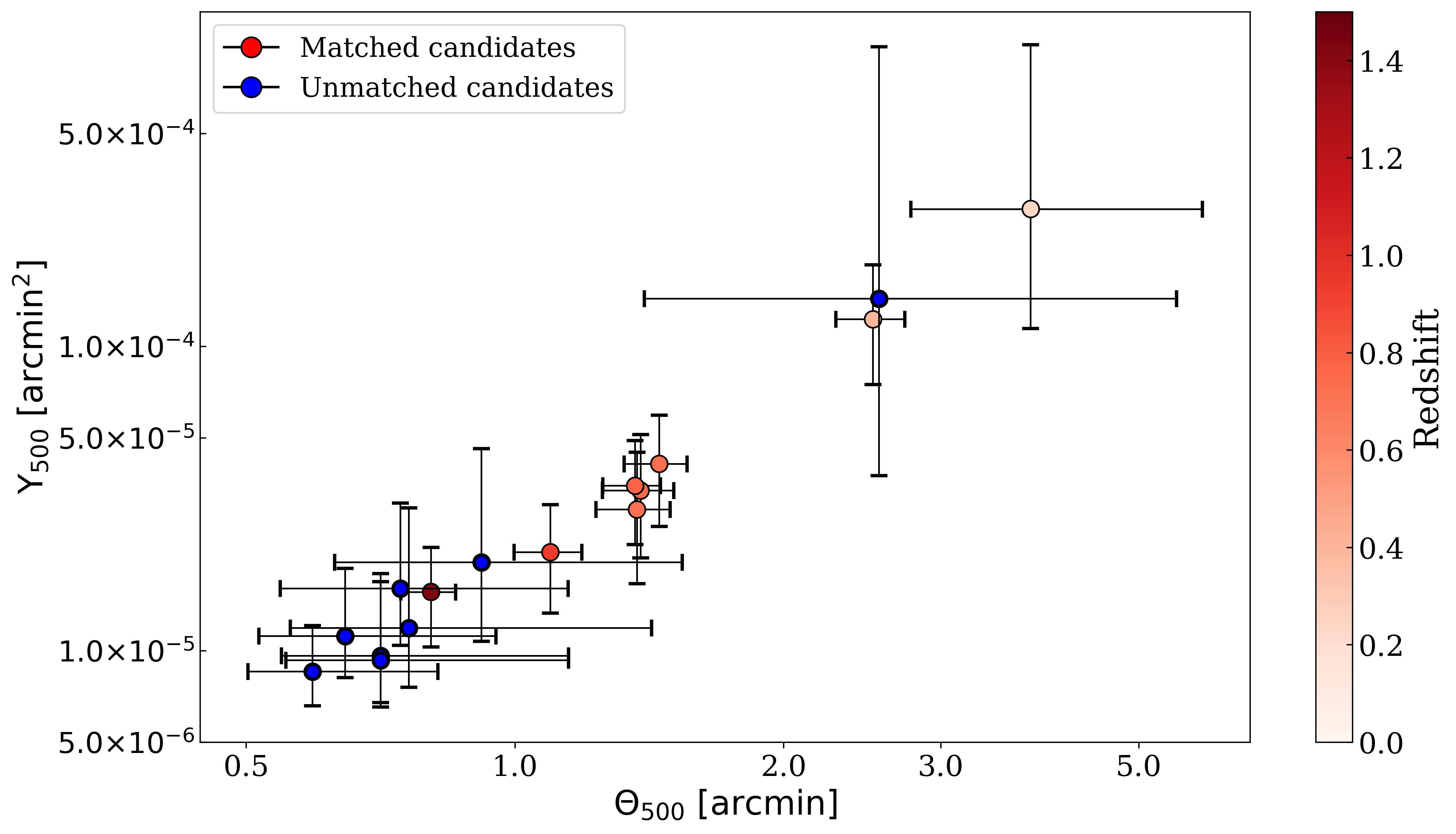}
	\caption{$\theta_{500}$- $Y_{500}$ 68\% confidence values, for each cluster candidate in the NIKA2 sample. Matched and unmatched candidates (see Sect.~\ref{Cluster_candidates}) are depicted in red and blue, respectively. The redshift of matched counterparts is illustrated with the red colorbar.}
	\label{fig:theta500_Y500_final}
\end{figure}

\indent As an illustration, we present in Fig.~\ref{fig:MCMC_fit_z} examples of the results of the fitting procedure at the map level for two cluster candidates. NK2-CL J100004.4+021148.4 (top) has a known redshift $z=0.94$, which is used in the MCMC analysis via a Gaussian prior as discussed before. By contrast, for NK2-CL J100045.8+020514.3 (bottom) we found no counterpart. On the left, we present the 1.2\,mm map centered on the cluster candidate, which is dominated by dusty point sources and has negligible tSZ contribution. Then, from left to right we present the NIKA2 2\,mm cutout map, the best-fit model, and residuals. We observe that the model is a good fit to the data, and the residual maps are consistent with the local noise level, including point sources. We also notice that the latter are much fainter at 2\,mm (as expected for dusty sources), and do not significantly affect the fit. \\
\indent We show in Fig.~\ref{fig:theta500_Y500_final} the estimated tSZ signal and angular size of each cluster candidate in the $Y_{500}-\theta_{500}$ plane. The 68\% confidence values for $Y_{500}$ and $\theta_{500}$ and uncertainties can be found in Table~\ref{tbl:candidates_table}. Posterior likelihood distributions in the $Y_{500}-\theta_{500}$ plane are shown in Fig.~\ref{corner_plots}, presented in Append.~\ref{app:mcmcresults}. We observe that cluster candidates for which a cluster counterpart has been identified, and thus a redshift estimate obtained, present much smaller scatter than unmatched candidates. A tighter redshift prior provides better constraints, as seen from the uncertainties given in Table~\ref{tbl:candidates_table}. However, at low redshift, where candidates are expected to have a large angular size, constraining their tSZ flux, mass, and size becomes challenging because much of the signal is filtered out (e.g., candidate NK2-CL J100011.9+021256.5, which has a counterpart at $z=0.24$).

\begin{figure*}[!t]
	\centering
	\includegraphics[width=0.48\textwidth]{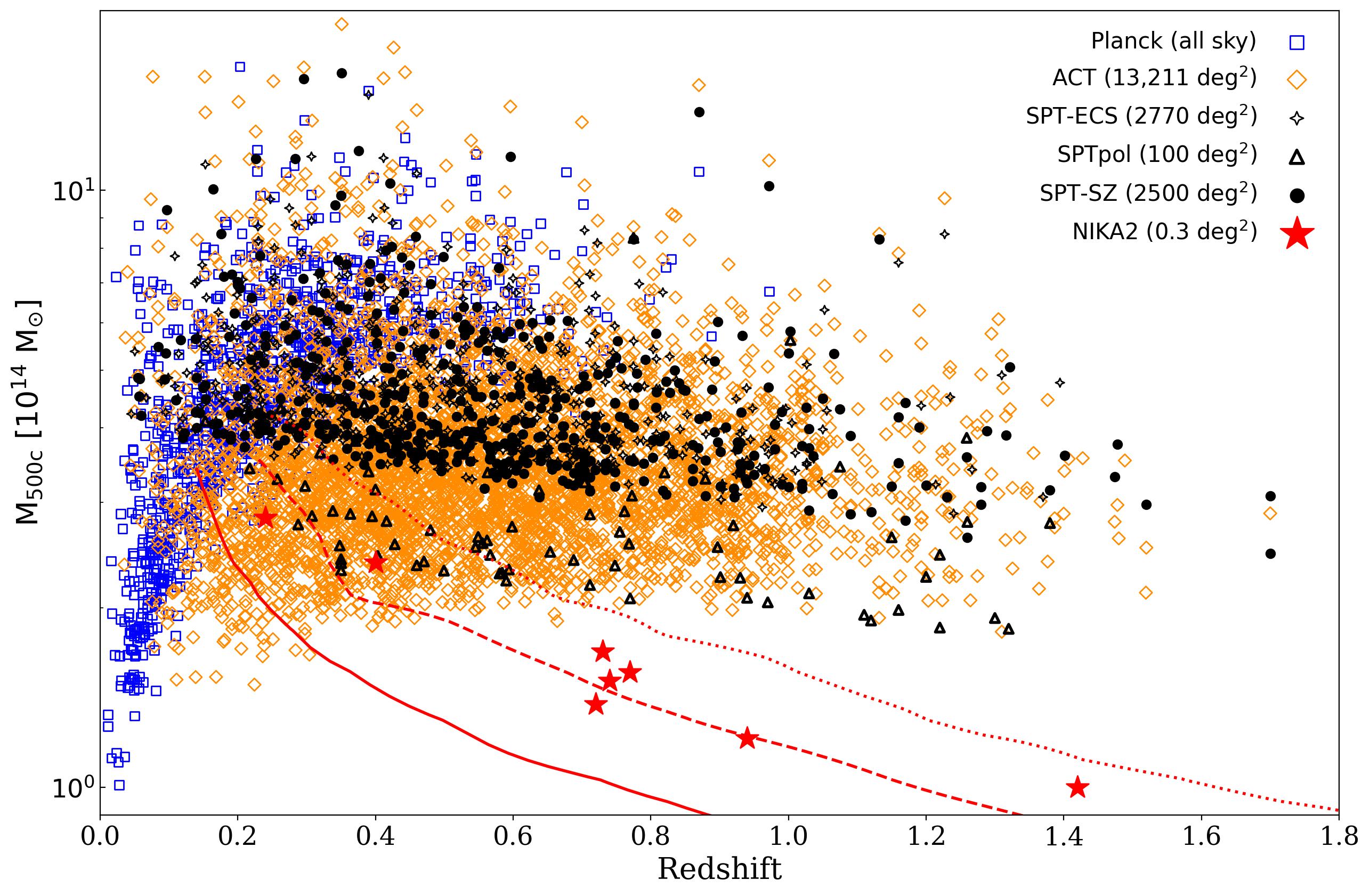}
    \includegraphics[width=0.48\textwidth]{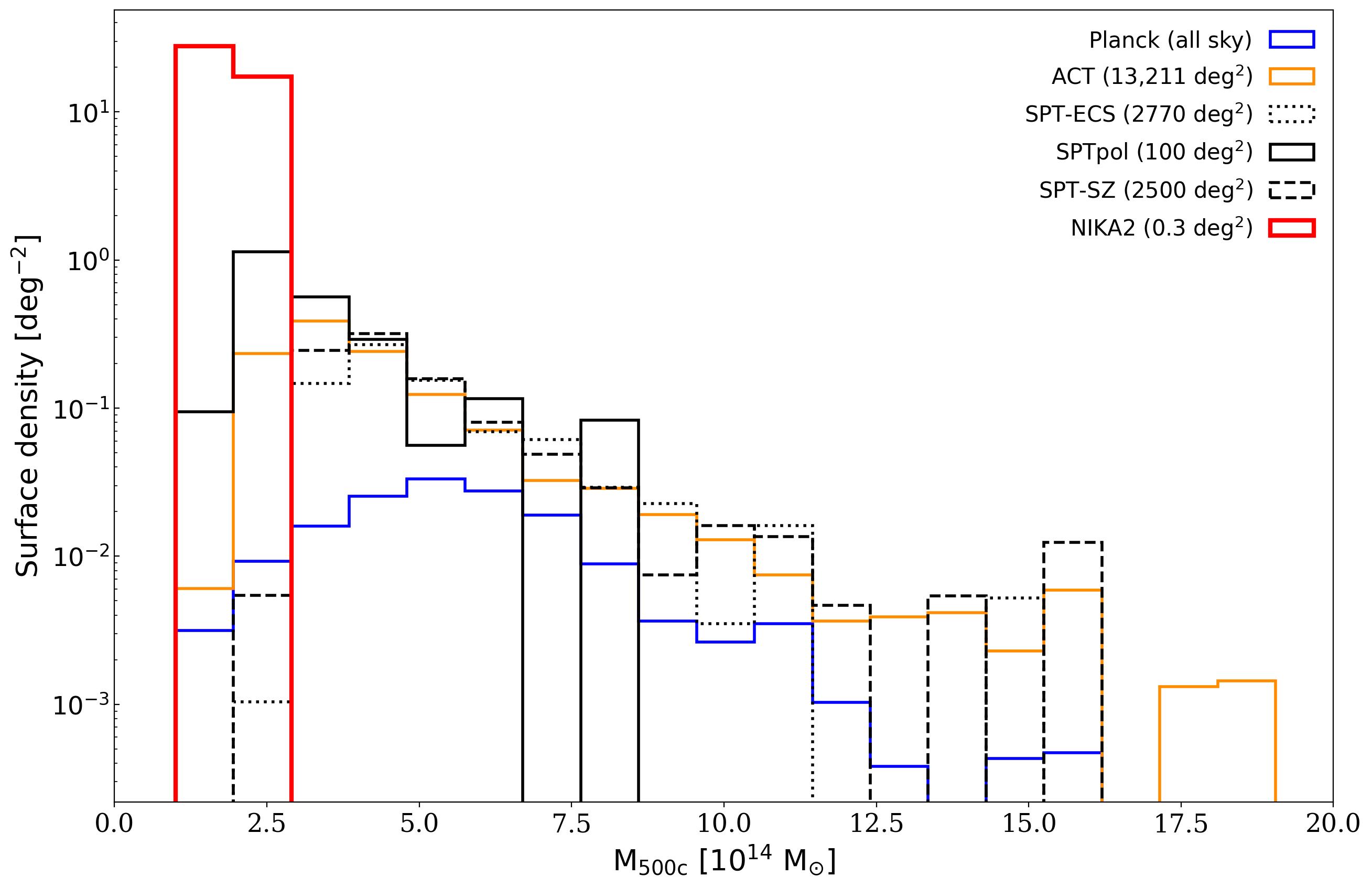}
	\caption{(Left) NIKA2 matched cluster sample in the mass-redshift plane.  Cluster samples from other blind tSZ surveys such as PSZ2 \citep{Planck_XXVII}, ACT DR5 \citep{ACT}, SPT-ECS \citep{SPT-ECS}, SPTpol \citep{SPTpol}, and SPT-SZ \citep{SPT-SZ} are shown for comparison. The 20\%, 50\%, and 80\% completeness contours are shown as solid, dashed, and dotted red lines, respectively. (Right) Cluster surface density as a function of cluster mass. We see that a dedicated NIKA2 survey would complement the currently available tSZ cluster catalogs quite well.
 }
	\label{fig:SZ_cluster_distrib}
\end{figure*}
\subsection{Cluster candidate mass estimates}

\label{mass_redshift}

\indent As discussed above, for matched cluster candidates we can use strong priors on the cluster redshift and then obtain reliable mass estimates from the MCMC analysis presented in Sect.~\ref{integrated_quantities}.
We stress that these mass estimates are only valid within the context of the assumed model (see Sect.~\ref{subsec:spatialclustermodel}), but are still very useful to understand the detection performance and capabilities of NIKA2. The left panel of Fig.~\ref{fig:SZ_cluster_distrib} shows the mass-redshift distribution for our matched sample and several other large SZ surveys. We also show the 20 \%, 50\%, and 80\% completeness of our NIKA2 survey (see Sect.~\ref{Simulations}) contours as solid, dashed, and dotted red lines. The mass 68\% confidence values and uncertainties are also presented in Table~\ref{tbl:candidates_table}.
We see that our completeness contours are steeper at high redshifts compared to other surveys, as expected from NIKA2's compact beam. This is because the tSZ flux of distant (and compact) clusters is less affected by beam dilution, making it more sensitive to fainter, lower-mass clusters.\\
\indent The median mass of the NIKA2 cluster candidate sample is $M_{500} = 1.54^{+0.76}_{-0.30}\times10^{14}M_{\odot}$. Overall, confirmed NIKA2 detections correspond to intermediate and high-redshift, low-mass clusters, probing a region in the mass-redshift plane where clusters are not detected by other millimeter instruments and surveys. As the COSMOS region was selected for being relatively empty, we do not expect to find high-mass clusters. We observe that a dedicated NIKA2 cluster survey would significantly extend current tSZ cluster surveys toward low-mass and high-redshift clusters, as shown in the right panel of Fig.~\ref{fig:SZ_cluster_distrib}.

\section{Summary and conclusion}
\label{Conclusion}

The detection of low-mass, high-redshift galaxy clusters is challenging due to their compact nature, but they provide great insights into deviations from self-similarity at lower cluster masses \citep{Deviations_self_similarity}, and the formation of large-scale structures.\\
\indent In this work, we have presented the first blind detection of galaxy clusters at mm wavelength via the tSZ effect at high angular resolution (18.5$^{\prime \prime}$) using the NIKA2 camera. We used the existing 195 hours of NIKA2 2\,mm data of the COSMOS field, which is part of the N2CLS Large Program. We constructed an adapted 2\,mm map of the region and then applied matched-filtering techniques to optimally extract the tSZ signal and identify cluster candidates.\\
\indent The NIKA2 sample consists of 16 cluster candidates with $\mathrm{S/N}>4$. Using realistic simulations of the dataset and of the clusters signal, we estimated the completeness and purity of the sample. Purity reaches 60\% for S/N > 5. Completeness is above 80\% for cluster masses $M_{500}>2\times 10^{14} M_{\odot}$. For eight of the cluster candidates, we found a counterpart by searching the SIMBAD, NED, or VizieR databases. Out of these eight clusters, seven are matched with optical and infrared clusters, and one with an X-ray cluster. The median redshift of the sample is $z\sim0.74$, with values ranging from $0.24<z<1.42$, with two detections at high redshift $z> 0.9$. \\
\indent We used available photometric and spectroscopic redshift catalogs to complete the previous search. For most of the matched candidates, their redshift estimates match with a peak in the galaxies redshift distribution. Some unmatched candidates also show a well-defined peak at intermediate or high redshift. \\
\indent The integrated tSZ emission $Y_{500}$ and angular size $\theta_{500}$ of each cluster candidate were estimated using the universal pressure profile and scaling relations from \cite{A10}. For the eight cluster candidates with redshift estimates, we derived reliable mass estimates. The median mass of the sample is $M_{500} = 1.54\times10^{14}M_{\odot}$, spanning from 1.0 to $3.1\times10^{14}M_{\odot}$ at intermediate to high redshift.\\
\indent It was then highlighted that the NIKA2 camera enables the detection of high-redshift and low-mass clusters thanks to its very high angular resolution and sensitivity. It is thus possible to explore a region in the mass-redshift plane that is not covered by current large cluster surveys based on tSZ. At sub 20 arcsec scales, the CMB contribution is negligible (see Sect.~\ref{Detection_simulations}) and dusty galaxies can be resolved on a finer scale than the SZ signal \citep{N2CLS_count,ponthieu2025}. This allows us to use single band match-filtering instead of a more standard multifrequency approach in CMB experiments. 
However, the latter would help reduce possible residual radio source contamination. By contrast, we also notice reduced S/N compared to larger and deeper CMB surveys. This is consistent with the fact that, in general, very large millimeter telescopes typically have a FoV of only a few tens of arcmins and are not designed for survey-mode observations. 

\section*{Data availability}

Table~\ref{tbl:candidates_table} is available at the CDS via anonymous ftp to \url{http://cdsarc.u-strasbg.fr/} (\url{130.79.128.5}) or via \url{http://cdsweb.u-strasbg.fr/cgi-bin/qcat?J/A+A/}.
The N2CLS final maps and catalogs are available online on the survey's home page: \url{https://data.lam.fr/n2cls/home}.

\begin{acknowledgements}

We thank Anna Niemiec for useful discussions on optical data images. This work is based on observations carried out under project numbers 192-16 with the IRAM 30m telescope. IRAM is supported by INSU/CNRS (France), MPG (Germany) and IGN (Spain). We would like to thank the IRAM staff for their support during the observation campaigns. The NIKA2 dilution cryostat has been designed and built at the Institut N\'eel. In particular, we acknowledge the crucial contribution of the Cryogenics Group, and in particular Gregory Garde, Henri Rodenas, Jean-Paul Leggeri, Philippe Camus. This work has been partially funded by the Foundation Nanoscience Grenoble and the LabEx FOCUS ANR-11-LABX-0013. This work is supported by the French National Research Agency under the contracts "MKIDS", "NIKA" and ANR-15-CE31-0017 and in the framework of the "Investissements d’avenir” program (ANR-15-IDEX-02). This work has been supported by the GIS KIDs. This work has benefited from the support of the European Research Council Advanced Grant ORISTARS under the European Union’s Seventh Framework Programme (Grant Agreement no. 291294). A. R. acknowledges financial support from the Italian Ministry of University and Research - Project Proposal CIR01\_00010. R. A. acknowledges support from the Programme National Cosmology et Galaxies (PNCG) of CNRS/INSU with INP and IN2P3, co-funded by CEA and CNES. R. A. was supported by the French government through the France 2030 investment plan managed by the National Research Agency (ANR), as part of the Initiative of Excellence of Université Côte d'Azur under reference number ANR-15-IDEX-01. M.M.E. acknowledges the support of the French Agence Nationale de la Recherche (ANR), under grant ANR-22-CE31-0010. M.D.P. acknowledges PRIN-MIUR grant 20228B938N "Mass and selection biases of galaxy clusters: a multi-probe approach" funded by the European Union Next generation EU, Mission 4 Component 1  CUP C53D2300092 0006. A. Maury acknowledges support the funding from the European Research Council (ERC) under the European Union’s Horizon 2020 research and innovation programme (Grant agreement No. 101098309 - PEBBLES).
This work made use of Astropy, a community-developed core Python package and an ecosystem of tools and resources for astronomy \citep{astropy:2013, astropy:2018, astropy:2022}. This research made use of Photutils, an Astropy package for
detection and photometry of astronomical sources \citep{photutils}.
\end{acknowledgements}

\bibliography{bibliography}

\begin{appendix} 

\onecolumn
\section{Cluster candidate sample}
\label{app:clustercandidatesample}

We present in Fig.~\ref{cutout}  zoom-in views of the COSMOS NIKA2 2\,mm map centered at the location of each of the cluster candidates in Table~\ref{tbl:candidates_table}.  We can distinguish negative signal in the center of each of the maps, as expected for the tSZ effect in a cluster of galaxies. We can also observe some positive point sources corresponding to those discussed in detail in \citet{N2CLS_count}.

\begin{figure*}[h!]
\centering
\begin{subfigure}{0.32\textwidth}
  \includegraphics[width = 1\textwidth]{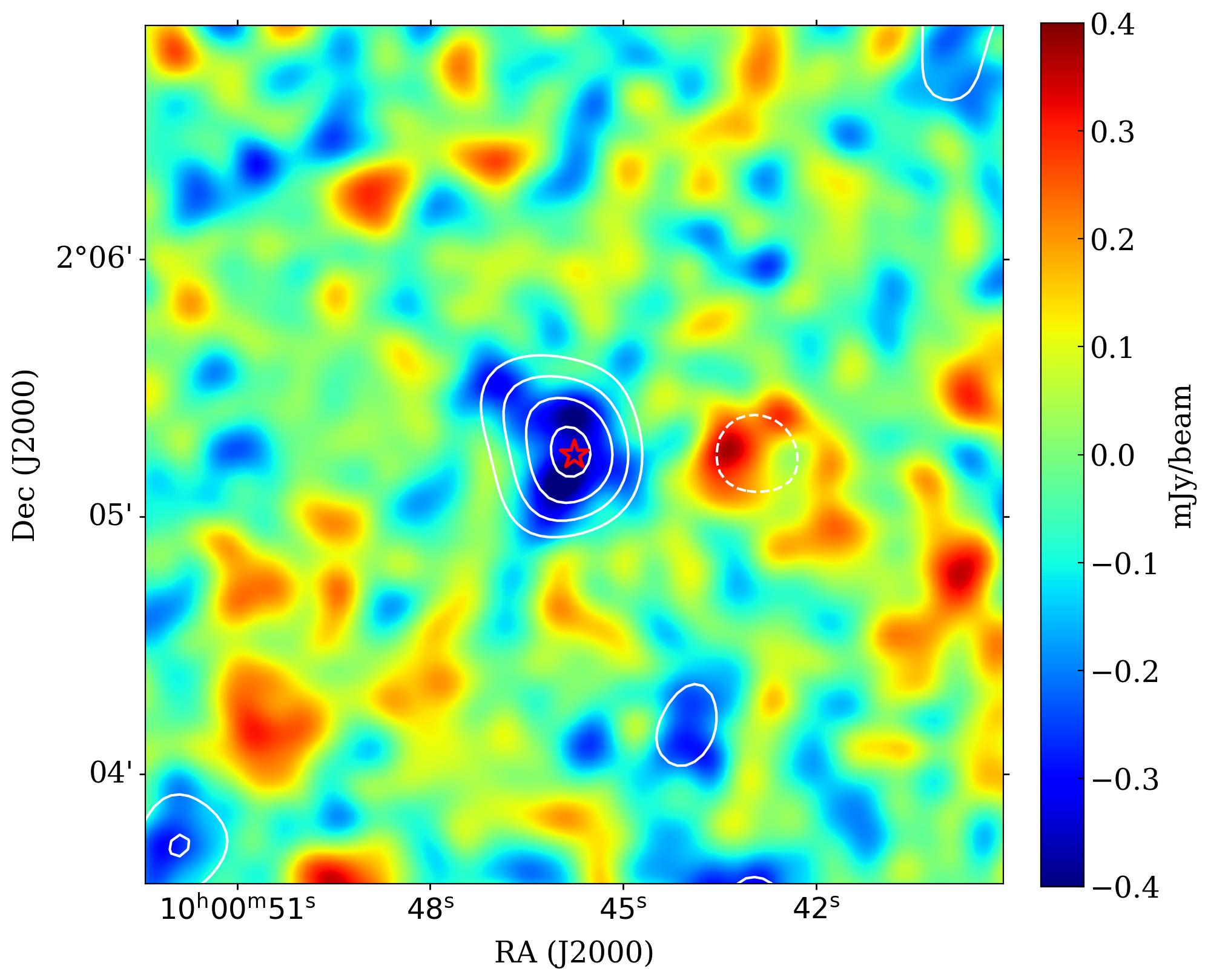}
  \caption{NK2-CL J100045.8+020514.3}
\end{subfigure}\hfill
\begin{subfigure}{0.32\textwidth}
  \includegraphics[width = 1\textwidth]{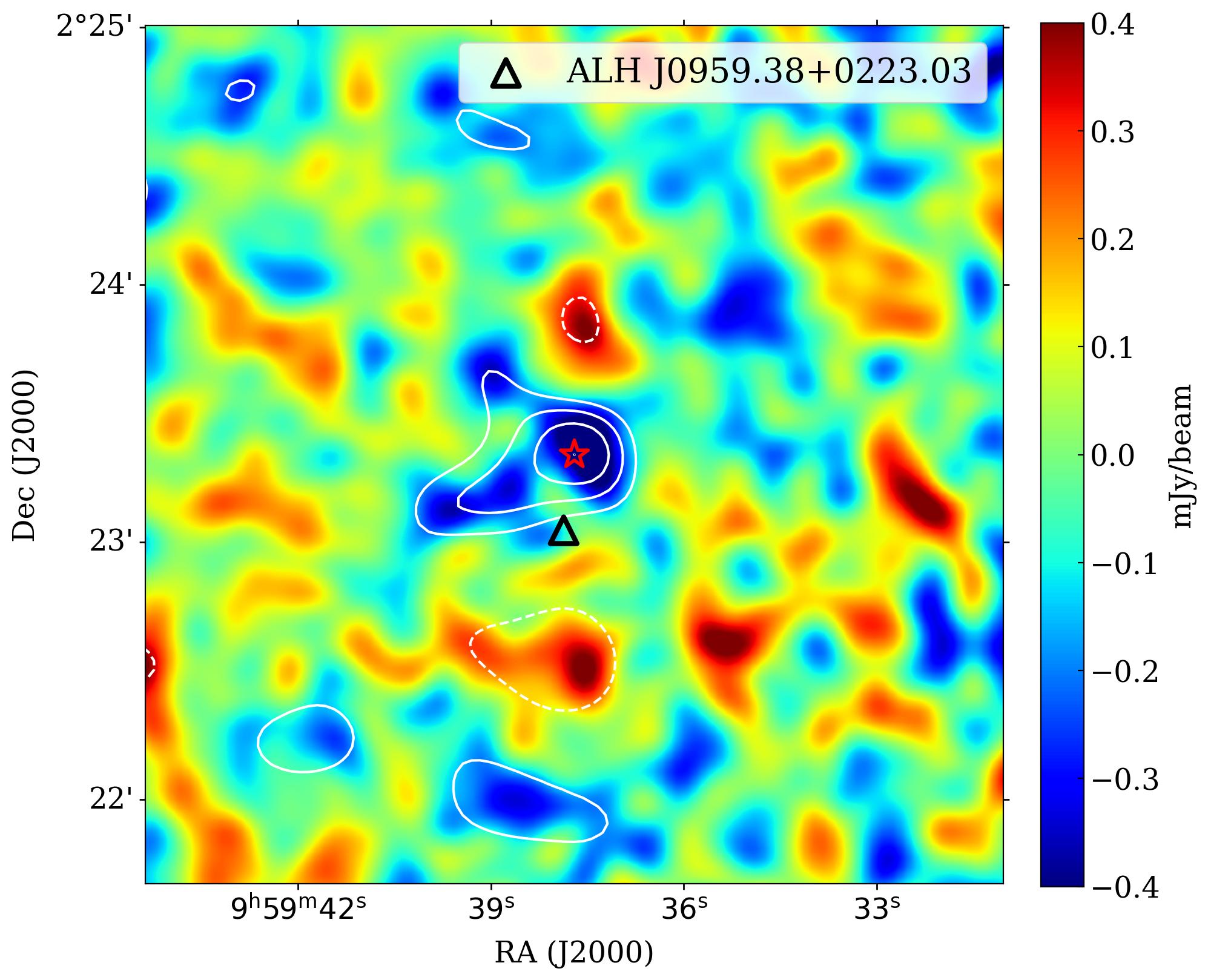}
  \caption{NK2-CL J095937.7+022320.4 ($z$=0.74)}
\end{subfigure}\hfill
\begin{subfigure}{0.32\textwidth}
  \includegraphics[width = 1\textwidth]{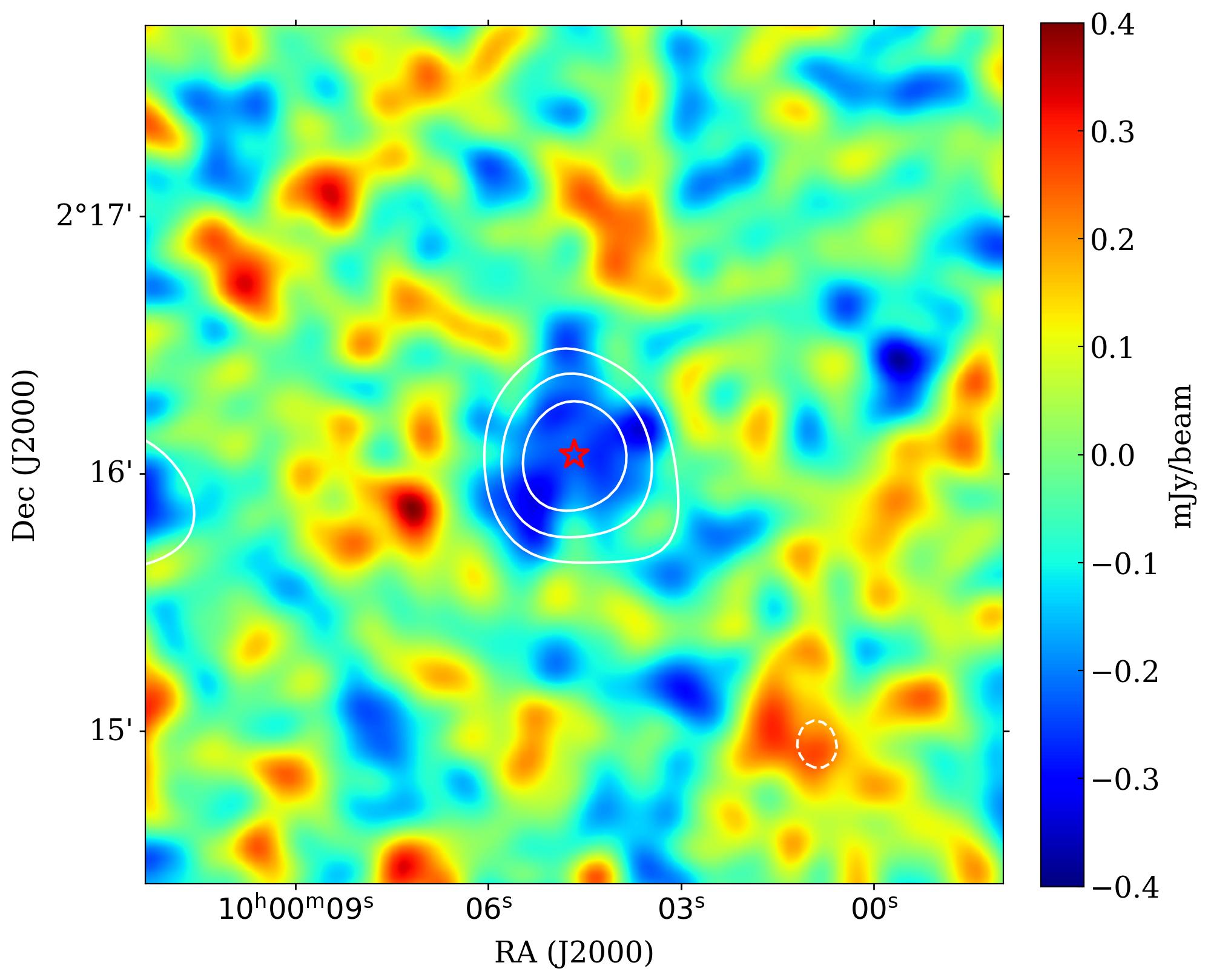}
  \caption{NK2-CL J100004.7+021604.4}
\end{subfigure}
\medskip
\begin{subfigure}{0.32\textwidth}
  \includegraphics[width = 1\textwidth]{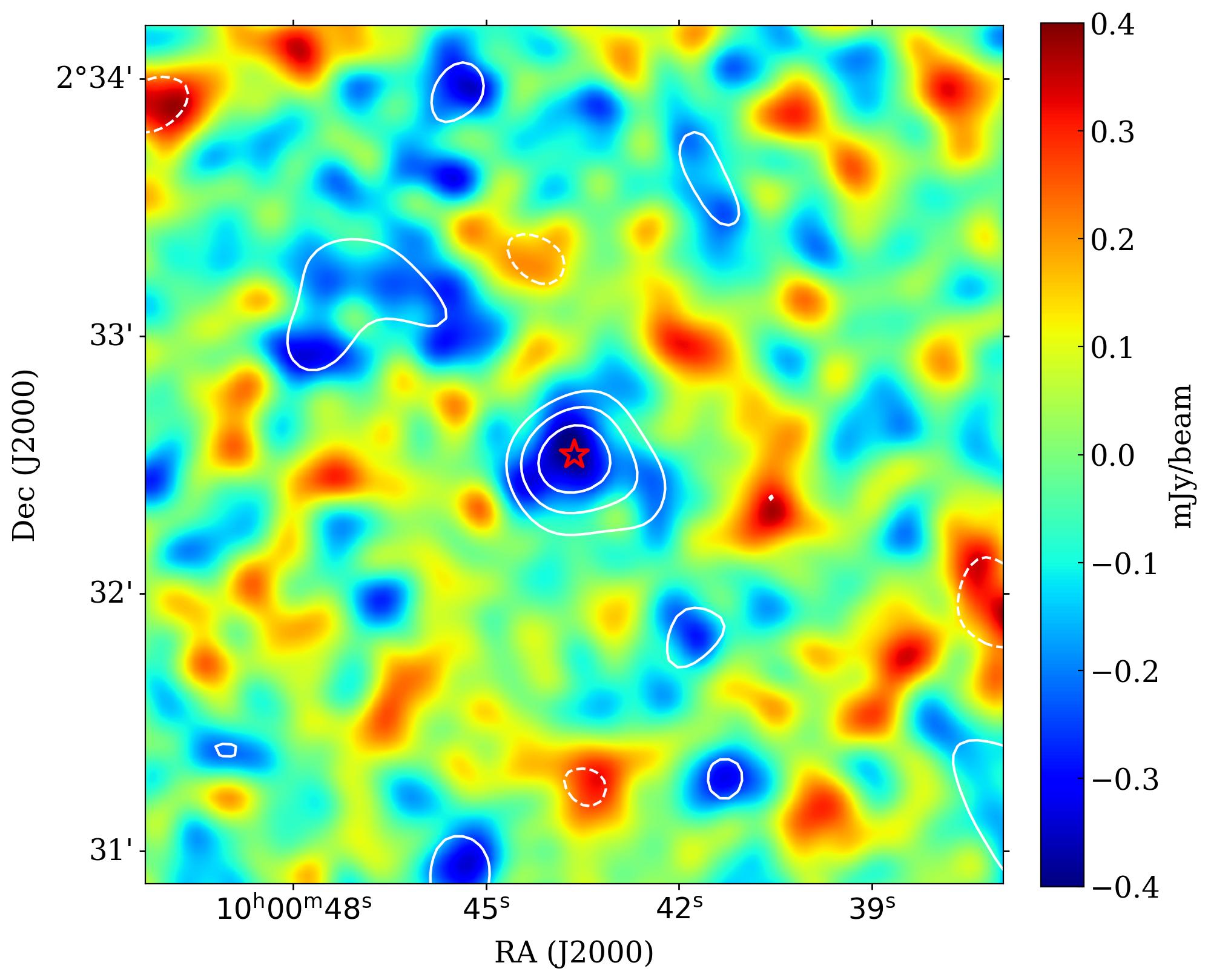}
  \caption{NK2-CL J100043.6+023232.4}
\end{subfigure}\hfill
\begin{subfigure}{0.32\textwidth}
  \includegraphics[width = 1\textwidth]{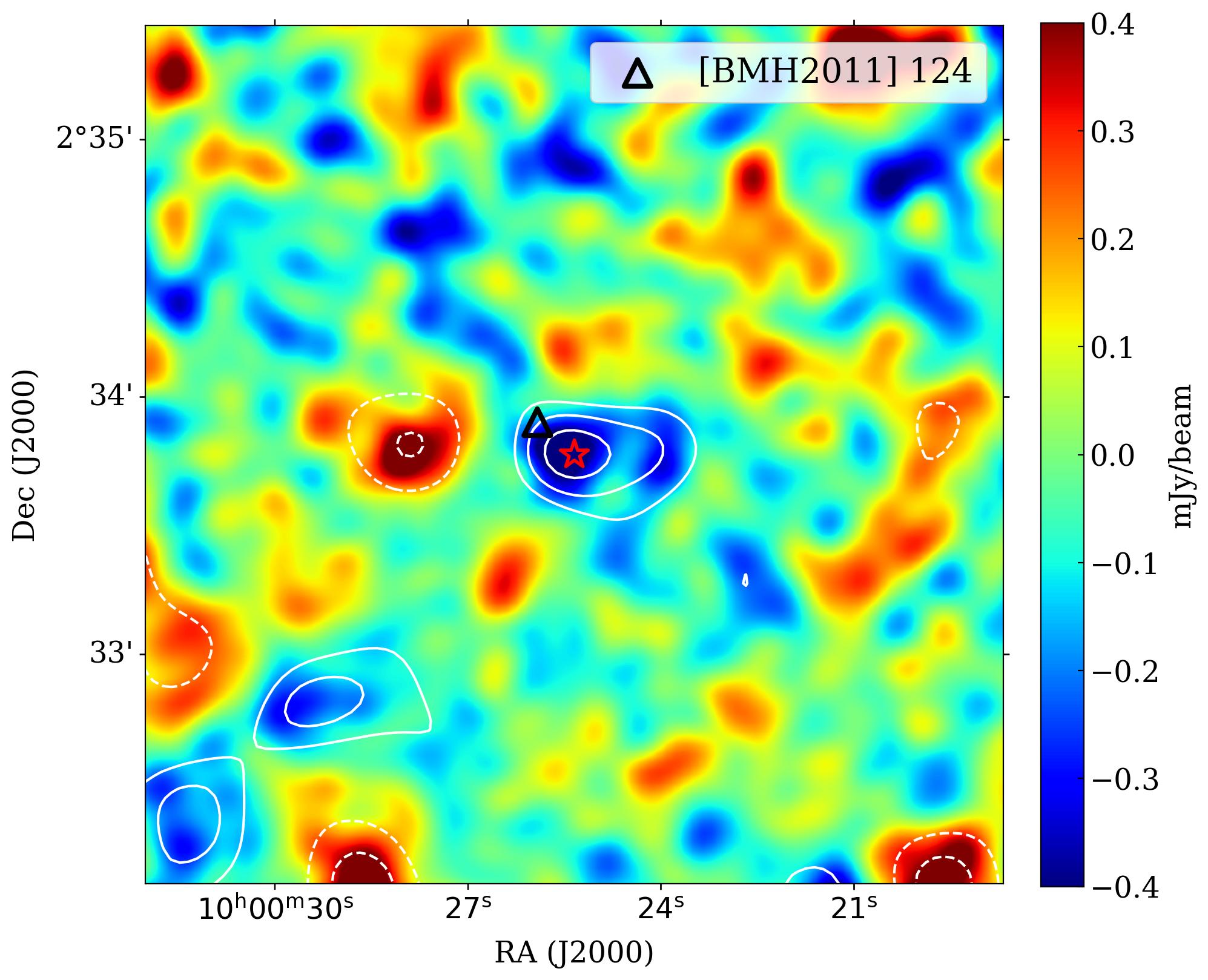}
  \caption{NK2-CL J100025.3+023346.4 ($z$=0.72)}
\end{subfigure}\hfill
\begin{subfigure}{0.32\textwidth}
  \includegraphics[width = 1\textwidth]{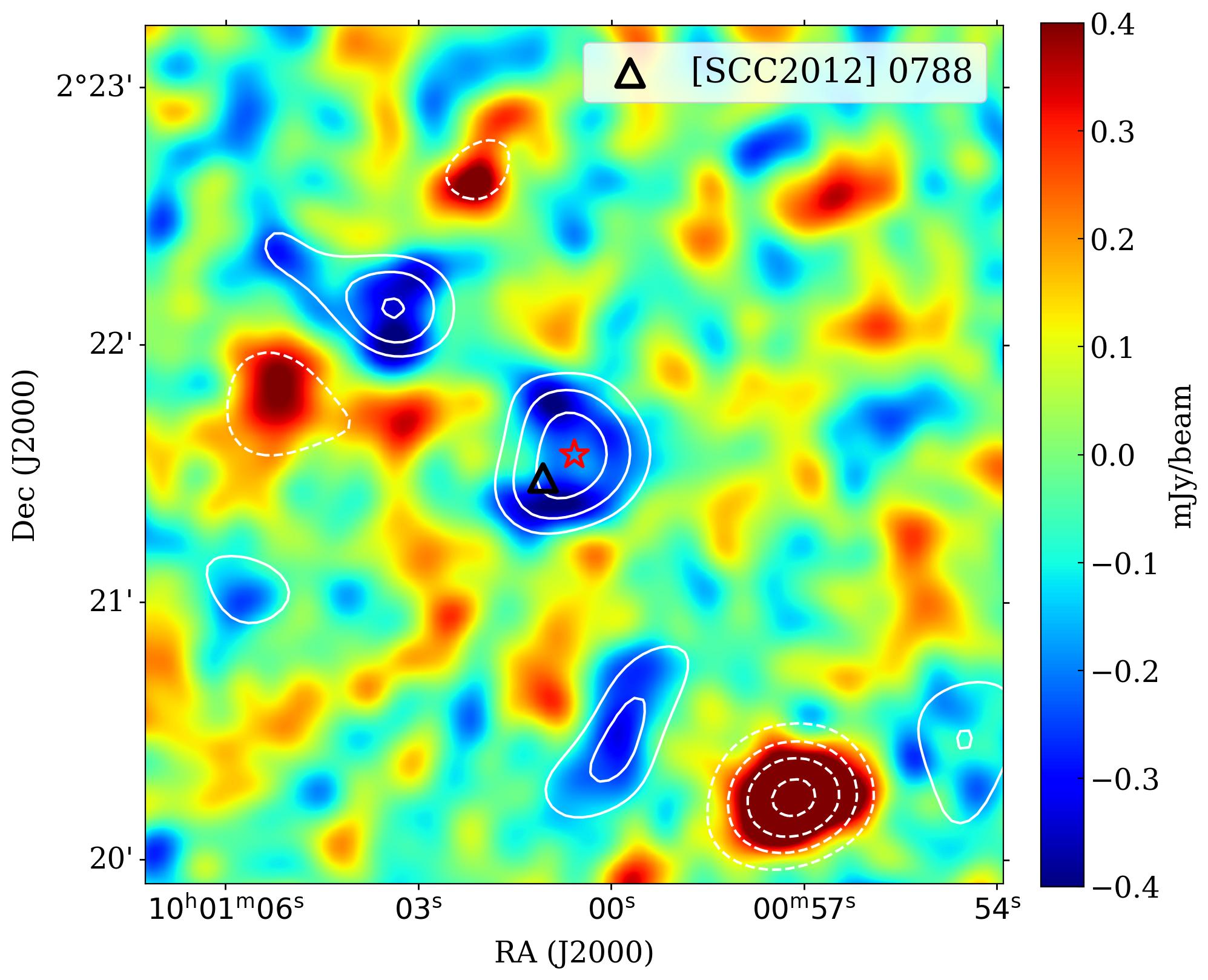}
  \caption{NK2-CL J100100.6+022134.4 ($z$=0.77)}
\end{subfigure}
\medskip
\begin{subfigure}{0.32\textwidth}
  \includegraphics[width = 1\textwidth]{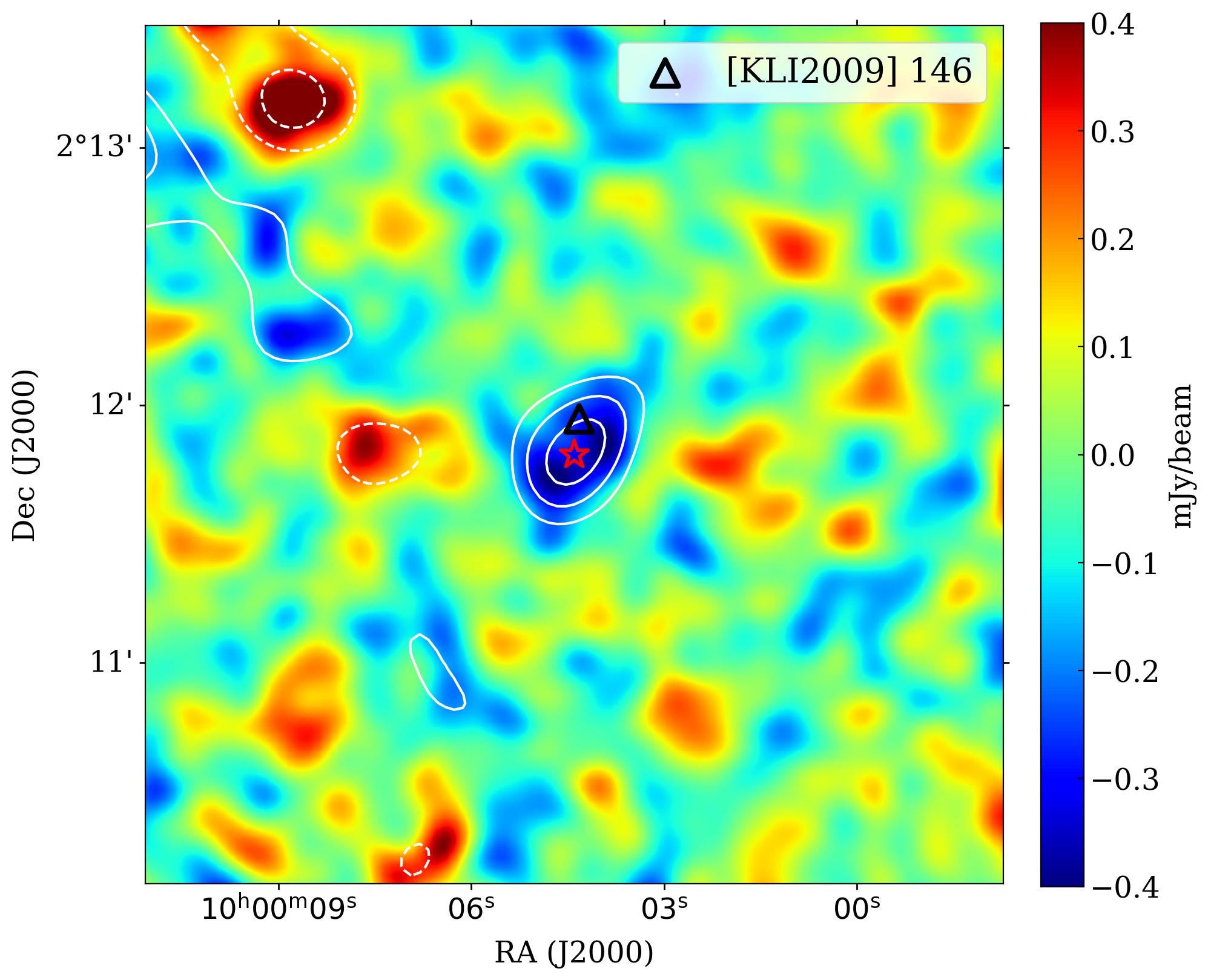}
  \caption{NK2-CL J100004.4+021148.4 ($z$=0.94)}
\end{subfigure}\hfill
\begin{subfigure}{0.32\textwidth}
  \includegraphics[width = 1\textwidth]{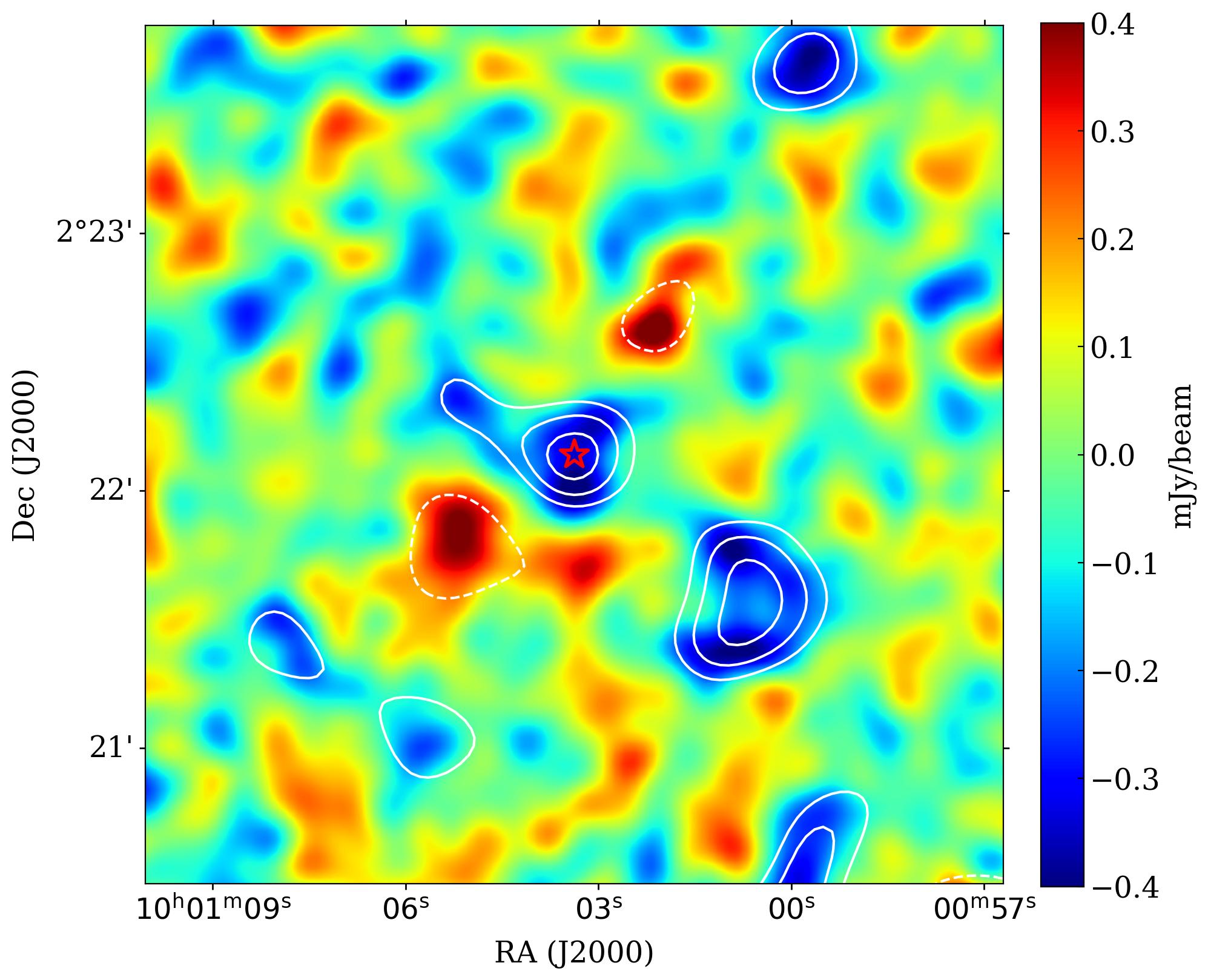}
  \caption{NK2-CL J100103.4+022208.4}
\end{subfigure}\hfill
\begin{subfigure}{0.32\textwidth}
  \includegraphics[width = 1\textwidth]{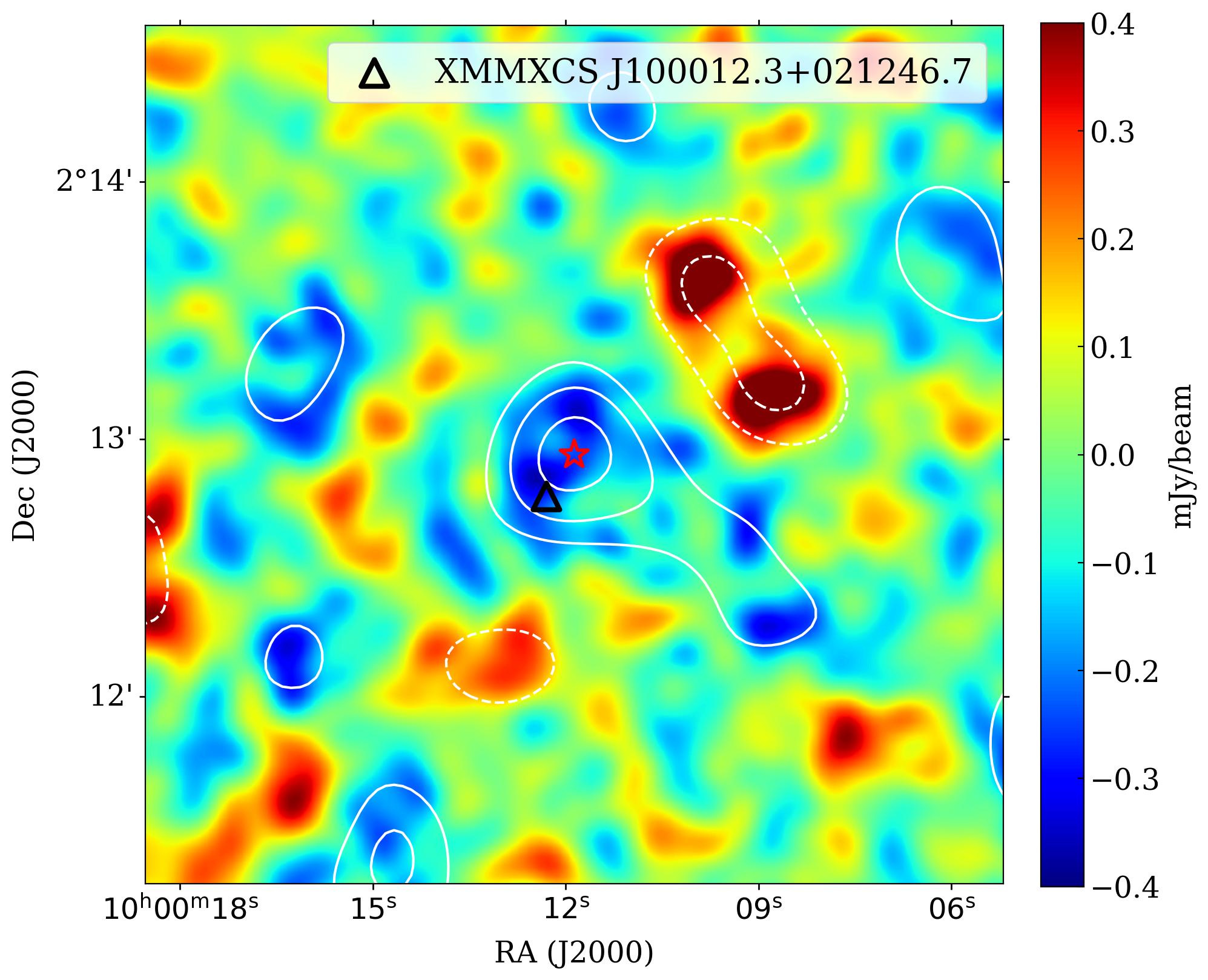}
  \caption{NK2-CL J100011.9+021256.5 ($z$=0.24)}
\end{subfigure}
\caption{$200\arcsec\times200\arcsec$ cutouts of the NIKA2 COSMOS 2\,mm map around each cluster candidate. The center of the candidate (red star) shows clear negative signal. Clusters found in the literature are shown as black triangles. For each candidate, we show as white contours the S/N levels in the matched-filtered map where its detection S/N is maximum (see Sect.~\ref{Sample identification}). The contours start at $\pm 2\sigma$ and are spaced by $\pm 1\sigma$. The maps have been smoothed with a 10$\arcsec$ gaussian kernel for display purposes.}
\label{cutout}
\end{figure*}

\newpage
\begin{figure*}\ContinuedFloat
\centering
\begin{subfigure}{0.32\textwidth}
  \includegraphics[width = 1\textwidth]{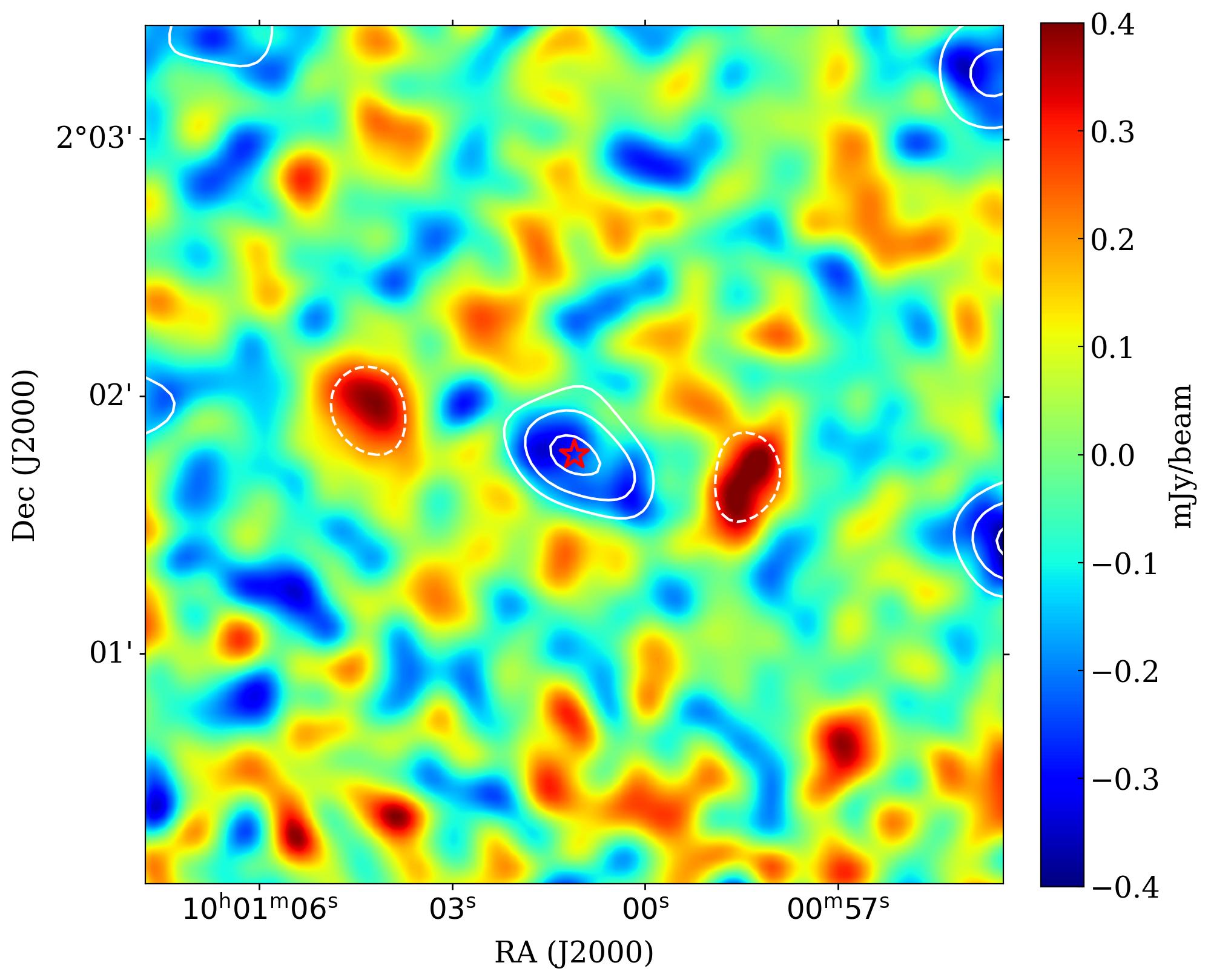}
  \caption{NK2-CL J100101.1+020146.6}
\end{subfigure}\hfill
\begin{subfigure}{0.32\textwidth}
  \includegraphics[width = 1\textwidth]{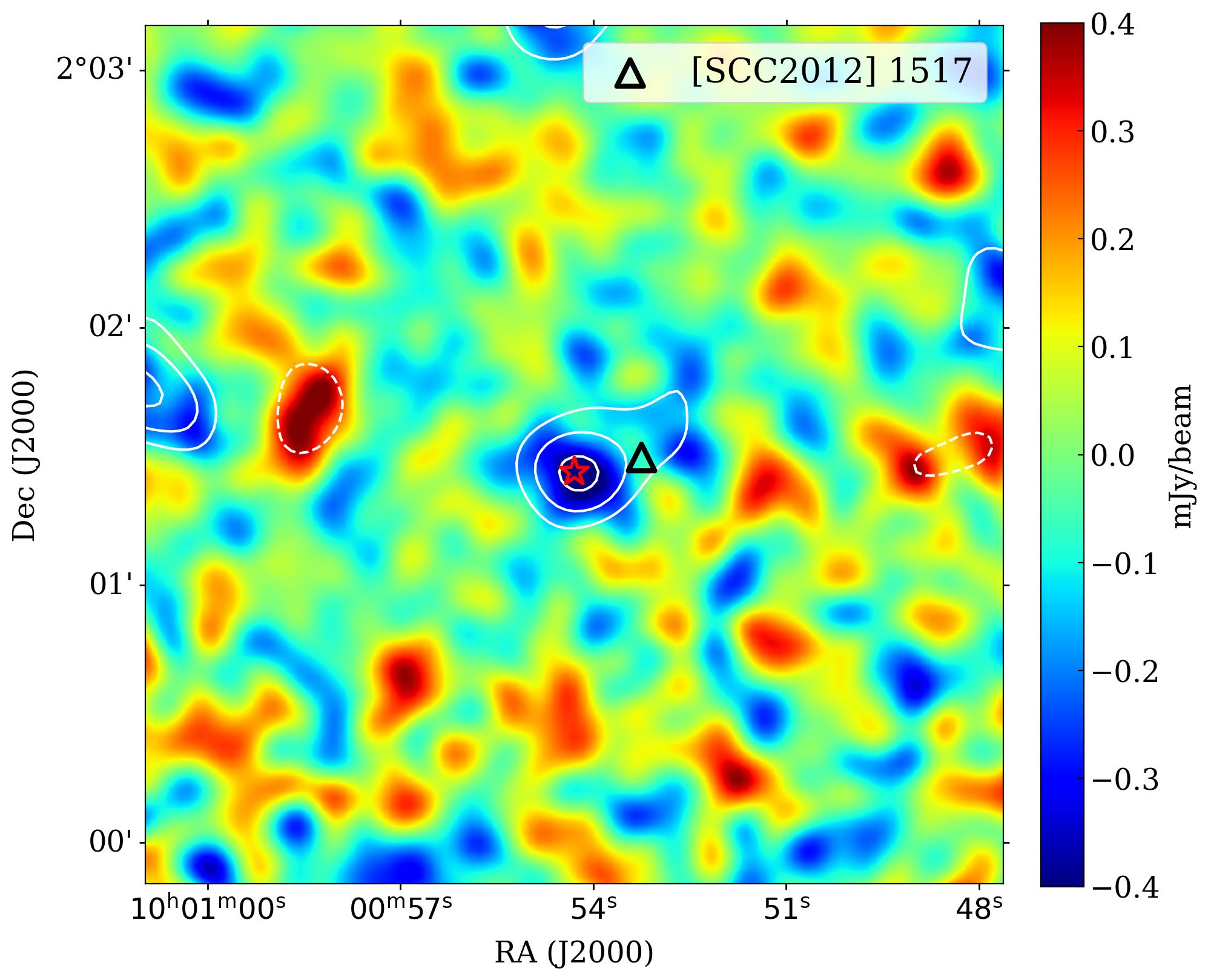}
  \caption{NK2-CL J100054.3+020126.4 ($z$=1.42)}
\end{subfigure}\hfill
\begin{subfigure}{0.32\textwidth}
  \includegraphics[width = 1\textwidth]{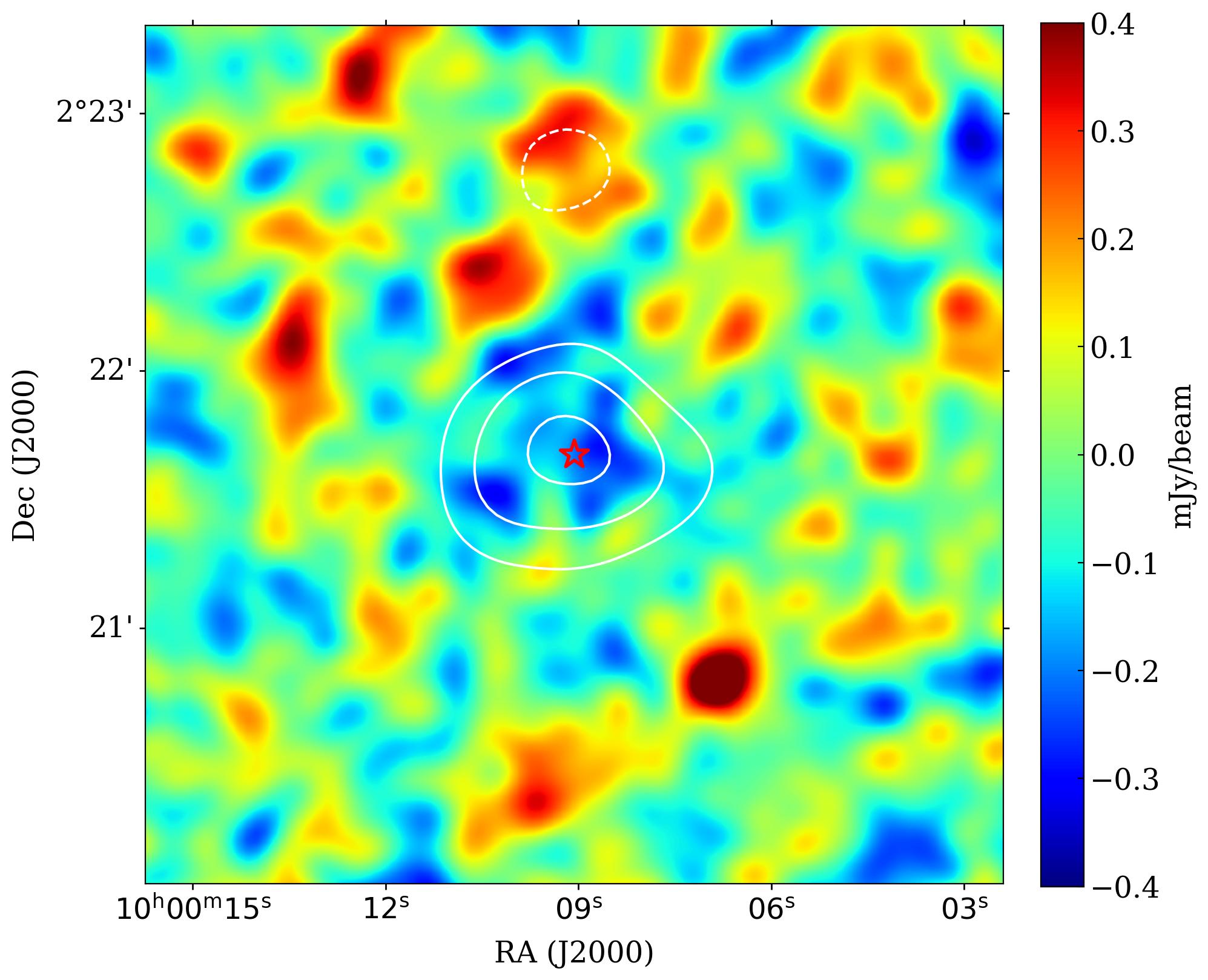}
  \caption{NK2-CL J100009.1+022140.3}
\end{subfigure}\hfill
\medskip
\begin{subfigure}{0.32\textwidth}
  \includegraphics[width = 1\textwidth]{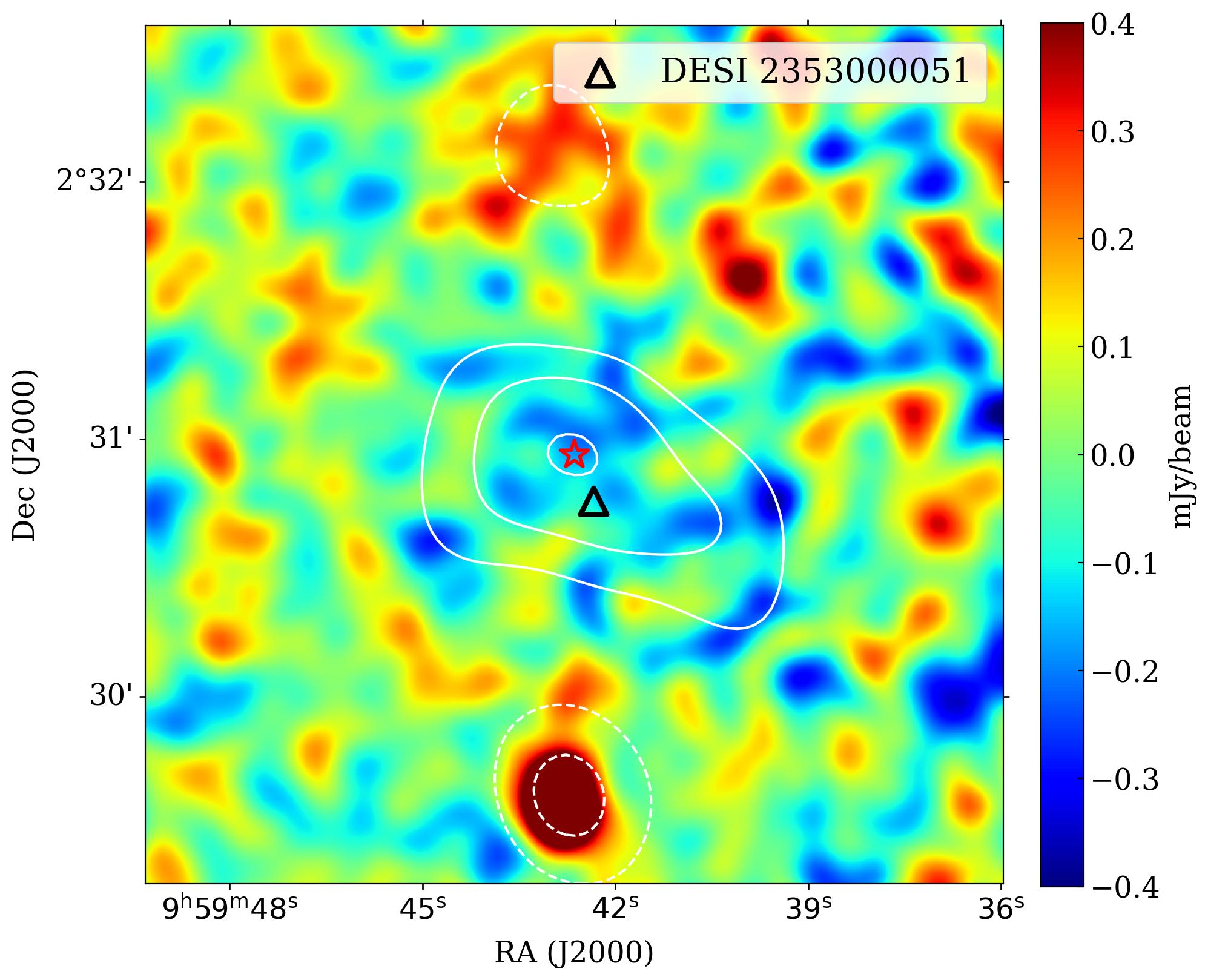}
  \caption{NK2-CL J095942.6+023056.5($z$=0.73)}
\end{subfigure}\hfill
\begin{subfigure}{0.32\textwidth}
  \includegraphics[width = 1\textwidth]{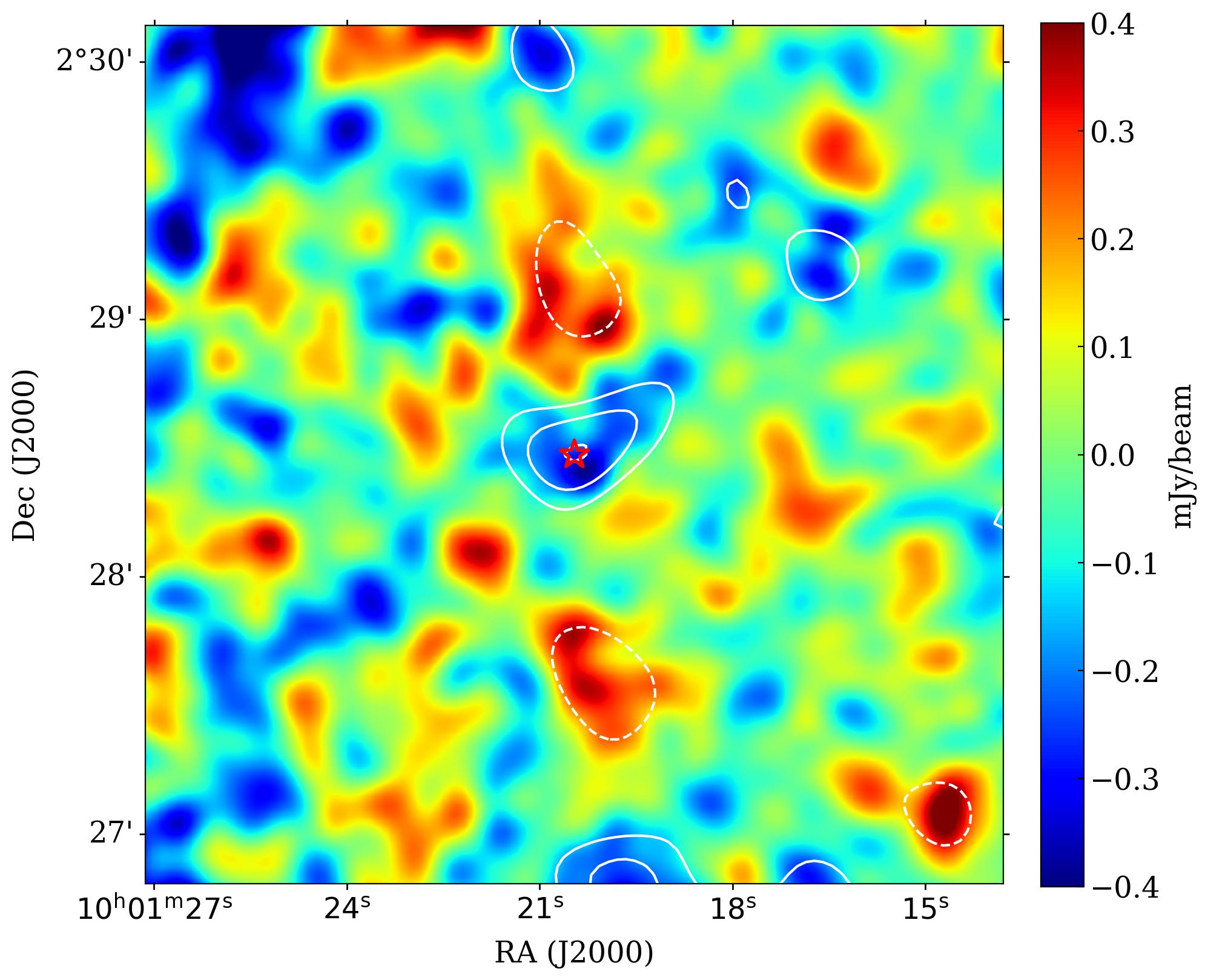}
  \caption{NK2-CL J100120.5+022828.2}
\end{subfigure}\hfill
\begin{subfigure}{0.32\textwidth}
  \includegraphics[width = 1\textwidth]{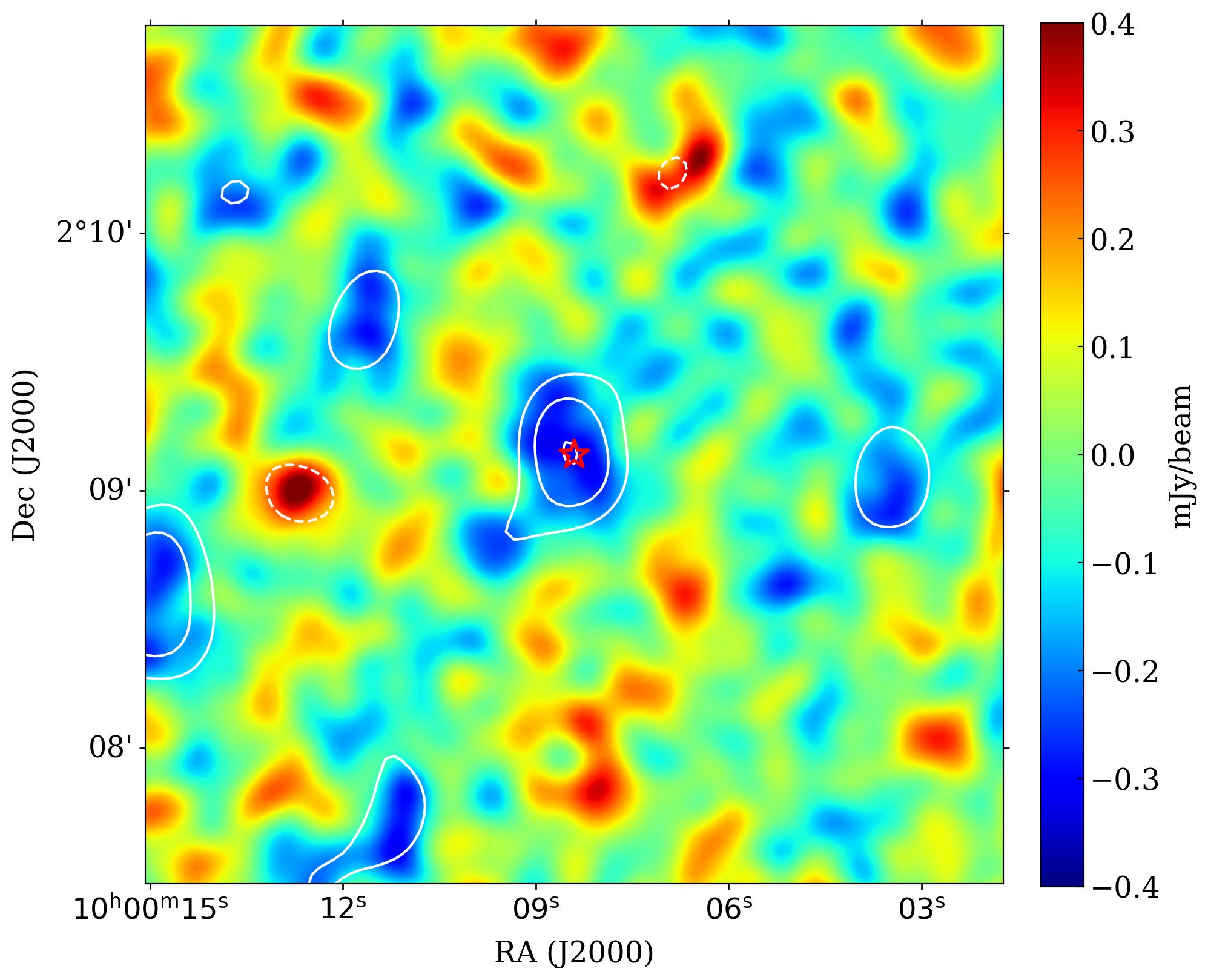}
  \caption{NK2-CL J100008.4+020908.3}
\end{subfigure}\hfill
\medskip
\begin{subfigure}{0.32\textwidth}
  \includegraphics[width = 1\textwidth]{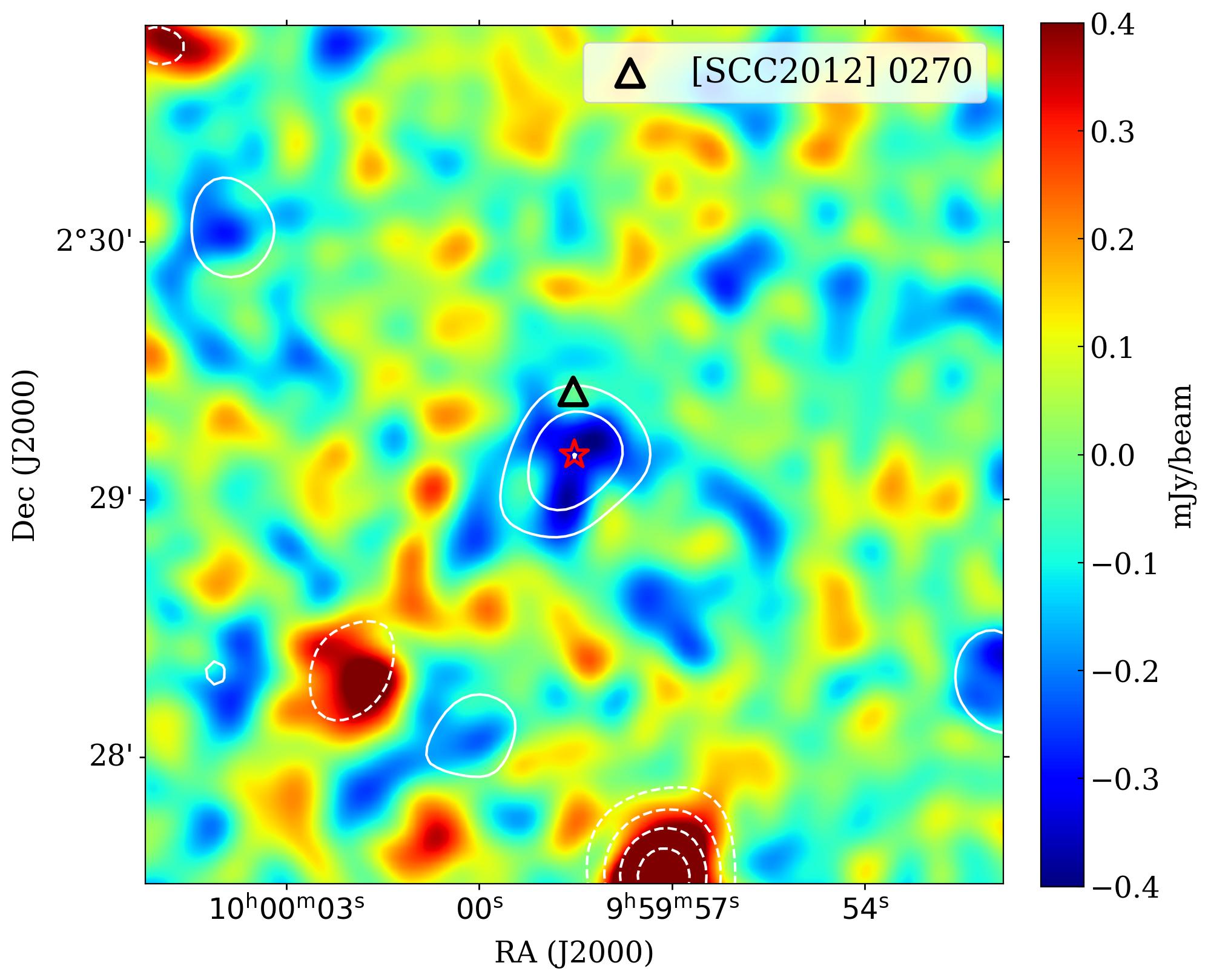}
  \caption{NK2-CL J095958.5+022910.4 ($z$=0.40)}
\end{subfigure}
\caption{continued.}
\end{figure*}

\clearpage
\section{Photometric and spectroscopic redshift histograms}
\label{app:redshift}

In this section, we present the photometric and spectroscopic redshift on patches of 1 arcmin radius centered at the position of the NIKA2 cluster candidates. For NIKA2 cluster candidates with redshift estimates in the literature, the value is represented as a vertical dashed line.

\begin{figure*}[h!]
\centering
\begin{subfigure}{0.3\textwidth}
  \includegraphics[width = 1\textwidth]{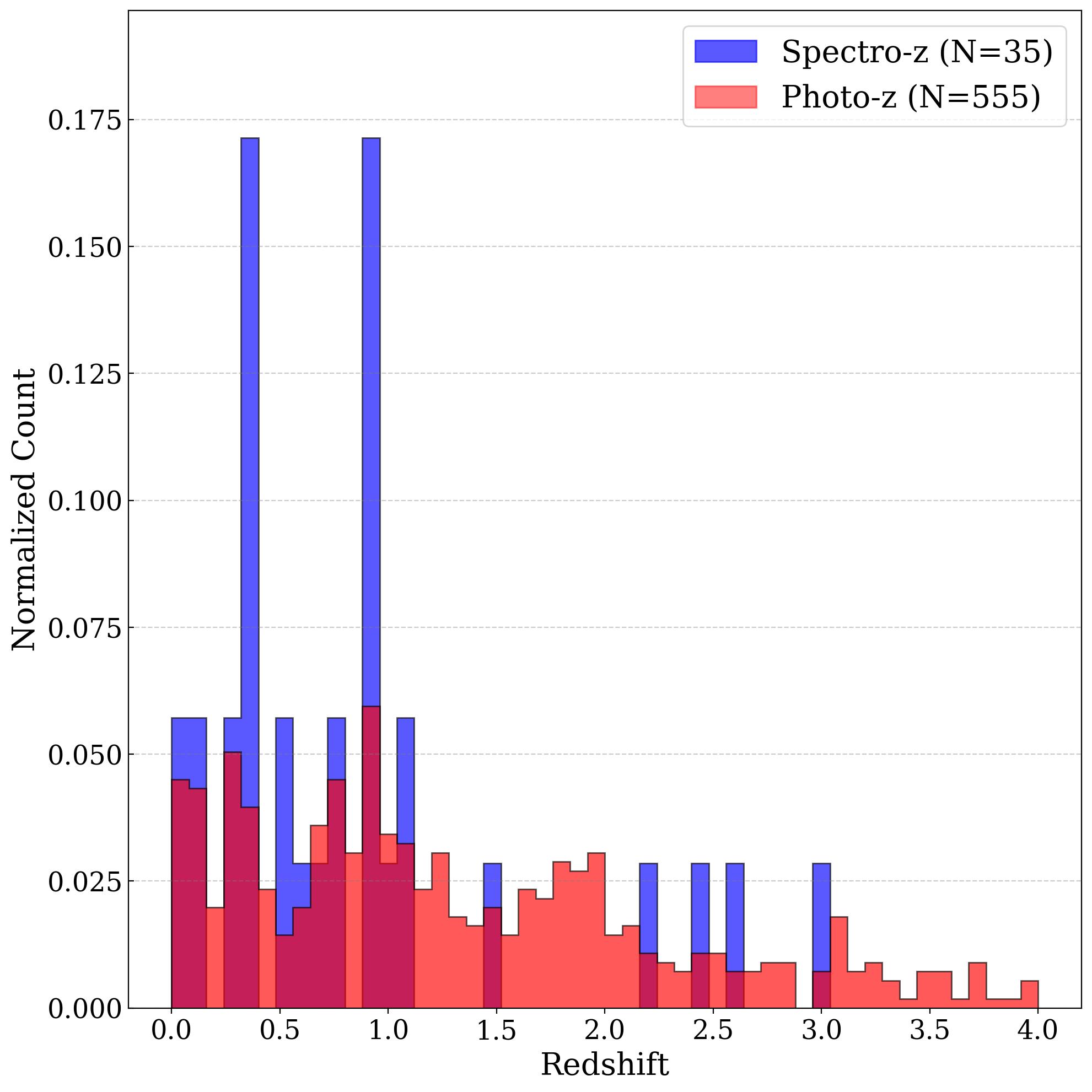}
  \caption{NK2-CL J100045.8+020514.3}
\end{subfigure}\hfill
\begin{subfigure}{0.3\textwidth}
  \includegraphics[width = 1\textwidth]{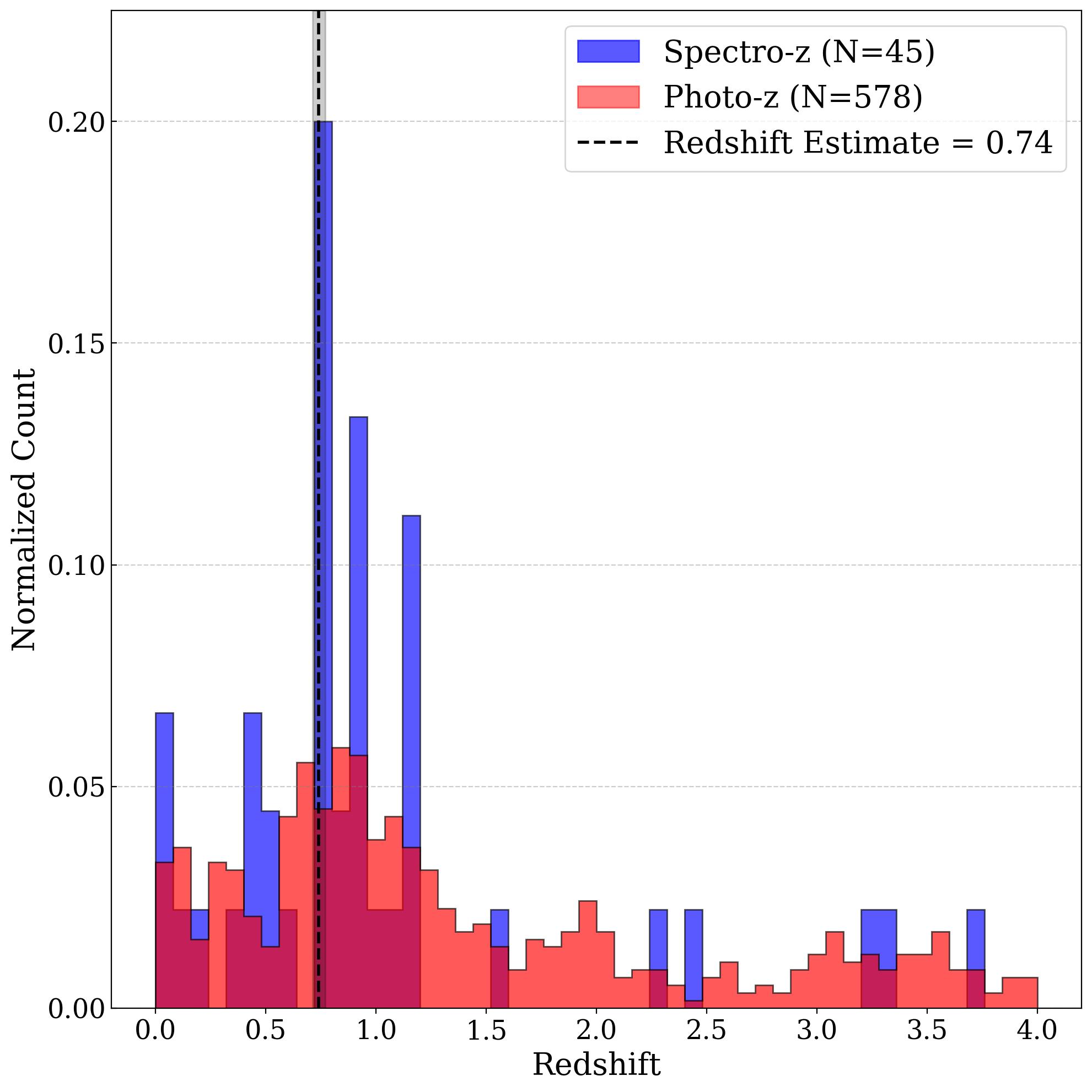}
  \caption{NK2-CL J095937.7+022320.4}
\end{subfigure}\hfill
\begin{subfigure}{0.3\textwidth}
  \includegraphics[width = 1\textwidth]{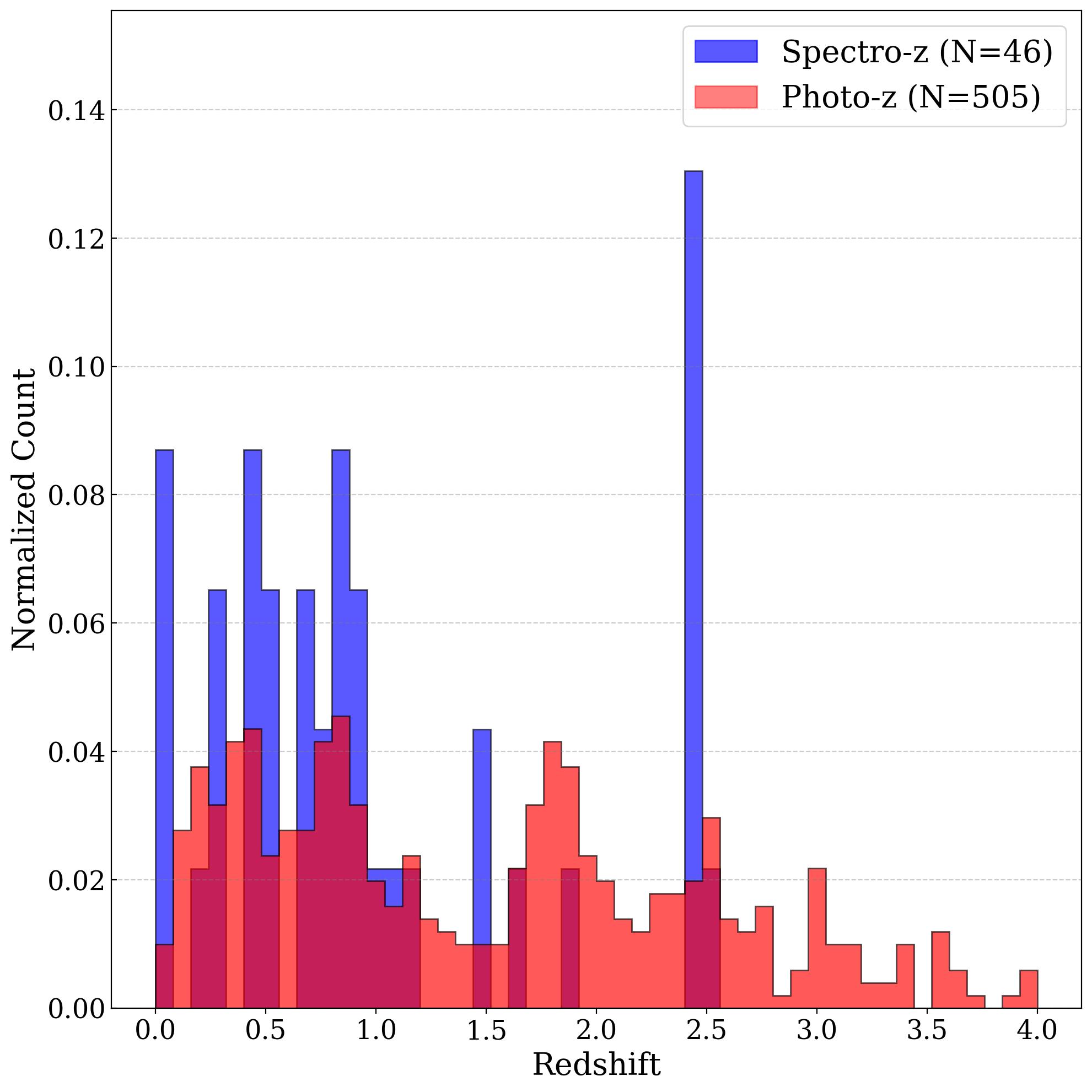}
  \caption{NK2-CL J100004.7+021604.4}
\end{subfigure}
\smallskip
\begin{subfigure}{0.3\textwidth}
  \includegraphics[width = 1\textwidth]{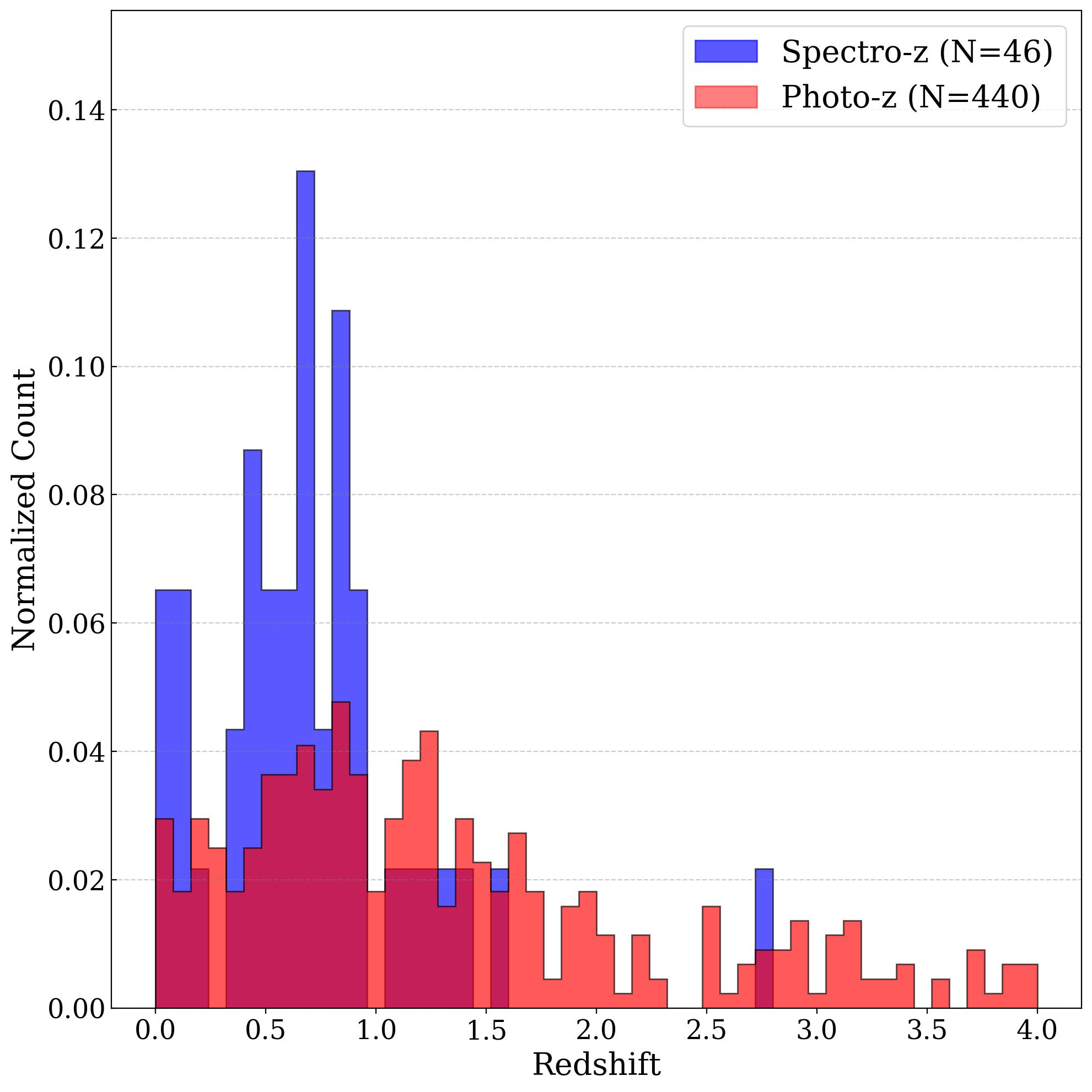}
  \caption{NK2-CL J100043.6+023232.4}
\end{subfigure}\hfill
\begin{subfigure}{0.3\textwidth}
  \includegraphics[width = 1\textwidth]{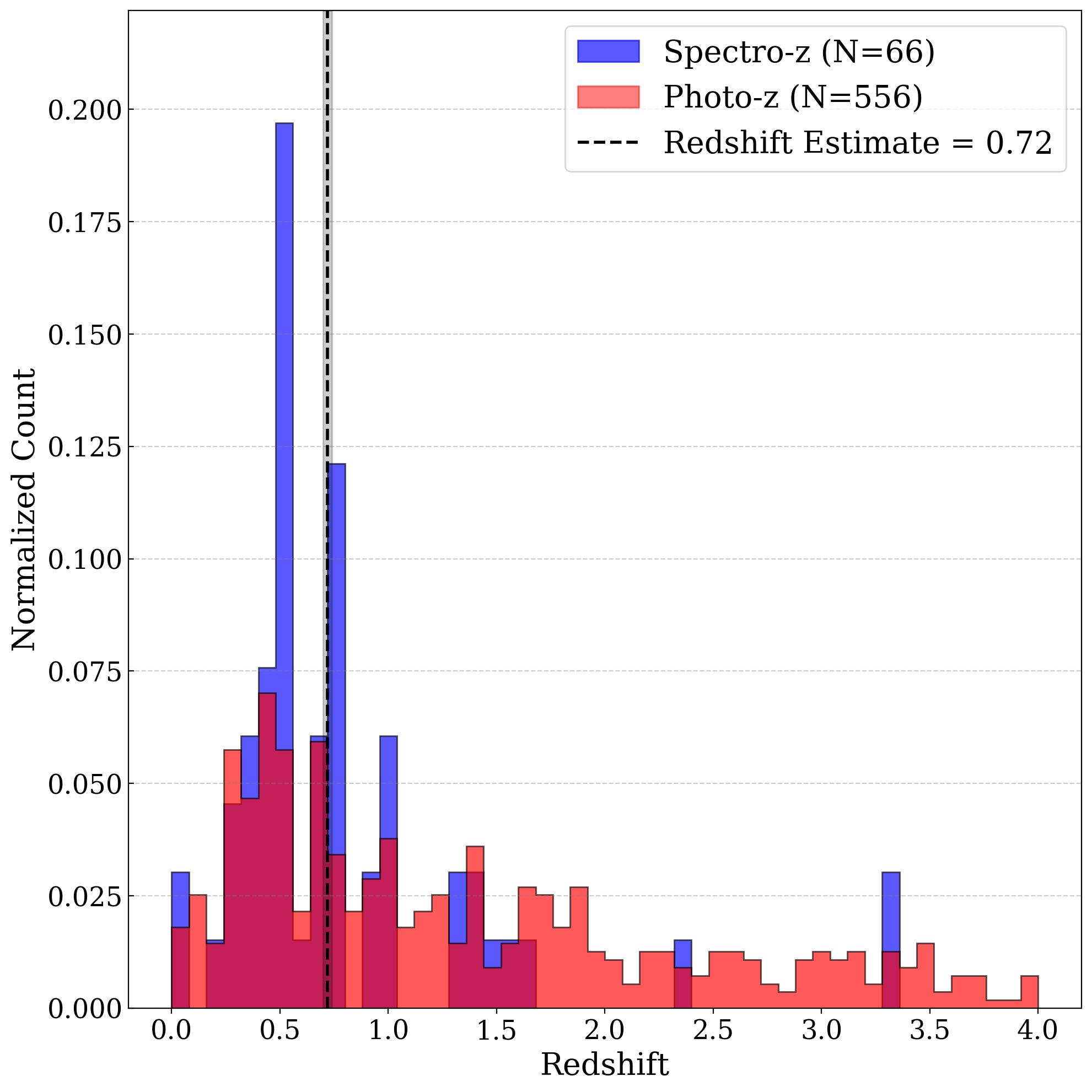}
  \caption{NK2-CL J100025.3+023346.4}
\end{subfigure}\hfill
\begin{subfigure}{0.3\textwidth}
  \includegraphics[width = 1\textwidth]{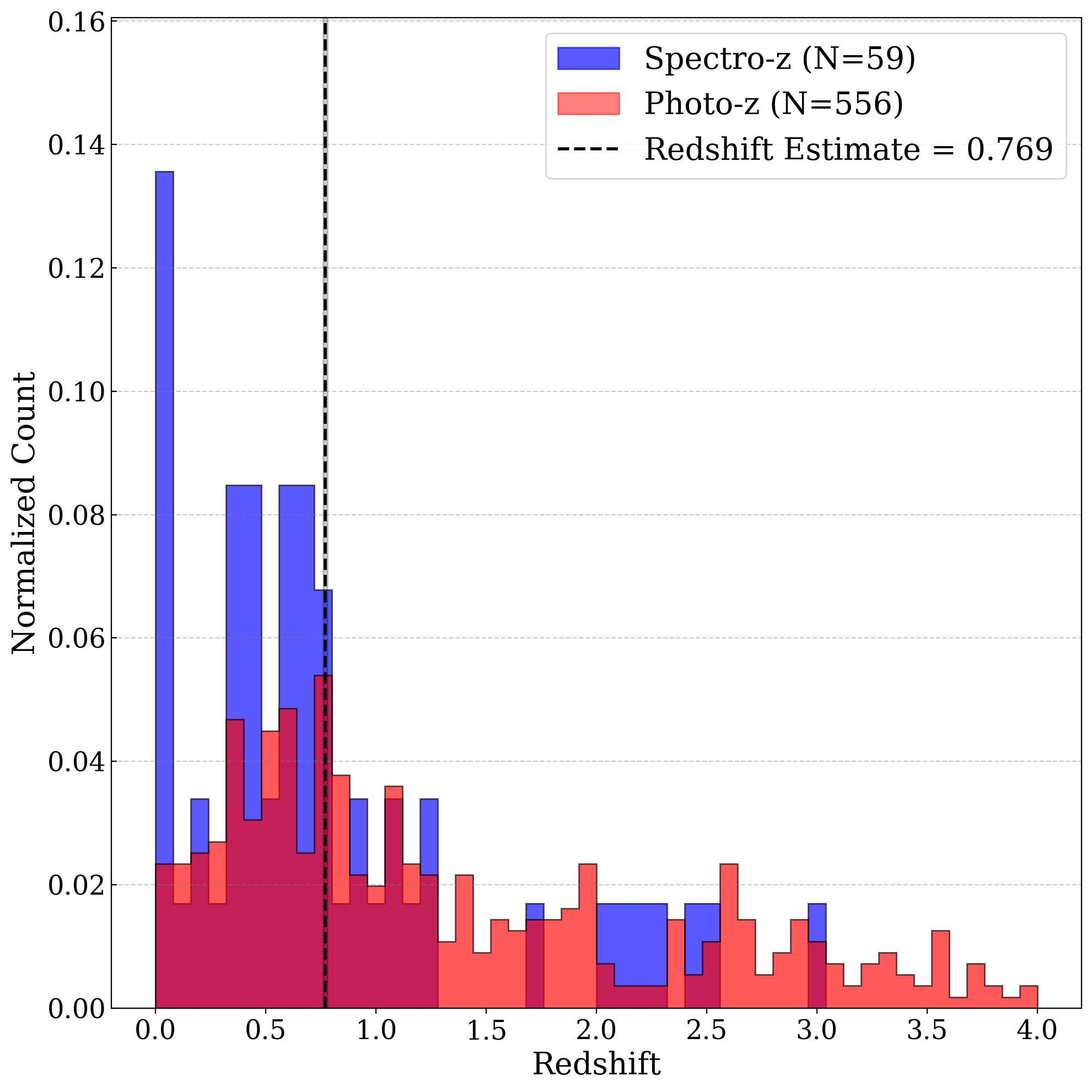}
  \caption{NK2-CL J100100.6+022134.4}
\end{subfigure}
\smallskip
\begin{subfigure}{0.3\textwidth}
  \includegraphics[width = 1\textwidth]{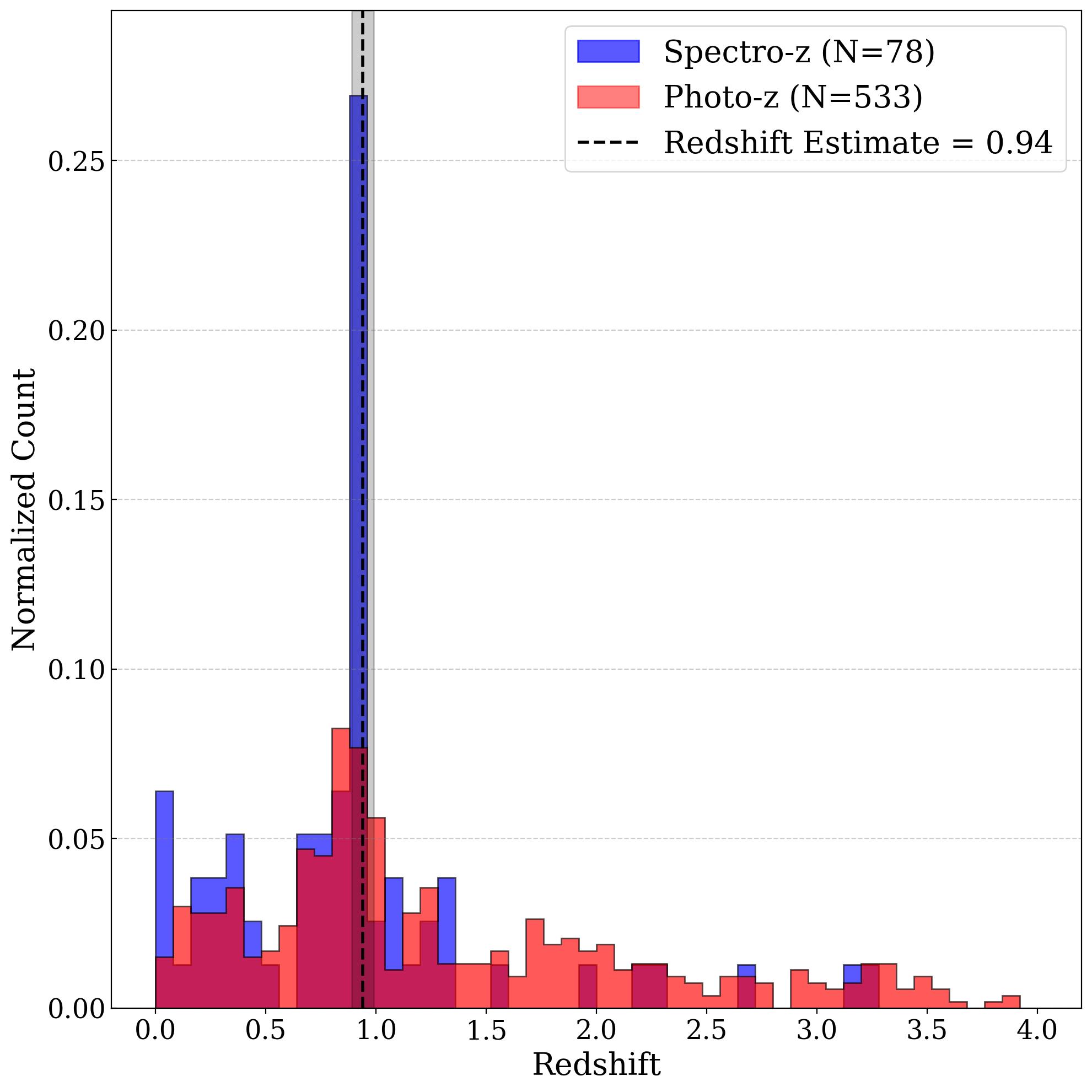}
  \caption{NK2-CL J100004.4+021148.4 }
\end{subfigure}\hfill
\begin{subfigure}{0.3\textwidth}
  \includegraphics[width = 1\textwidth]{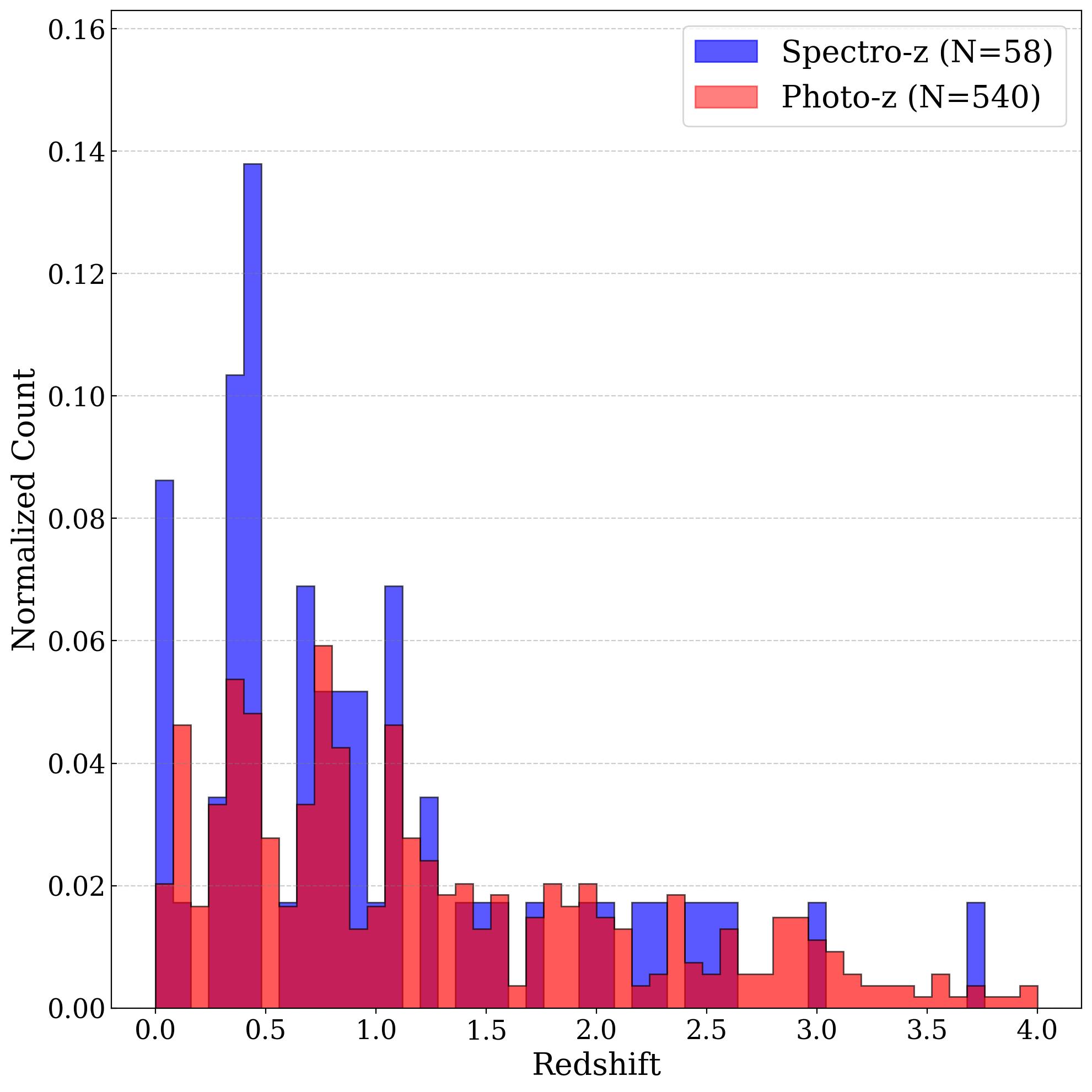}
  \caption{NK2-CL J100103.4+022208.4}
\end{subfigure}\hfill
\begin{subfigure}{0.3\textwidth}
  \includegraphics[width = 1\textwidth]{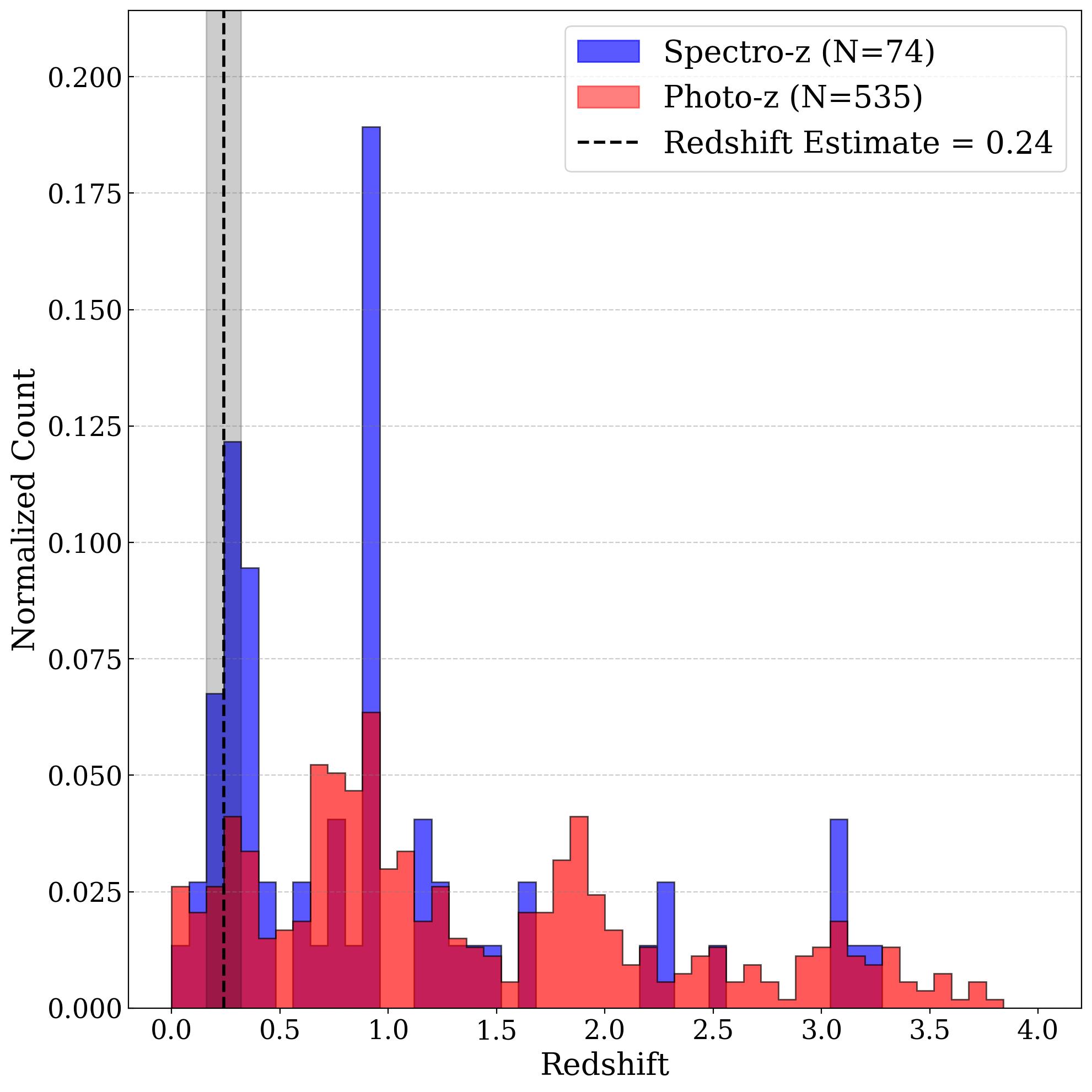}
  \caption{NK2-CL J100011.9+021256.5}
\end{subfigure}
\caption{Normalized spectroscopic (blue) and photometric (red) redshift counts within a 1 arcmin radius around each cluster candidate. The total number of galaxies, $N$, within this radius is indicated in the legend of the figure for each redshift type. The number of galaxies per redshift bin is obtained by multiplying $N$ by the normalized counts. If the NIKA2 cluster candidate has a redshift estimate (see Table~\ref{tbl:candidates_table}), its redshift is represented by a black dashed line.}
\label{redshift_hist}
\end{figure*}

\newpage
\begin{figure*}\ContinuedFloat
\centering
\begin{subfigure}{0.3\textwidth}
  \includegraphics[width = 1\textwidth]{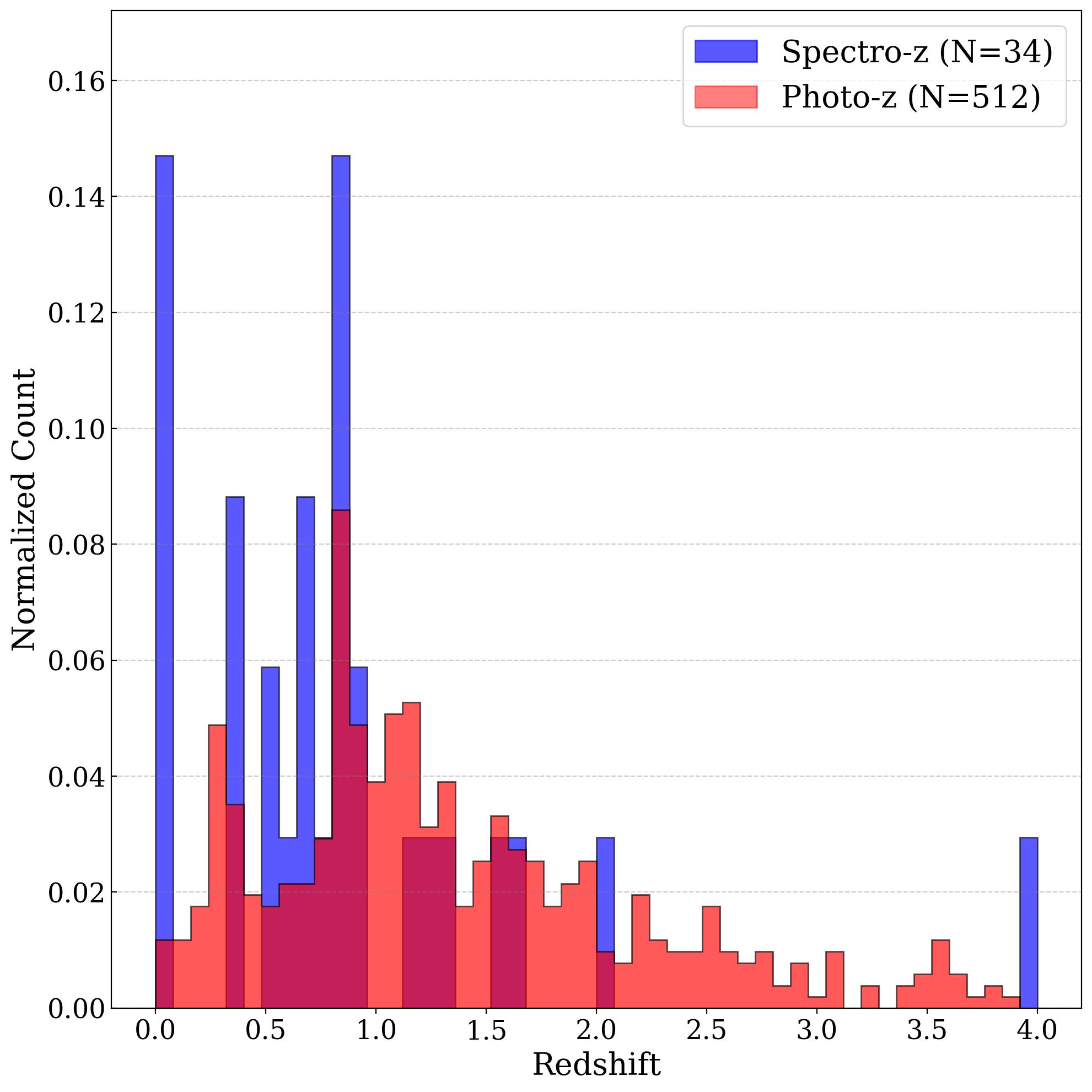}
  \caption{NK2-CL J100101.1+020146.6}
\end{subfigure}\hfill
\begin{subfigure}{0.3\textwidth}
  \includegraphics[width = 1\textwidth]{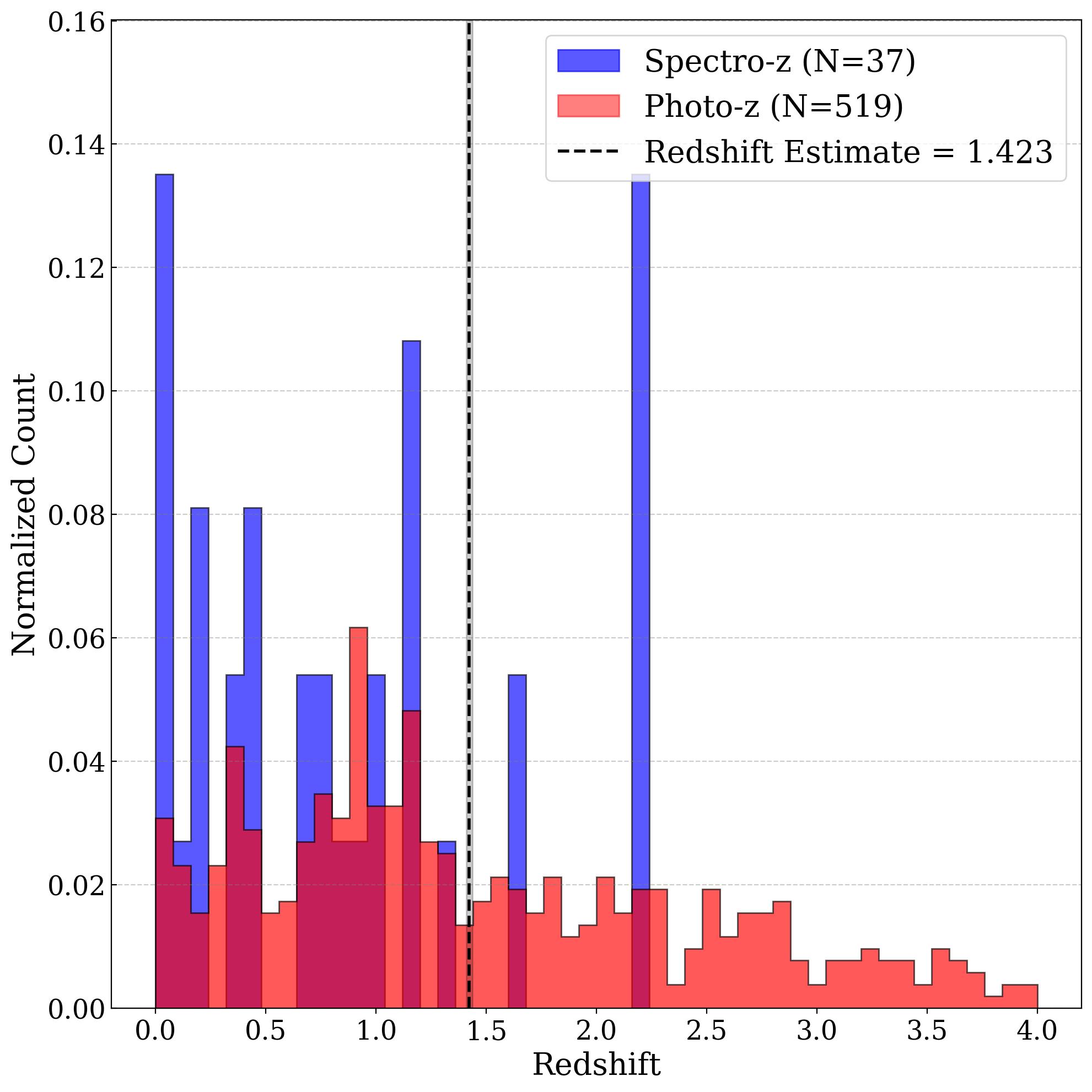}
  \caption{NK2-CL J100054.3+020126.4}
\end{subfigure}\hfill
\begin{subfigure}{0.3\textwidth}
  \includegraphics[width = 1\textwidth]{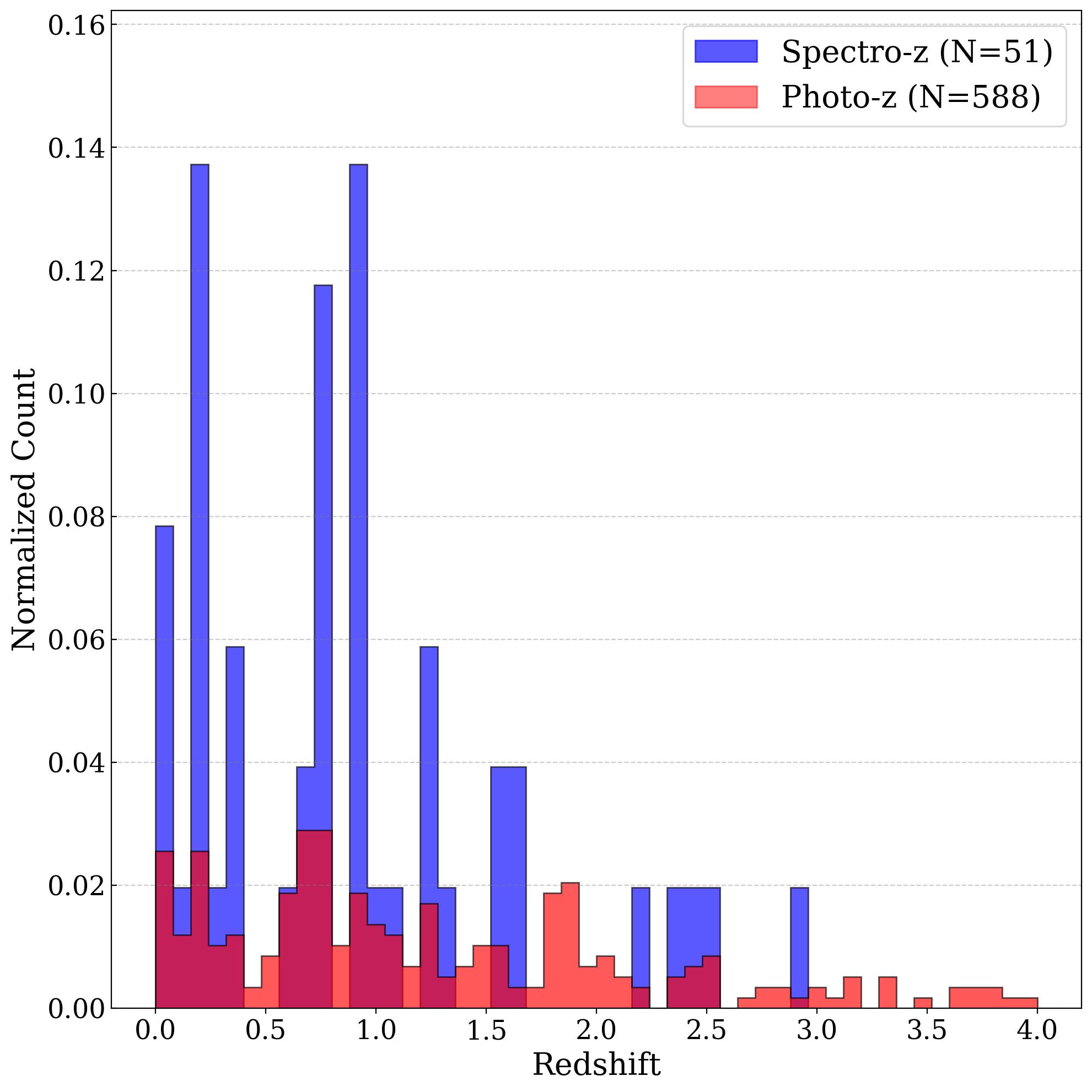}
  \caption{NK2-CL J100009.1+022140.3}
\end{subfigure}
\smallskip
\begin{subfigure}{0.3\textwidth}
  \includegraphics[width = 1\textwidth]{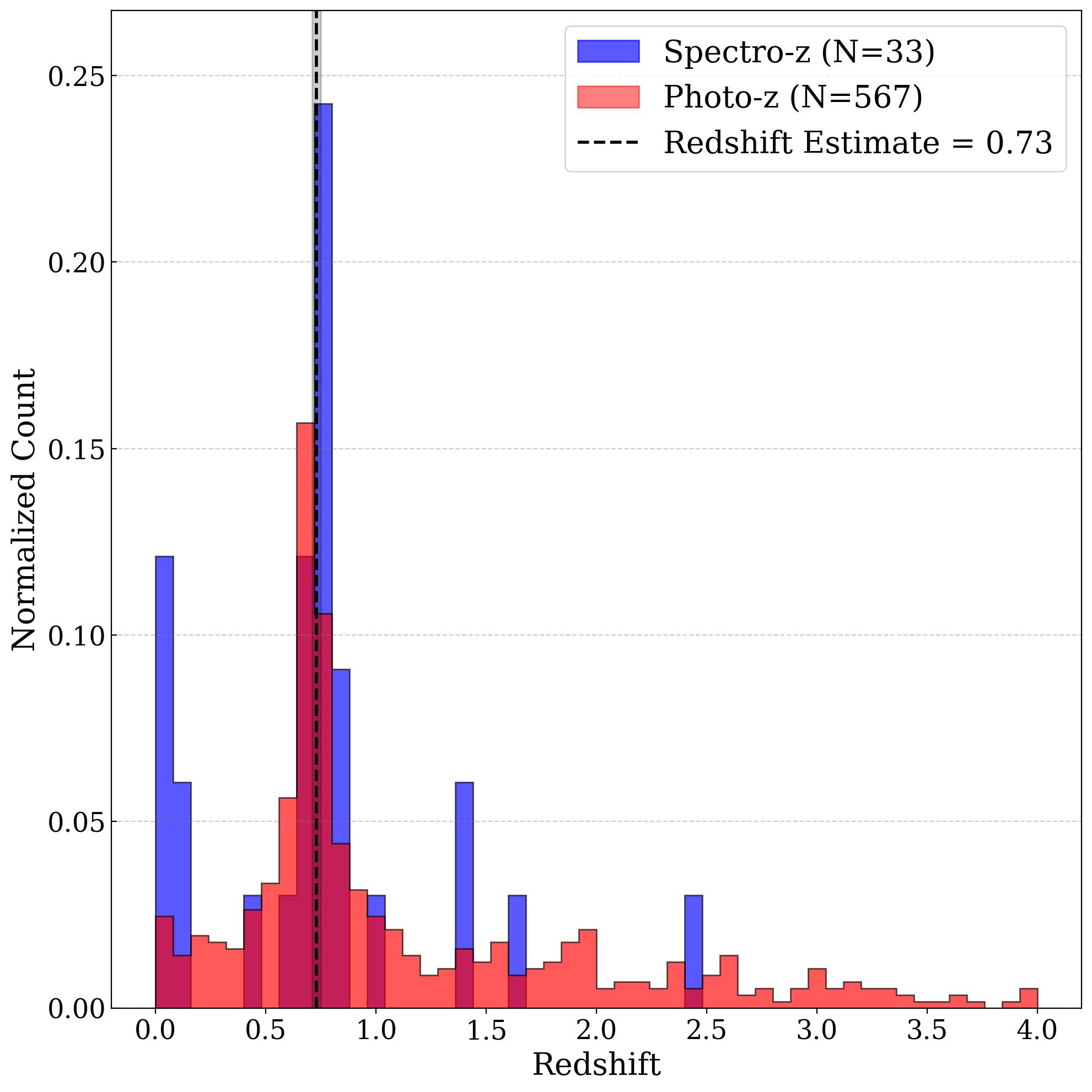}
  \caption{NK2-CL J095942.6+023056.5}
\end{subfigure}\hfill
\begin{subfigure}{0.3\textwidth}
  \includegraphics[width = 1\textwidth]{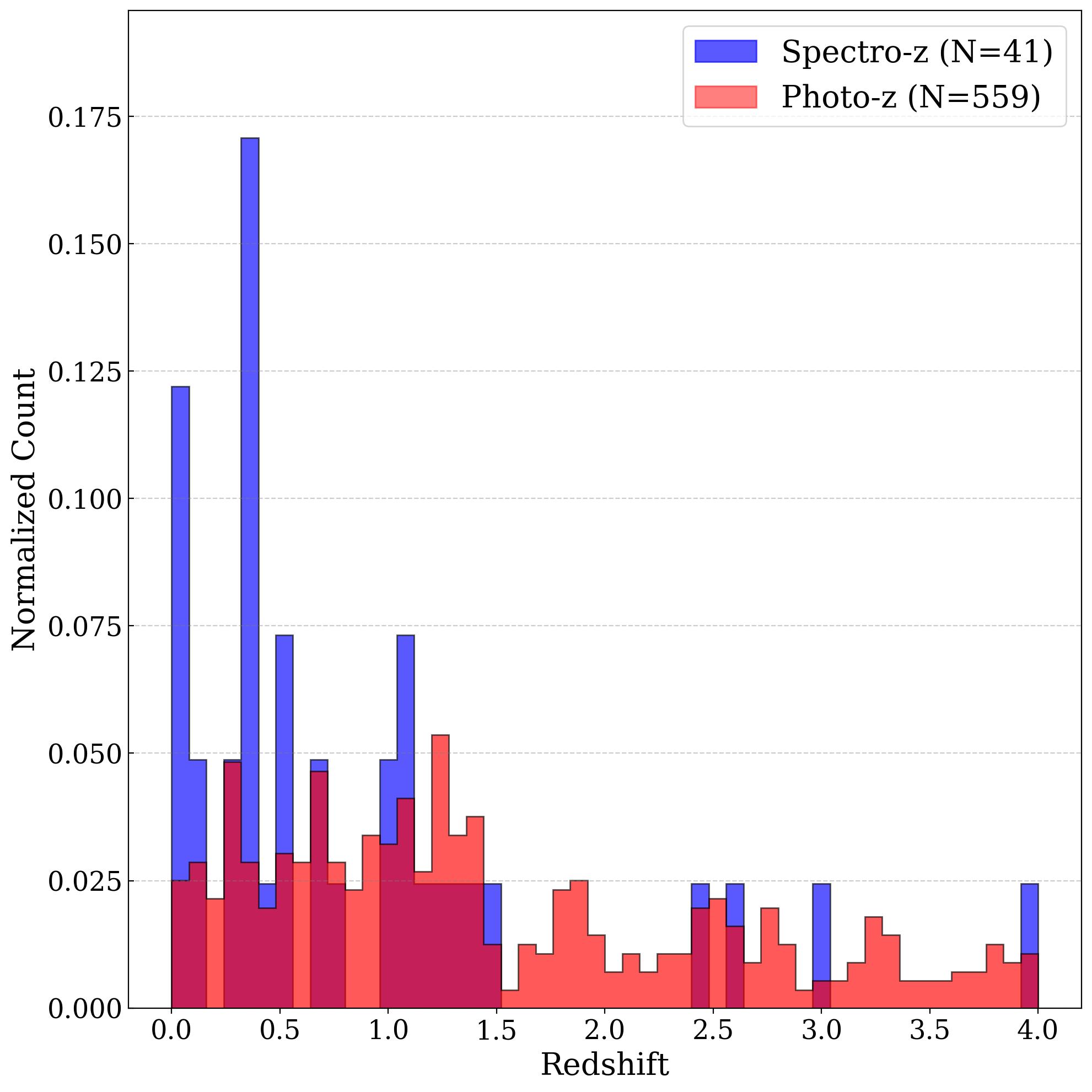}
  \caption{NK2-CL J100120.5+022828.2}
\end{subfigure}\hfill
\begin{subfigure}{0.3\textwidth}
  \includegraphics[width = 1\textwidth]{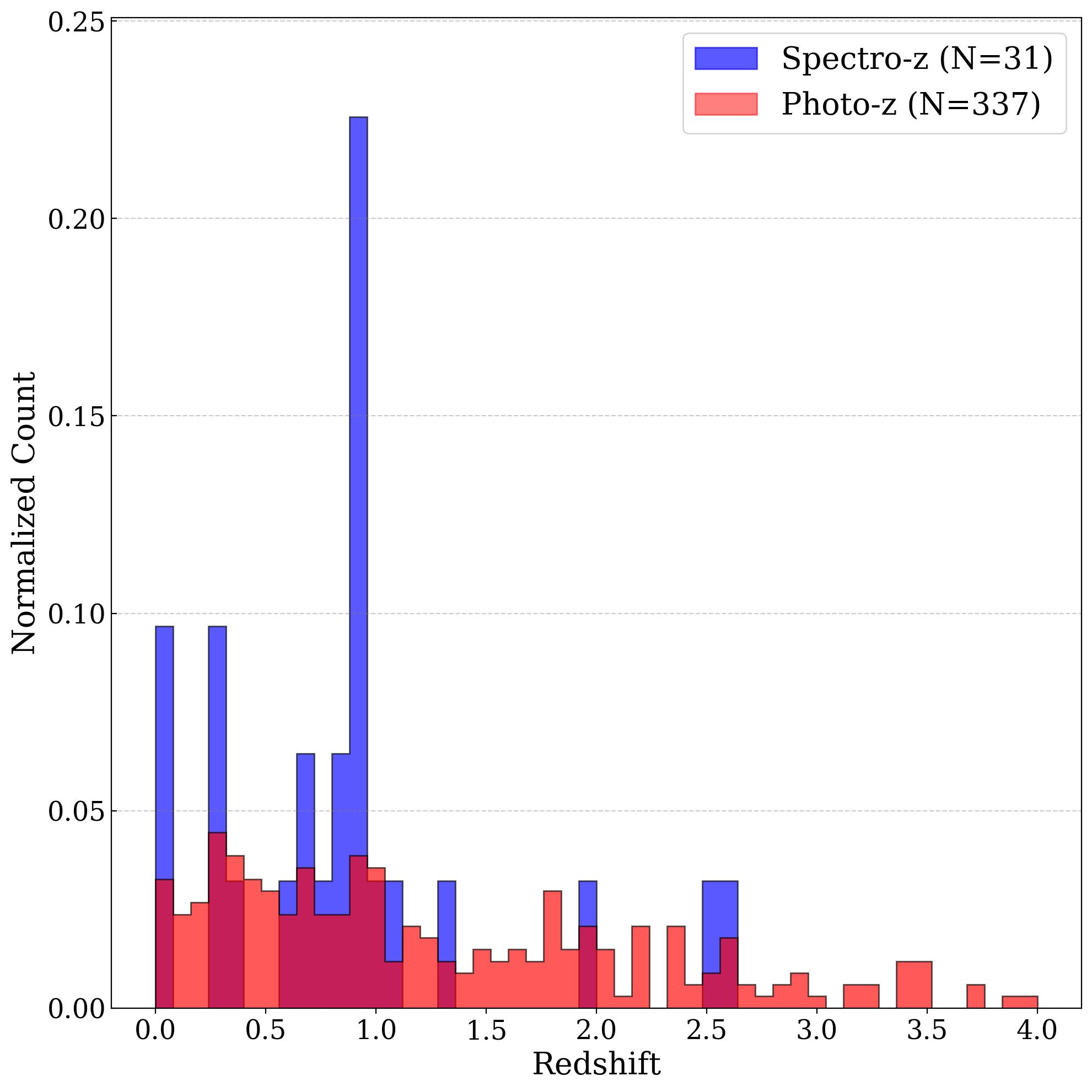}
  \caption{NK2-CL J100008.4+020908.3}
\end{subfigure}
\smallskip
\begin{subfigure}{0.3\textwidth}
  \includegraphics[width = 1\textwidth]{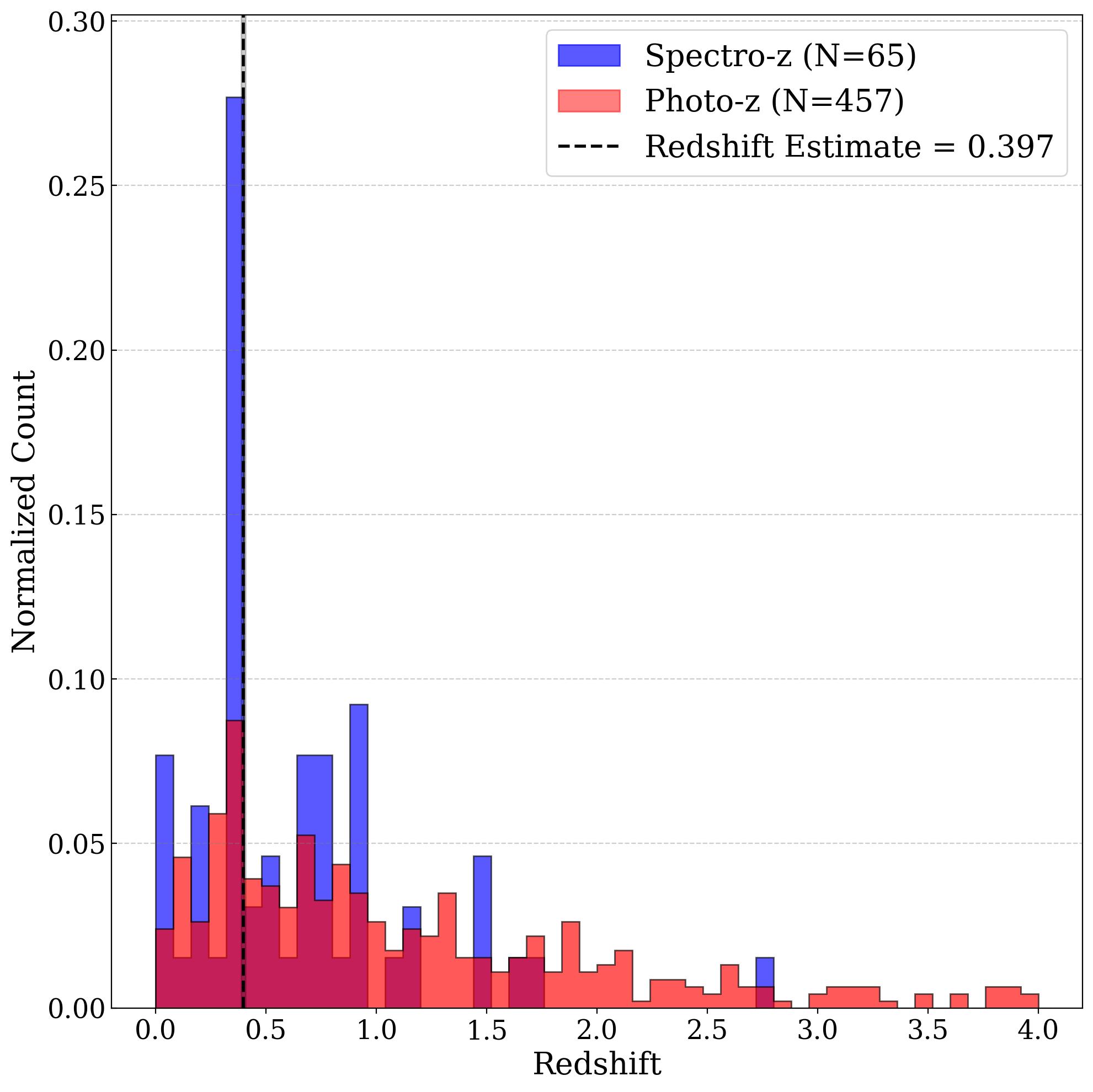}
  \caption{NK2-CL J095958.5+022910.4}
\end{subfigure}

\caption{continued.}
\end{figure*}

\clearpage
\section{Results of the MCMC analysis}
\label{app:mcmcresults}

We show the main results of the fitting procedure for all cluster candidates presented in Sect.~\ref{integrated_quantities}.  We present the posterior likelihood distribution in the $\theta_{500}-Y_{500}$ plane.

\begin{figure*}[h!]
\centering
\begin{subfigure}{0.3\textwidth}
  \includegraphics[width = 1\textwidth]{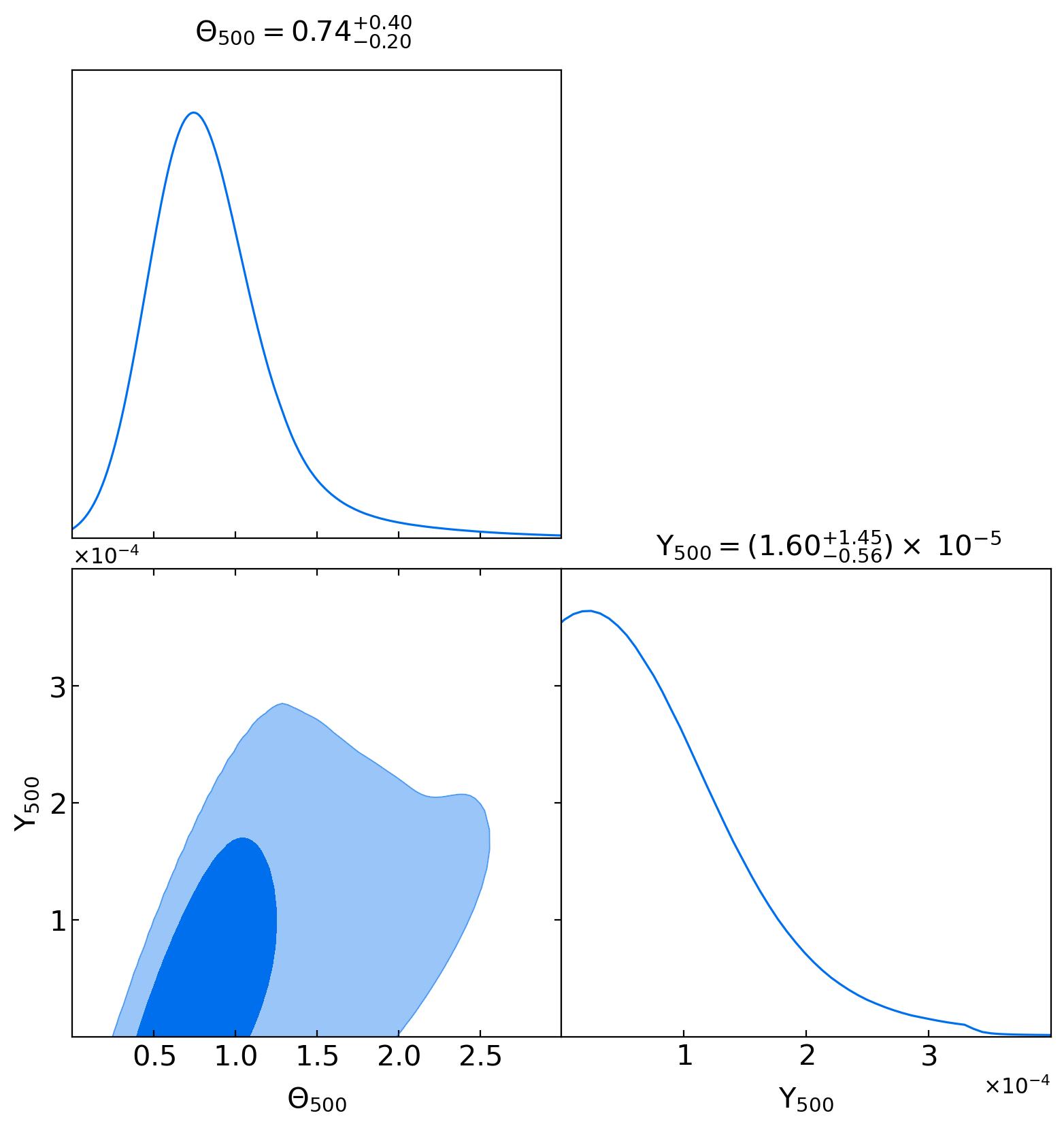}
  \caption{NK2-CL J100045.8+020514.3}
\end{subfigure}\hfill
\begin{subfigure}{0.3\textwidth}
  \includegraphics[width = 1\textwidth]{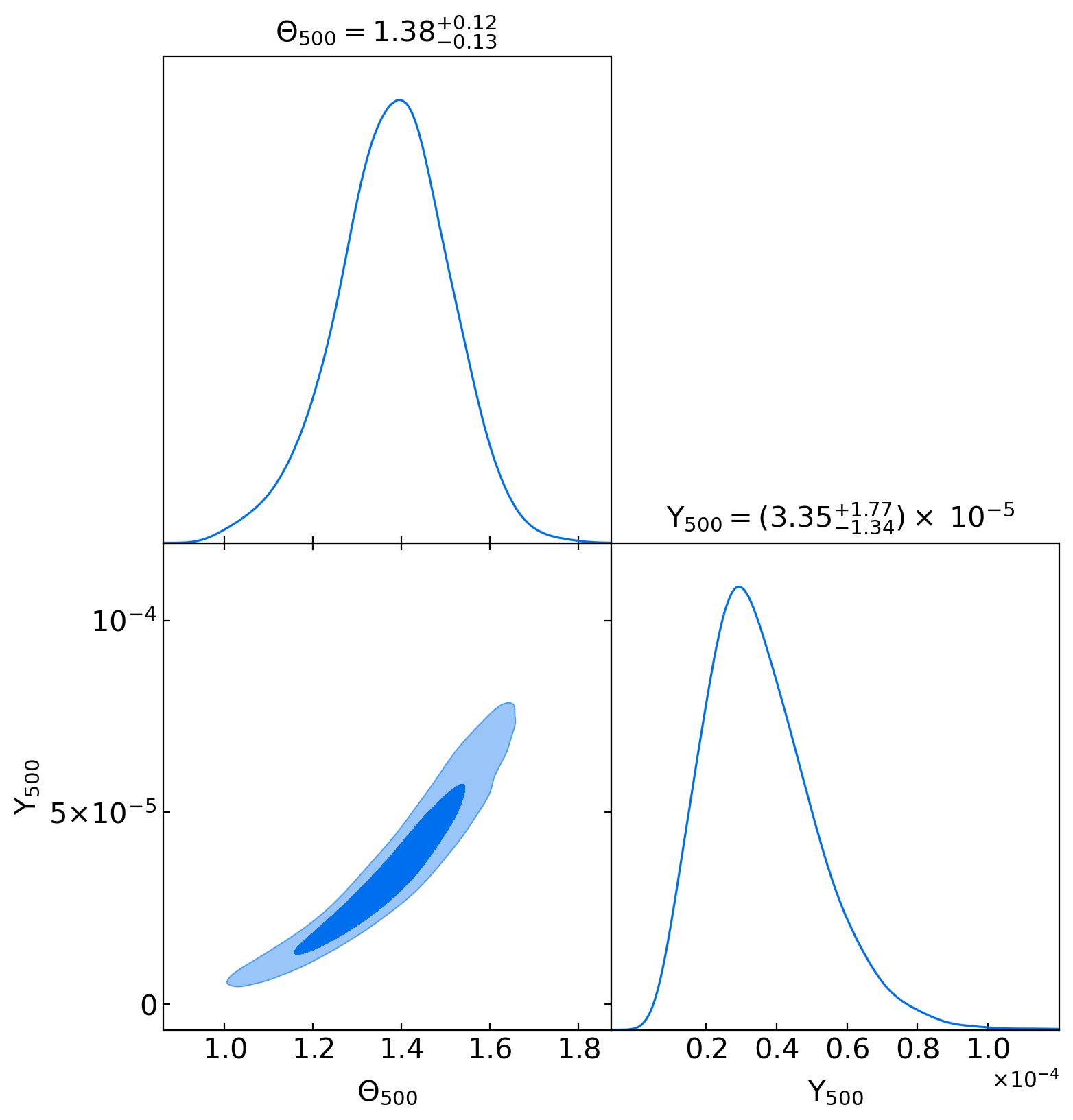}
  \caption{NK2-CL J095937.7+022320.4}
\end{subfigure}\hfill
\begin{subfigure}{0.3\textwidth}
  \includegraphics[width = 1\textwidth]{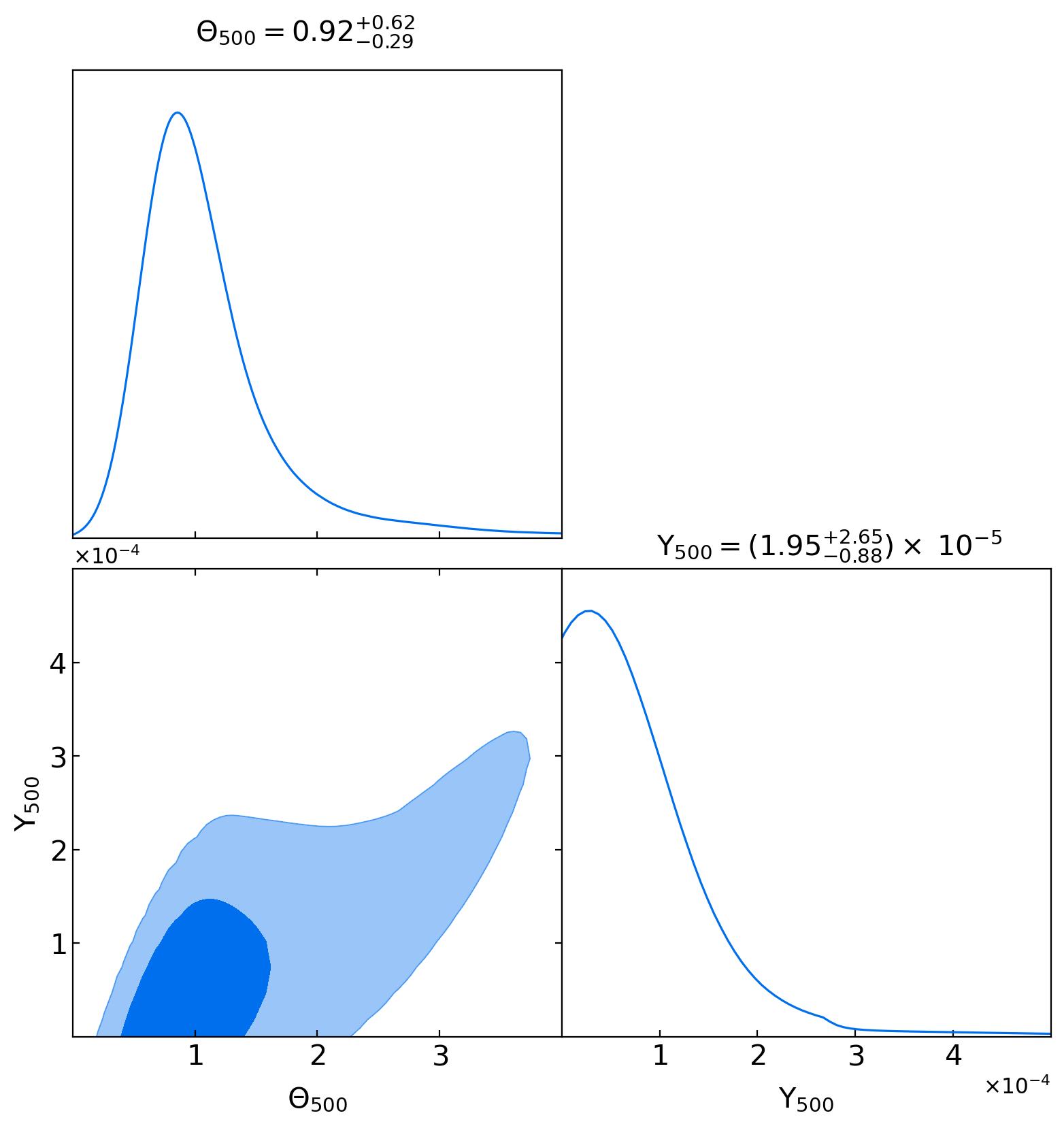}
  \caption{NK2-CL J100004.7+021604.4}
\end{subfigure}

\smallskip

\begin{subfigure}{0.3\textwidth}
  \includegraphics[width = 1\textwidth]{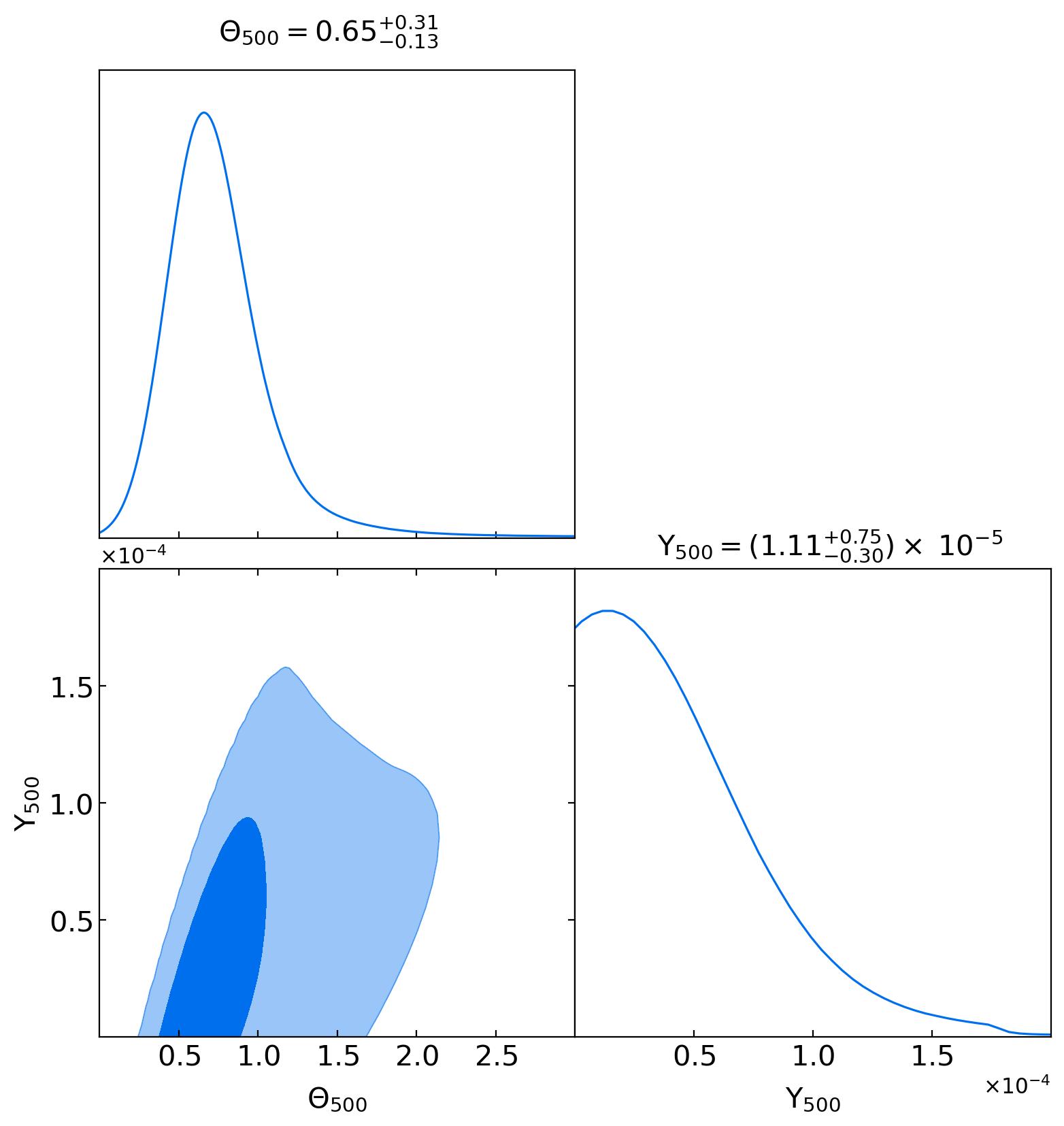}
  \caption{NK2-CL J100043.6+023232.4}
\end{subfigure}\hfill
\begin{subfigure}{0.3\textwidth}
  \includegraphics[width = 1\textwidth]{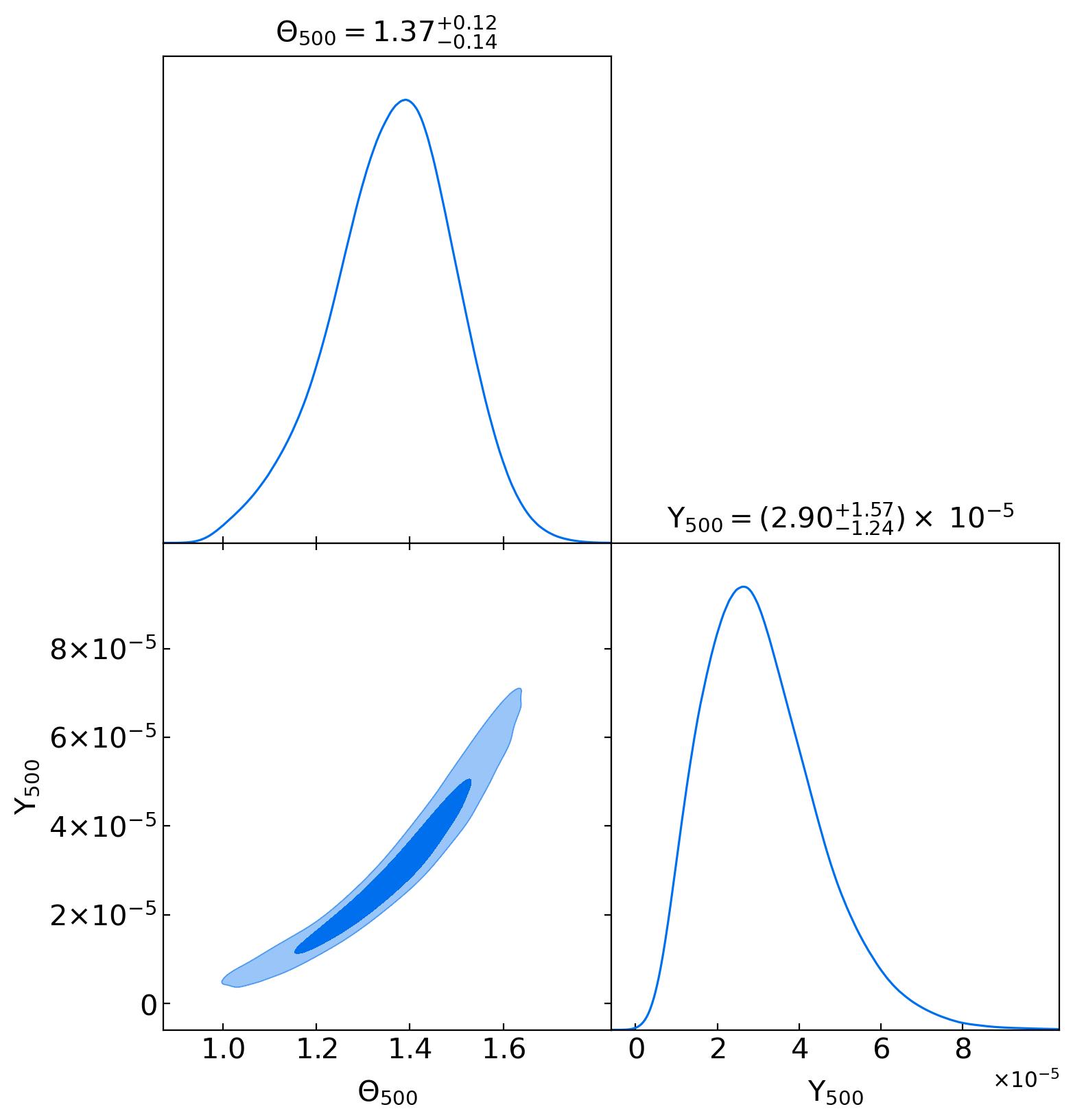}
  \caption{NK2-CL J100025.3+023346.4}
\end{subfigure}\hfill
\begin{subfigure}{0.3\textwidth}
  \includegraphics[width = 1\textwidth]{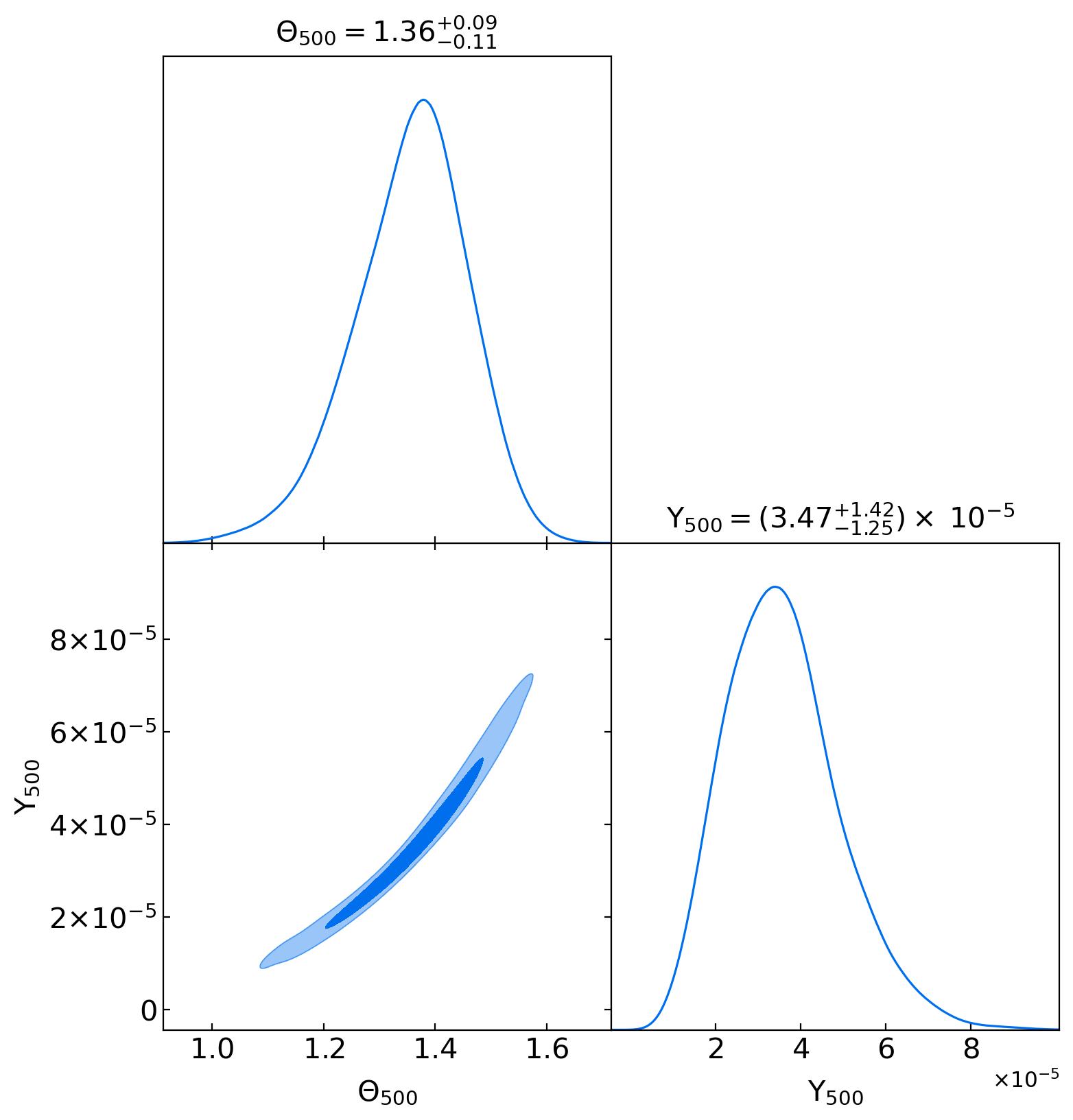}
  \caption{NK2-CL J100100.6+022134.4}
\end{subfigure}

\smallskip

\begin{subfigure}{0.3\textwidth}
  \includegraphics[width = 1\textwidth]{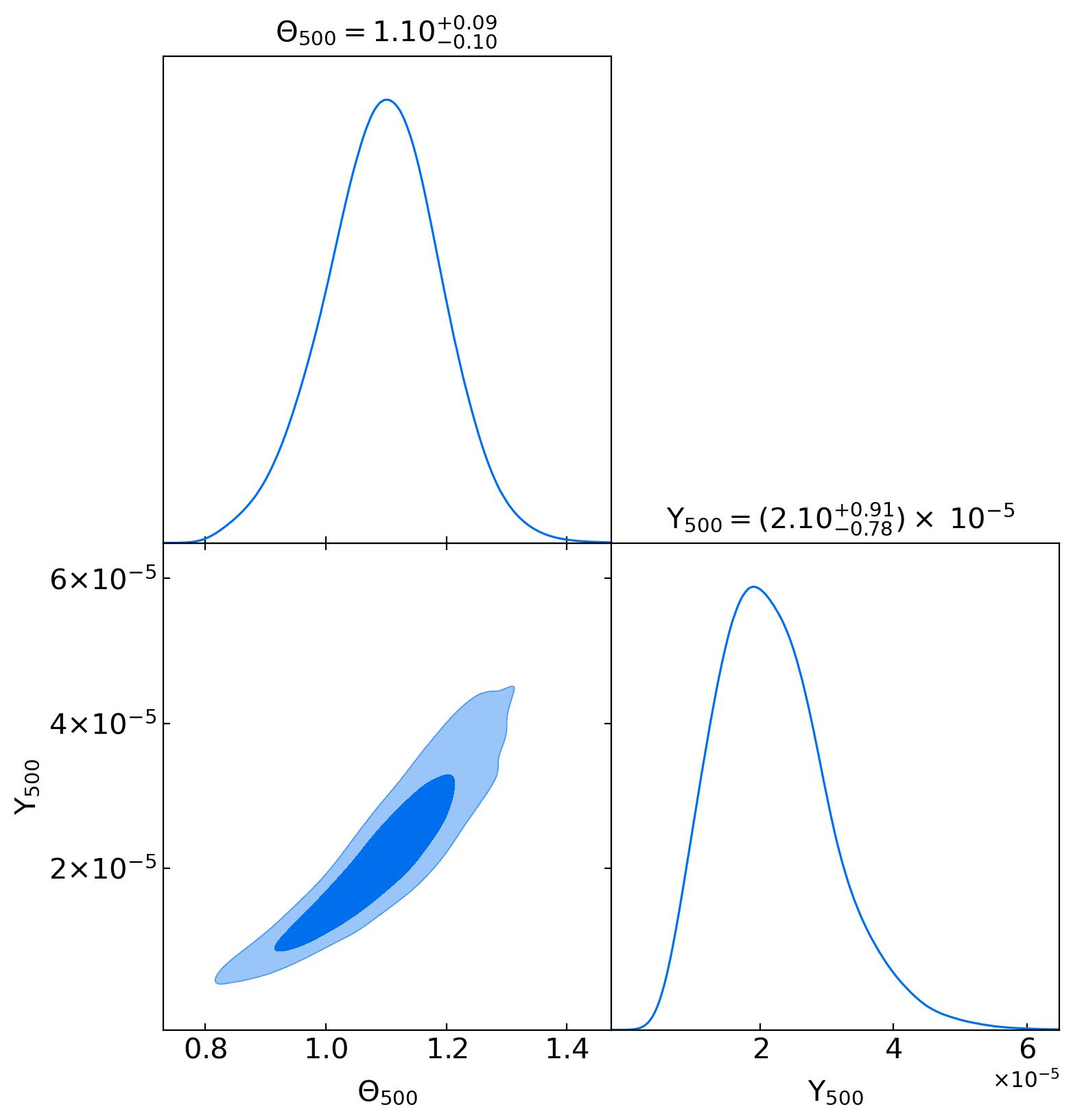}
  \caption{NK2-CL J100004.4+021148.4}
\end{subfigure}\hfill
\begin{subfigure}{0.3\textwidth}
  \includegraphics[width = 1\textwidth]{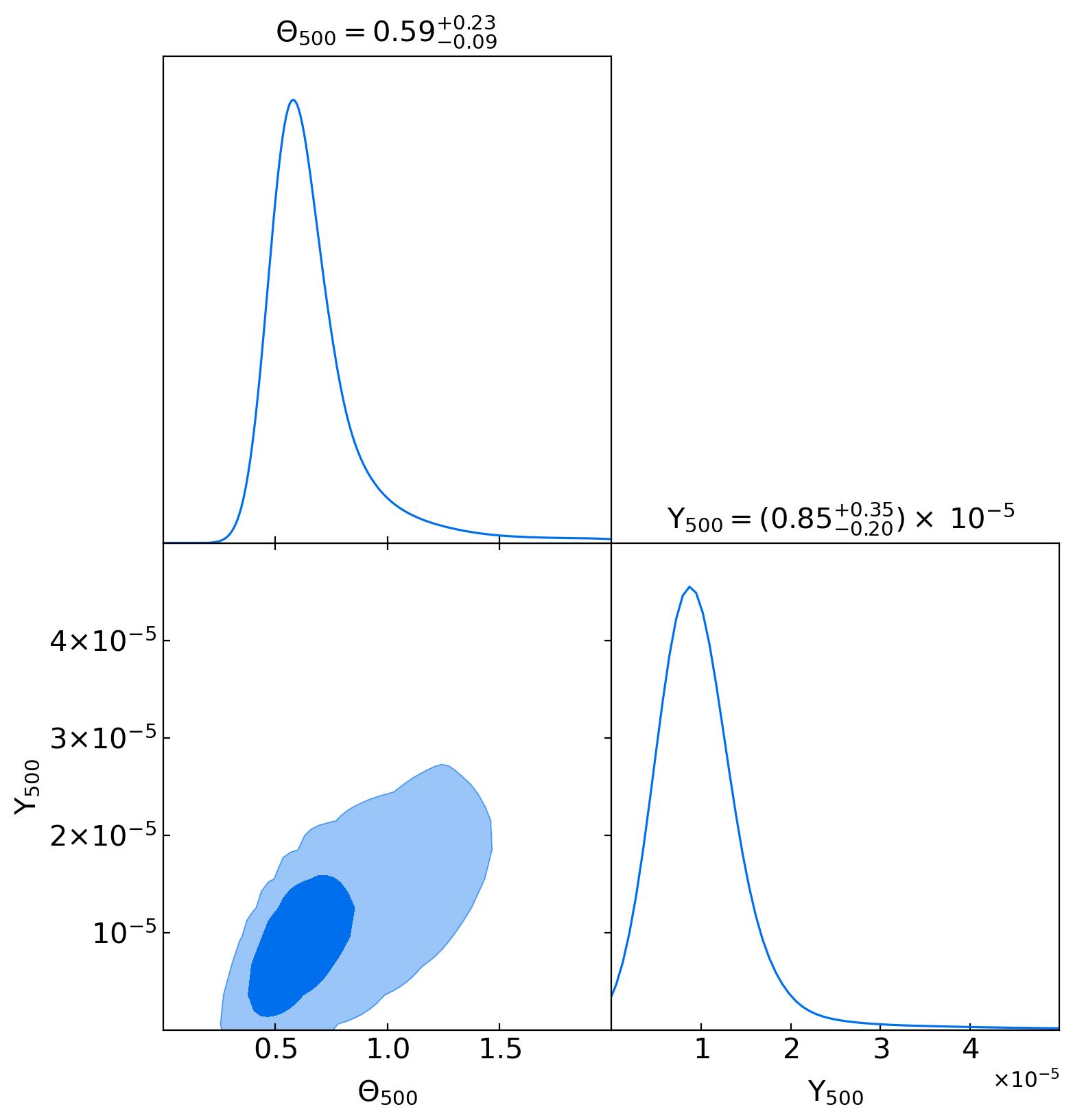}
  \caption{NK2-CL J100103.4+022208.4}
\end{subfigure}\hfill
\begin{subfigure}{0.3\textwidth}
  \includegraphics[width = 1\textwidth]{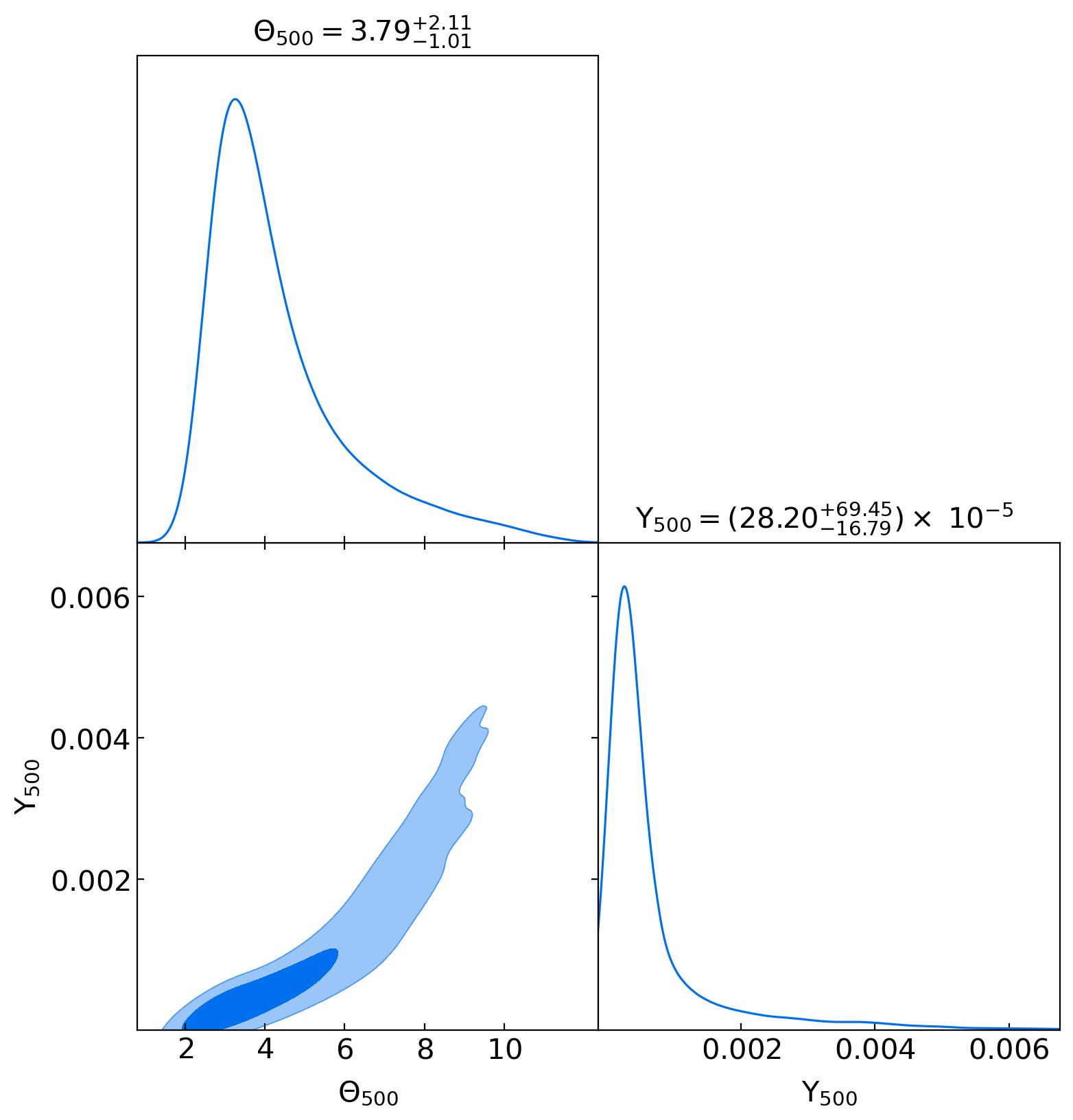}
  \caption{NK2-CL J100011.9+021256.5}
\end{subfigure}
\caption{Corner plot of the $\theta_{500}-Y_{500}$ posterior maximum likelihood distribution for each of the cluster candidate. A more detailed description of the fitting procedure can be found in Sect.~\ref{integrated_quantities}. The 68\% confidence values and uncertainties are given in the figure. We see that the posterior distributions are much better constrained for cluster candidates with a redshift estimate, as expected from the tighter redshift prior.}
\label{corner_plots}
\end{figure*}

\newpage

\begin{figure*}\ContinuedFloat
\centering
\begin{subfigure}{0.3\textwidth}
  \includegraphics[width = 1\textwidth]{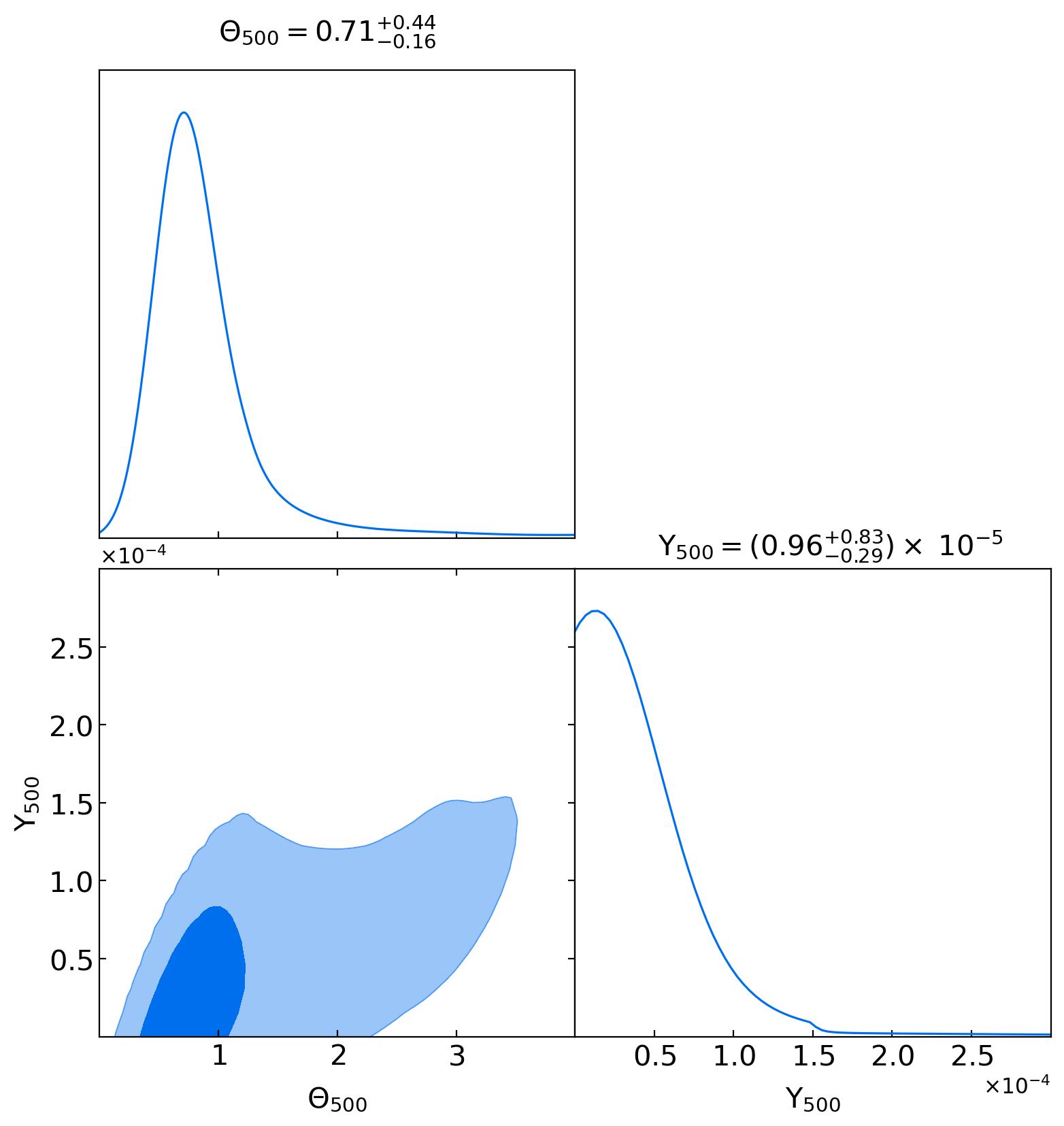}
  \caption{NK2-CL J100101.1+020146.6}
\end{subfigure}\hfill
\begin{subfigure}{0.3\textwidth}
  \includegraphics[width = 1\textwidth]{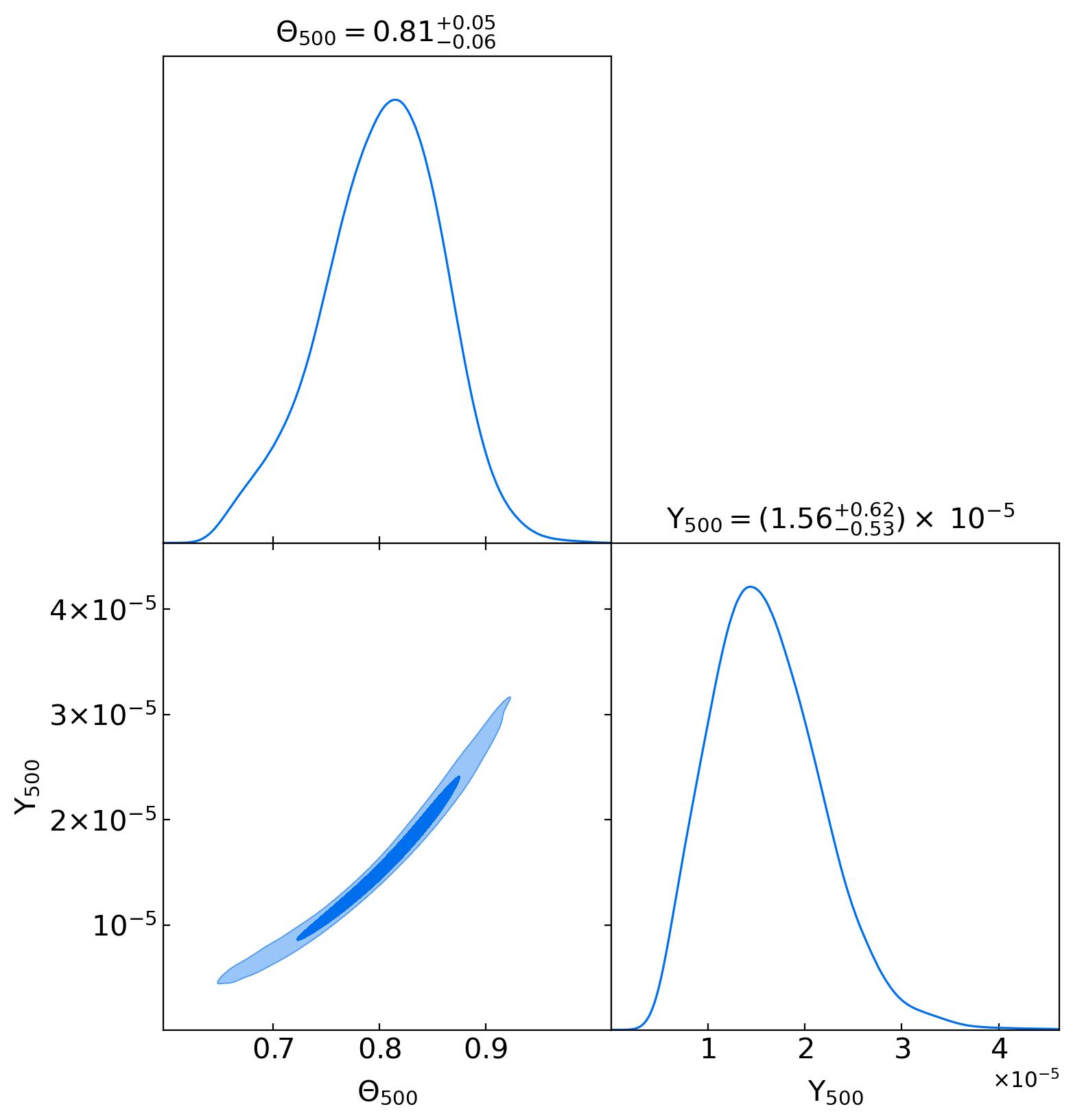}
  \caption{NK2-CL J100054.3+020126.4}
\end{subfigure}\hfill
\begin{subfigure}{0.3\textwidth}
  \includegraphics[width = 1\textwidth]{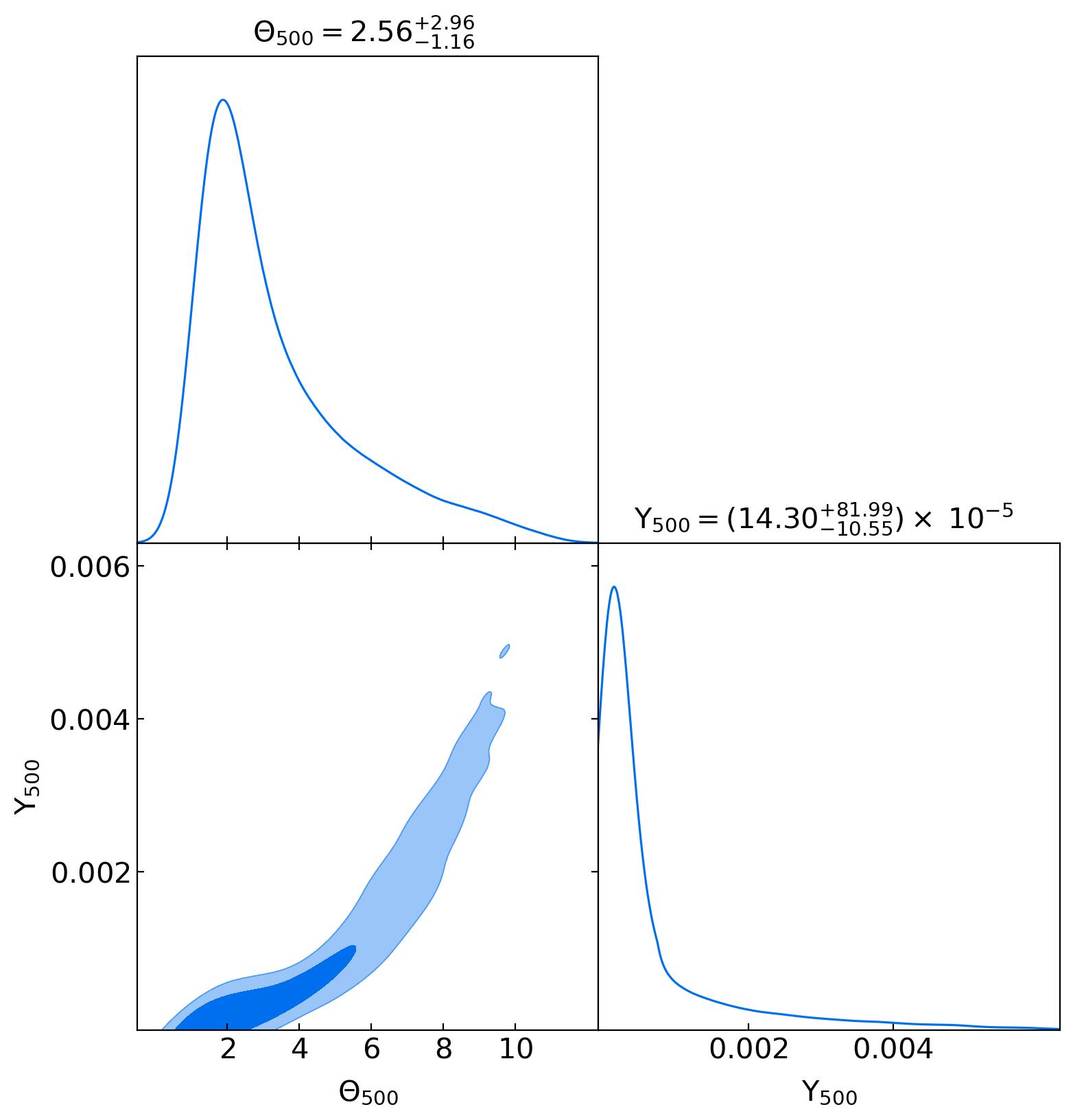}
  \caption{NK2-CL J100009.1+022140.3}
\end{subfigure}\hfill
\smallskip
\begin{subfigure}{0.3\textwidth}
  \includegraphics[width = 1\textwidth]{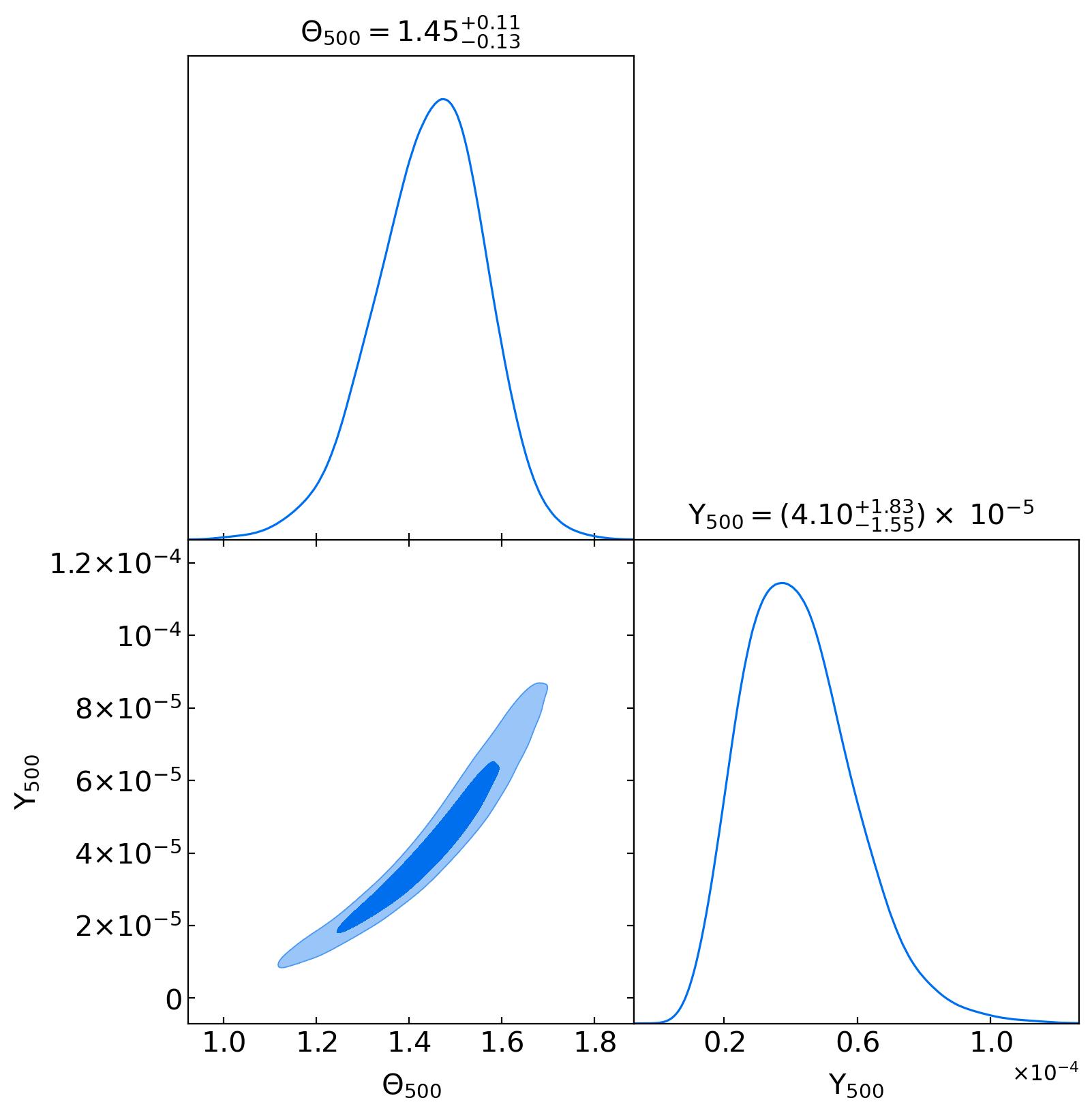}
  \caption{NK2-CL J095942.6+023056.5}
\end{subfigure}\hfill
\begin{subfigure}{0.3\textwidth}
  \includegraphics[width = 1\textwidth]{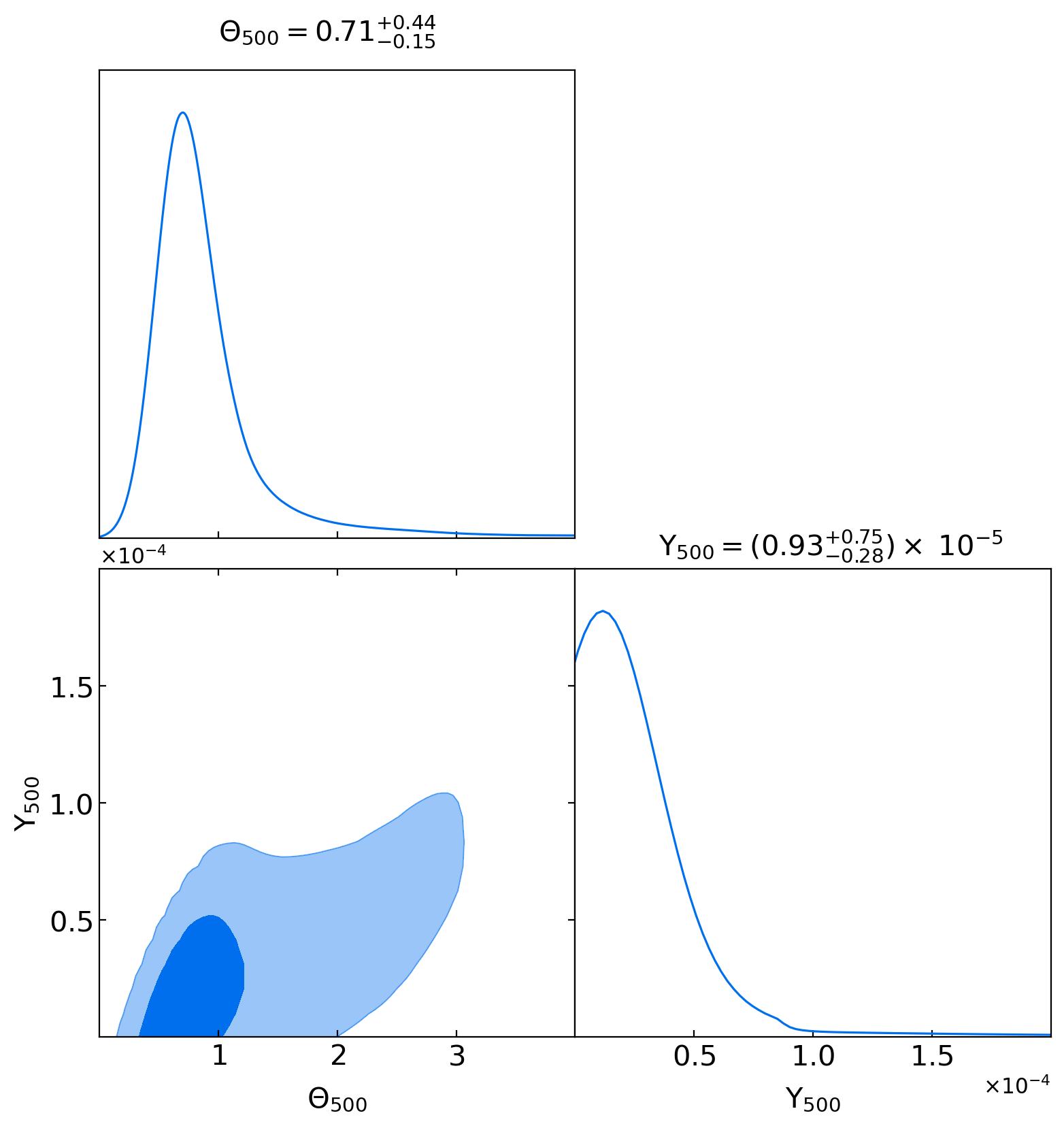}
  \caption{NK2-CL J100120.5+022828.2}
\end{subfigure}\hfill
\begin{subfigure}{0.3\textwidth}
  \includegraphics[width = 1\textwidth]{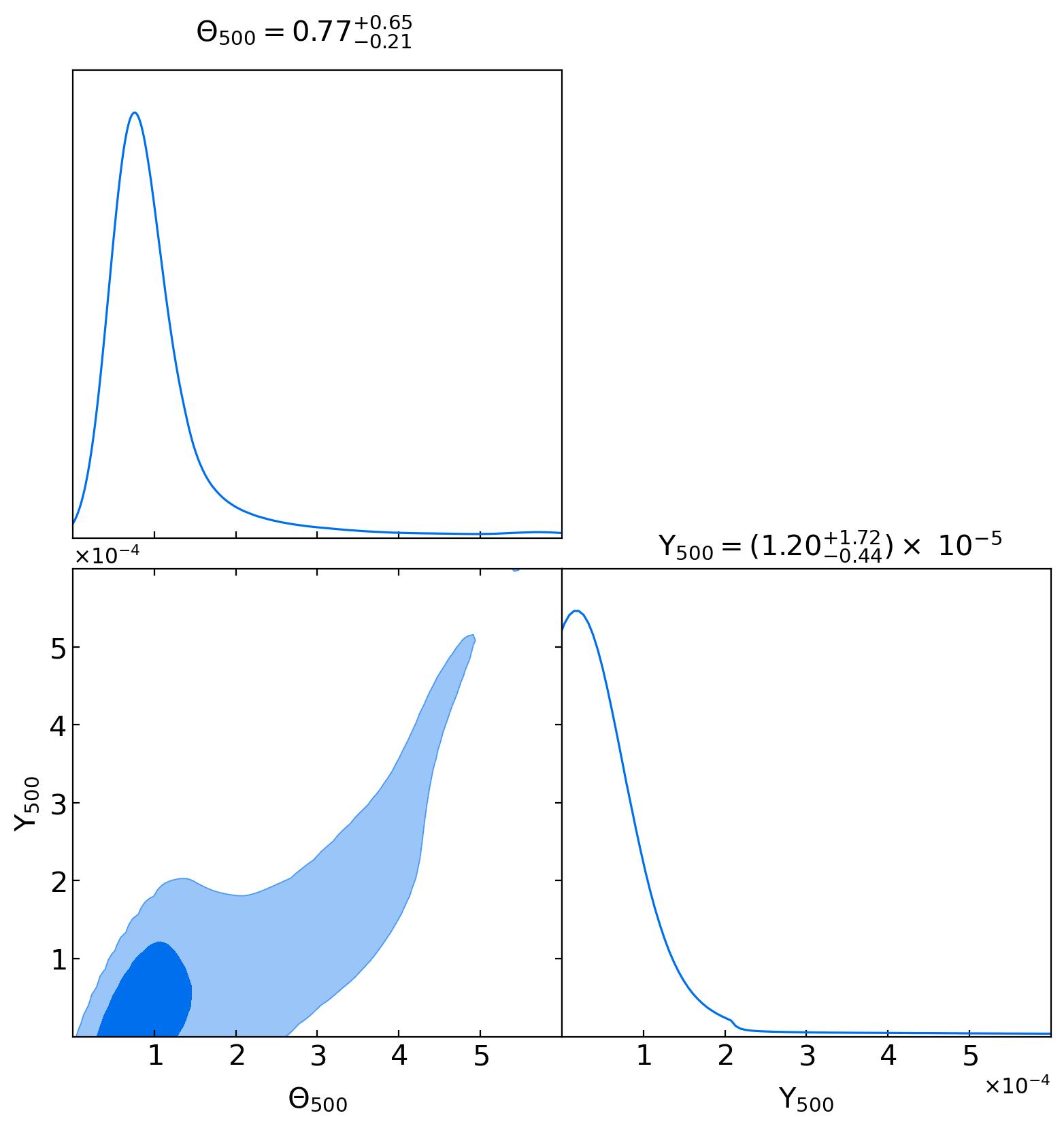}
  \caption{NK2-CL J100008.4+020908.3}
\end{subfigure}\hfill
\smallskip
\begin{subfigure}{0.3\textwidth}
  \includegraphics[width = 1\textwidth]{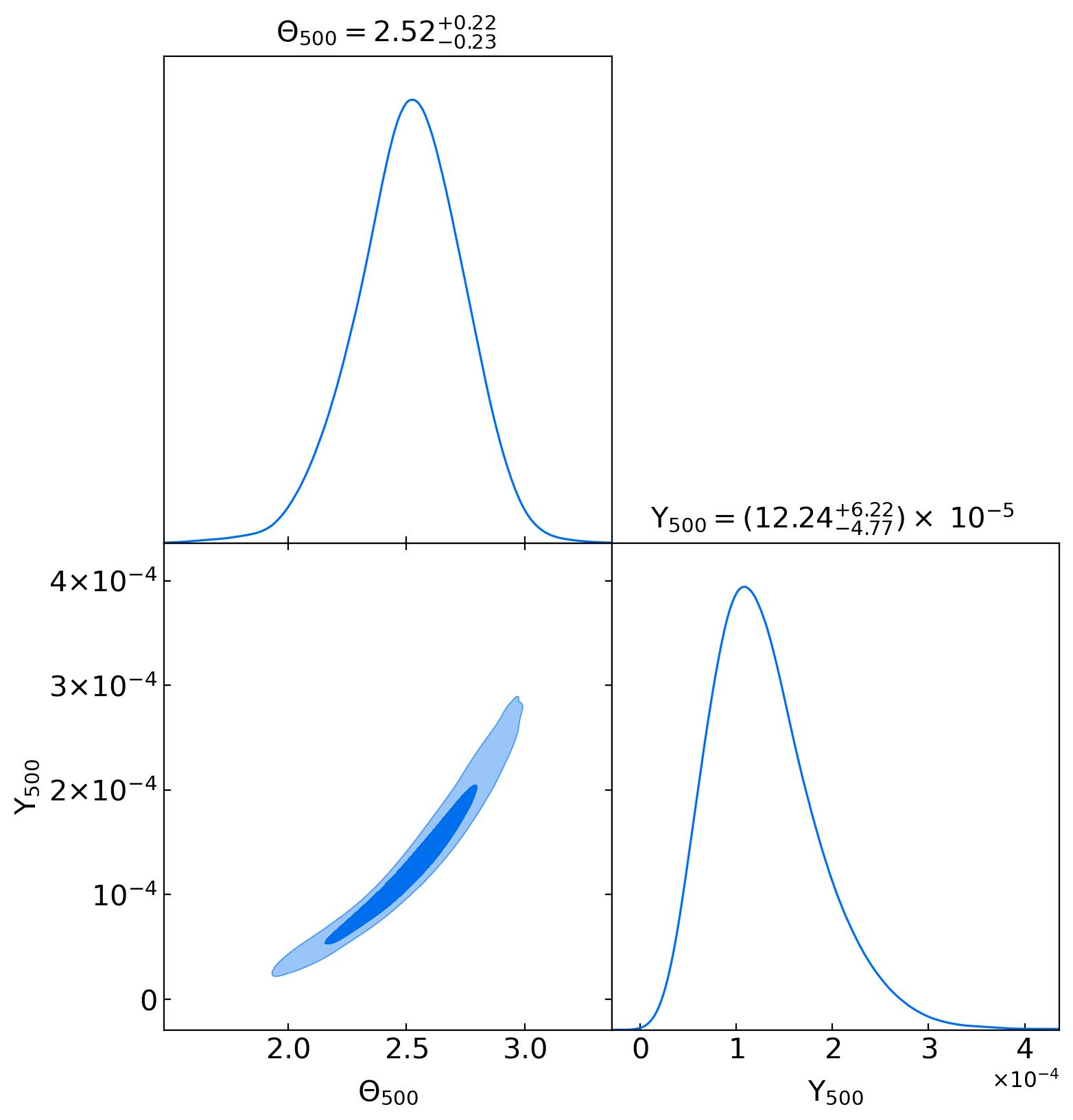}
  \caption{NK2-CL J095958.5+022910.4}
  
\end{subfigure}
\caption{continued.}
\end{figure*}

\end{appendix}

\end{document}